\documentclass[%
reprint,
superscriptaddress,
nofootinbib,
nobibnotes,
 amsmath,amssymb,
 aps,
 prd,
 longbibliography,
 twocolumn,
]{revtex4-2}

\usepackage{hyperref}
\hypersetup{
    colorlinks=true,
    linkcolor=blue,
    filecolor=magenta,      
    urlcolor=cyan,
    pdftitle={Methods},
    pdfpagemode=FullScreen,
    }

\usepackage{subcaption,epic}

\usepackage{tikz}
\newcommand{\rectangle}[2]{\tikz \fill (0,0) rectangle (#1,#2);}

\usepackage{graphicx}
\usepackage{dcolumn}
\usepackage{bm}
\usepackage{xcolor}
\usepackage{comment}

\usepackage[newcommands]{ragged2e}

\usepackage[labelformat=simple, margin=1.0mm]{subcaption}
\DeclareCaptionJustification{justified}{\justifying}

\begin{document}

\preprint{APS/123-QED}

\newcommand{\ANL}{Argonne National Laboratory (ANL), Lemont, IL, 60439, USA}
\newcommand{\Bern}{Universit{\"a}t Bern, Bern CH-3012, Switzerland}
\newcommand{\BNL}{Brookhaven National Laboratory (BNL), Upton, NY, 11973, USA}
\newcommand{\UCSB}{University of California, Santa Barbara, CA, 93106, USA}
\newcommand{\Cambridge}{University of Cambridge, Cambridge CB3 0HE, United Kingdom}
\newcommand{\CIEMAT}{Centro de Investigaciones Energ\'{e}ticas, Medioambientales y Tecnol\'{o}gicas (CIEMAT), Madrid E-28040, Spain}
\newcommand{\Chicago}{University of Chicago, Chicago, IL, 60637, USA}
\newcommand{\Cincinnati}{University of Cincinnati, Cincinnati, OH, 45221, USA}
\newcommand{\CSU}{Colorado State University, Fort Collins, CO, 80523, USA}
\newcommand{\Columbia}{Columbia University, New York, NY, 10027, USA}
\newcommand{\Edinburgh}{University of Edinburgh, Edinburgh EH9 3FD, United Kingdom}
\newcommand{\FNAL}{Fermi National Accelerator Laboratory (FNAL), Batavia, IL 60510, USA}
\newcommand{\Granada}{Universidad de Granada, Granada E-18071, Spain}
\newcommand{\IIT}{Illinois Institute of Technology (IIT), Chicago, IL 60616, USA}
\newcommand{\ICL}{Imperial College London, London SW7 2AZ, United Kingdom}
\newcommand{\Indiana}{Indiana University, Bloomington, IN 47405, USA}
\newcommand{\KSU}{Kansas State University (KSU), Manhattan, KS, 66506, USA}
\newcommand{\Lancaster}{Lancaster University, Lancaster LA1 4YW, United Kingdom}
\newcommand{\LANL}{Los Alamos National Laboratory (LANL), Los Alamos, NM, 87545, USA}
\newcommand{\Louisiana}{Louisiana State University, Baton Rouge, LA, 70803, USA}
\newcommand{\Manchester}{The University of Manchester, Manchester M13 9PL, United Kingdom}
\newcommand{\MIT}{Massachusetts Institute of Technology (MIT), Cambridge, MA, 02139, USA}
\newcommand{\Michigan}{University of Michigan, Ann Arbor, MI, 48109, USA}
\newcommand{\MSU}{Michigan State University, East Lansing, MI 48824, USA}
\newcommand{\Minnesota}{University of Minnesota, Minneapolis, MN, 55455, USA}
\newcommand{\Nankai}{Nankai University, Nankai District, Tianjin 300071, China}
\newcommand{\NMSU}{New Mexico State University (NMSU), Las Cruces, NM, 88003, USA}
\newcommand{\Oxford}{University of Oxford, Oxford OX1 3RH, United Kingdom}
\newcommand{\Pitt}{University of Pittsburgh, Pittsburgh, PA, 15260, USA}
\newcommand{\QMUL}{Queen Mary University of London, London E1 4NS, United Kingdom}
\newcommand{\Rutgers}{Rutgers University, Piscataway, NJ, 08854, USA}
\newcommand{\SLAC}{SLAC National Accelerator Laboratory, Menlo Park, CA, 94025, USA}
\newcommand{\SDSMT}{South Dakota School of Mines and Technology (SDSMT), Rapid City, SD, 57701, USA}
\newcommand{\Maine}{University of Southern Maine, Portland, ME, 04104, USA}
\newcommand{\Syracuse}{Syracuse University, Syracuse, NY, 13244, USA}
\newcommand{\TelAviv}{Tel Aviv University, Tel Aviv, Israel, 69978}
\newcommand{\UTA}{University of Texas, Arlington, TX, 76019, USA}
\newcommand{\Tufts}{Tufts University, Medford, MA, 02155, USA}
\newcommand{\VTech}{Center for Neutrino Physics, Virginia Tech, Blacksburg, VA, 24061, USA}
\newcommand{\Warwick}{University of Warwick, Coventry CV4 7AL, United Kingdom}

\affiliation{\ANL}
\affiliation{\Bern}
\affiliation{\BNL}
\affiliation{\UCSB}
\affiliation{\Cambridge}
\affiliation{\CIEMAT}
\affiliation{\Chicago}
\affiliation{\Cincinnati}
\affiliation{\CSU}
\affiliation{\Columbia}
\affiliation{\Edinburgh}
\affiliation{\FNAL}
\affiliation{\Granada}
\affiliation{\IIT}
\affiliation{\ICL}
\affiliation{\Indiana}
\affiliation{\KSU}
\affiliation{\Lancaster}
\affiliation{\LANL}
\affiliation{\Louisiana}
\affiliation{\Manchester}
\affiliation{\MIT}
\affiliation{\Michigan}
\affiliation{\MSU}
\affiliation{\Minnesota}
\affiliation{\Nankai}
\affiliation{\NMSU}
\affiliation{\Oxford}
\affiliation{\Pitt}
\affiliation{\QMUL}
\affiliation{\Rutgers}
\affiliation{\SLAC}
\affiliation{\SDSMT}
\affiliation{\Maine}
\affiliation{\Syracuse}
\affiliation{\TelAviv}
\affiliation{\UTA}
\affiliation{\Tufts}
\affiliation{\VTech}
\affiliation{\Warwick}

\author{P.~Abratenko} \affiliation{\Tufts}
\author{O.~Alterkait} \affiliation{\Tufts}
\author{D.~Andrade~Aldana} \affiliation{\IIT}
\author{L.~Arellano} \affiliation{\Manchester}
\author{J.~Asaadi} \affiliation{\UTA}
\author{A.~Ashkenazi}\affiliation{\TelAviv}
\author{S.~Balasubramanian}\affiliation{\FNAL}
\author{B.~Baller} \affiliation{\FNAL}
\author{A.~Barnard} \affiliation{\Oxford}
\author{G.~Barr} \affiliation{\Oxford}
% ...

\collaboration{The MicroBooNE Collaboration}
\thanks{microboone\_info@fnal.gov}\noaffiliation

\title{Data-driven model validation for neutrino-nucleus cross section measurements }

\date{\today}

\begin{abstract}
Neutrino-nucleus cross section measurements are needed to improve interaction modeling to meet the precision needs of neutrino experiments in efforts to measure oscillation parameters and search for physics beyond the Standard Model. We review the difficulties associated with modeling neutrino-nucleus interactions that lead to a dependence on event generators in oscillation analyses and cross section measurements alike. We then describe data-driven model validation techniques intended to address this model dependence. The method relies on utilizing various goodness-of-fit tests and the correlations between different observables and channels to probe the model for defects in the phase space relevant for the desired analysis. These techniques shed light on relevant mis-modeling, allowing it to be detected before it begins to bias the cross section results. We compare more commonly used model validation methods which directly validate the model against alternative ones to these data-driven techniques and show their efficacy with fake data studies. These studies demonstrate that employing data-driven model validation in cross section measurements represents a reliable strategy to produce robust results that will stimulate the desired improvements to interaction modeling.
\end{abstract}

\maketitle

\section{Introduction}\label{sec:intro}

The desire to measure neutrino-nucleus cross sections is motivated by the needs of modern neutrino experiments. Moving forward, precision measurements of muon-to-electron neutrino oscillations~\cite{T2K:2023smv,NOvA:2021nfi,DUNE:2020ypp,Hyper-KamiokandeProto-:2015xww} will enable the characterization of charge-parity violation in the neutrino sector~\cite{Nunokawa:2007qh}, the determination of the neutrino mass ordering~\cite{Qian:2015waa}, and searches for physics beyond the Standard Model. These oscillations are studied through measurements of neutrino-nucleus interactions, which represent the nucleus’s response to a neutrino probe~\cite{Diwan:2016gmz}. In order to interpret these measurements properly and to disentangle any new physics from background Standard Model processes~\cite{DUNE:2020ypp,Hyper-KamiokandeProto-:2015xww}, this data must be accompanied by precise modeling of neutrino-nucleus scattering in the $\sim$GeV energy region~\cite{Formaggio:2012cpf} benchmarked with rigorous cross section measurements.

Neutrino-nucleus interactions present a challenging theoretical problem. They involve both the electroweak force and the strong force, which is non-perturbative in the relevant energy regime~\cite{Gross:2022hyw}, all within the complex multi-body environment of the nucleus. This results in an incomplete theoretical description of neutrino-nucleus interactions in the $\sim$GeV regime. However, these challenges do not necessarily prevent the success of accelerator-based neutrino oscillation experiments~\cite{T2K:2023smv,NOvA:2021nfi}. As long as the nucleus’s response to the neutrino probe can be described with sufficient detail, the desired precision can still be achieved in oscillation measurements. For this purpose, experiments utilize event generators, which simulate neutrino interactions through a collection of effective models constructed to explain different modes of neutrino-nucleon interactions~\cite{Formaggio:2012cpf} and are used to estimate event selection efficiencies and detector responses. In this light, neutrino-nucleus interaction cross section measurements are calibration points which help ensure that simulations provide a robust description of nature.

However, this naturally raises the question of model dependence in cross section measurements, which likewise utilize event generators to correct for backgrounds, efficiencies, finite resolution, and biases in the reconstruction of kinematic quantities. The process of ``extracting" or ``unfolding" the cross section, which constitutes mapping the reconstructed distributions onto physics quantities, assumes that the model captures the true value of these corrections within its uncertainties. Though this is required in order to obtain a robust result, the validity of this assumption is not known a priori, and must be verified in any cross section measurement. 

To address this issue of model dependence, we propose utilizing a data-driven model validation procedure to test whether the model, together with its uncertainties, can describe the data in a self-consistent manner. The validation is based upon constructing a variety of data-driven tests in order to identify mis-modeling relevant to the desired cross section measurement. When the model passes validation, it suggests that the data is a suitable realization of the range of possibilities afforded by the model's uncertainties. We demonstrate that, in general,  when this condition is met, any bias introduced in the cross section extraction will be within the quoted uncertainties of the measurement, thereby building confidence in a robust result. The MicroBooNE experiment has previously used these data-driven techniques in a variety of analyses~\cite{MicroBooNE:2021sfa,MicroBooNE:2023foc,numuCC0pNp_PRD,numuCC0pNp_PRL}. These results include measurements of visible kinematic variables, such as the energy and angle of the outgoing muon in a charged current muon neutrino ($\nu_\mu$CC) interaction, as well as measurements of derived quantities, such as the cross section as a function of the incoming neutrino energy or the energy transferred to the nucleus. We emphasize that these techniques can be employed equally to extract neutrino energy dependent cross sections and to extract cross sections as a function of visible variables.

This paper is organized as follows. In Sec.~\ref{sec:import}, we motivate why model validation is critical in all cross section measurements and explore where model dependence may arise. In Sec.~\ref{sec:validation}, we describe various techniques which may be used to detect relevant mis-modeling and general considerations for designing a sufficiently sensitive model validation procedure. In Sec.~\ref{sec:fds}, we compare the usage of the fake data sets in two cases: one in evaluating the model uncertainties and the other one on the model validation procedure. We then present fake data studies (FDSs) as described in the latter case to demonstrate the efficacy of the data-driven model validation procedure. These points are then summarised in Sec.~\ref{sec:summary}.

\section{Importance of model validation}\label{sec:import}

\subsection{A Priori Information About Event Generators}
\label{sec:generators}
The complex nature of neutrino-nucleus scattering in the $\sim$GeV energy regime necessitates the use of effective models to describe these interactions. Event generators are formed from a collection of these effective models and are used by neutrino experiments to simulate neutrino interactions and interpret experimental data. A variety of generator codes exist, including $\texttt{GENIE}$~\cite{GENIE}, $\texttt{NEUT}$~\cite{neut}, $\texttt{NuWro}$~\cite{nuwro} and $\texttt{GiBUU}$~\cite{gibuu2}. Though these generators are built upon similar underlying theory and may even employ some of the same models, the details of the implementation and choice of model parameters can have a large impact on the generator's prediction and its ability to describe data. Despite their crucial role, each effective model generally suffers from not being able to describe the corresponding interaction modes across the complete phase space~\cite{GenCompare,golobal_tki}.  This poses issues for experiments because event generators are often required to be capable of describing the complete contents of the final state and to provide coverage over the entirety of the available phase space, but usually require substantial interaction uncertainties to do so. 

This motivates the need for data-driven inputs to inform the allowed parameters for these models and reduce their uncertainties. Experiments often supplement generators with tailored “tunes” to better represent their data~\cite{MicroBooNE:2021ccs,minerva_tune_pion,nova_tune,neut_tune,nu_A_tune_genie,nu_N_tune_genie}. The success of this strategy in fulfilling the requirement of generating complete final-state particle kinematics in neutrino-nucleus interactions has been demonstrated with the consistent $|\Delta m^2_{\mathrm{atm}}|$ values extracted from the accelerator neutrino oscillation experiments~\cite{T2K:2023smv,NOvA:2021nfi} and reactor antineutrino oscillation experiments~\cite{DayaBay:2022orm,RENO:2018dro}, in which the inverse $\beta$ decay process allows for a simpler and more precise reconstruction of neutrino energy. Nevertheless, because of their hybrid nature, event generators tend to better describe inclusive processes or phase spaces where sufficient data were accumulated. For other exclusive processes, which require a more detailed description of the hadronic final states, event generators are more likely to fall short.

This places experiments in the following situation. On one hand, it is unlikely the parameters available in current event generators are sufficient in describing interaction modes in the complete phase space. On the other hand, the conservative uncertainties assigned on these parameters partially mitigate this shortcoming but often lead to large systematic uncertainties. As such, cross section measurements that treat the reliance on event generators with care remain essential in stimulating improvements to simulation that will enable the desired level of precision in current and future neutrino experiments.

\subsection{Sources of Model Dependence}
Extracting cross section measurements from experimental data constitutes mapping reconstructed distributions to truth counterparts that can be more readily compared to external predictions. In this process, an overall model, generally consisting of flux, detector, and interaction models, is required to estimate the mapping from reconstructed quantities to true quantities. This leaves such measurements susceptible to model dependence, hence the need for model validation. 

To see more explicitly how such a dependence arises, consider the general equation describing the process of extracting a flux-averaged differential cross section, $\frac{d\sigma}{dx}$, as a function of a given truth variable $x$ from a measured distribution $n_a=d_a-b_a$, where, for the $a$th bin, $d_a$ is the number of selected events, $b_a$ is the background prediction, and $n_a$ is the estimated number of selected signal events. In generic terms, this equation takes the form
\begin{equation}
\label{eq:extract}
\left(\frac{d\sigma}{dx}\right)_\beta =
\frac{\Sigma_a U_{\beta a}\left(d_a-b_a\right)}
{\Phi\cdot T \cdot \varepsilon_\beta \cdot (\Delta x)_\beta},
\end{equation}
where $U_{\beta a}$ is the smearing matrix describing the predicted probability that an event in reconstructed bin $a$ belongs to truth bin $\beta$, $\varepsilon_\beta$ is the estimated selection efficiency for signal events in truth bin $\beta$, $(\Delta x)_\beta$ are the widths of the truth bins, and $T$ is the number of target nuclei. The total integrated flux prediction $\Phi$ is the integral of the predicted neutrino flux $\phi(E_\nu)$ over the entire neutrino energy spectrum
\begin{equation}
    \Phi = \int \phi(E_\nu)~dE_\nu .
\end{equation}
In the case of extracting the cross section as a function of the neutrino energy, $\sigma(E_\nu$), the integrated flux prediction acquires a dependence on the truth bin, $\Phi_\beta$, to account for the fact that any truth bin only corresponds to a specific subset of the neutrino flux spectrum.

From Eq.~\ref{eq:extract}, it is clear that cross section extraction depends on a model prediction for the neutrino flux, the rate of signal and background events, the selection efficiency, and detector effects that result in the imperfect reconstruction of kinematic quantities accounted for in the smearing matrix. This dependence on the overall model means that mis-modeling in the phase space relevant to the cross section extraction can introduce bias into the measurement. For measurements that must contend with low efficiencies, purities, or substantial detector effects, this phase space may extend beyond the measured one into background channels, where mis-modeling could lead to an incorrect background correction, or into channels and observables adjacent to the measurement where mis-modeling could suggest an incorrect efficiency estimation. The fundamental challenge here is that one does not know if the model used in the extraction can describe nature. Moreover, when the overall model is unable to describe nature, it is usually unclear if the discrepancy is due to detector, cross section, or flux effects and may even be due to a combination of a variety of sources of mis-modeling. In these cases, there may be a need for additional uncertainties beyond those inside the model, and therefore a form of model validation is required to verify that the existing uncertainties provide sufficient coverage of the data.

Model validation is especially important to any analysis that extracts cross sections as a function of the neutrino energy or energy transferred to the nucleus. These quantities are not directly observable and must be estimated from the measurement of the visible leptonic and hadronic energy. The way the unfolding maps from the reconstructed hadronic energy to the true energy transfer depends on the overall model, particularly the cross section model, to correct for the missing hadronic energy going to particles that cannot be reconstructed by the detector. Care must be taken to avoid introducing model dependence that biases these measurements beyond stated uncertainties, and a rigorous examination of the model is essential.

Visible kinematic variables do not entirely avoid model dependence either. Whenever a measurement relies upon mapping from reconstructed quantities to true quantities this mapping must be estimated through a model, motivating the need for model validation in these cases. 
In particular, the mapping for quantities like the available energy $E_{\mathrm{avail}}$~\cite{minerva_Eavail1,minerva_Eavail2,minerva_Eavail3,minerva_Eavail4,numuCC0pNp_PRD,low_nu,GenCompare}, often defined as the sum of reconstructable energy deposited by visible particles, serves as an alternative to the energy transfer but has a strong dependence on the accurate simulation of particles that deposit energy in the detector. Since different modeling and reconstruction failures may be present in different final states, the mapping from reconstructed to true $E_{\mathrm{avail}}$ requires a robust description of the complete contents of the hadronic final state, thereby making it susceptible to model dependence.

Similar forms of mis-modeling may impact measurements of visible kinematic variables that depend on the final state hadronic kinematics. Measurements of differential cross sections for more exclusive final states are also susceptible to mis-modeling as they generally impose a detection threshold based on the reconstruction performance of the detector. Because selection efficiencies can be a complex function of the particle kinematics and contents of the final state, substantially different modeling of nuclear effects between generators can lead to drastically different predictions for the number of above threshold particles~\cite{GenCompare}. This can have a large impact on the estimated selection efficiencies, background predictions, and bin migration effects, thereby introducing model dependence into these results.

\subsection{Real versus Nominal Flux}
Measurements of differential cross sections inherently depend on the mapping between neutrino energy and the visible kinematic variable when unfolding or generating predictions. Due to the broad energy spectrum of neutrino beams, cross section measurements are typically averaged over the integral of the entire flux spectrum, $\Phi$, as described in Eq.~\ref{eq:extract}. This may reduce the dependence on this mapping, but does not entirely remove it. For example, a 1~GeV neutrino cannot produce a muon with 2~GeV of kinetic energy. Therefore, despite being averaged over the entire flux, any given muon energy measurement bin naturally maps to a sub-range of the neutrino energy spectrum. This introduces model dependence related to the neutrino flux prediction. 
 
The extent of this model dependence depends on subtleties in the treatment of the flux and its related uncertainties. In the literature, this has been described as whether the measurement is averaged over an assumed well defined \textit{nominal} neutrino flux spectrum, or averaged over the unknown \textit{real} neutrino flux spectrum impingent on the detector~\cite{Koch:2020oyf}. When extracting cross sections in the real flux, unfolding amounts to correcting for detector effects with a presumed cross section and flux model. When extracting cross sections in a nominal flux, one corrects for detector effects but then also extrapolates from the (unknown) real flux to the nominal flux in the unfolding. This distinction is subtle but leads to important differences in the treatment of flux uncertainties. These uncertainties can be quite prominent in the context of accelerator-based neutrino cross section measurements, for which flux uncertainties are typically 5-10\%, though can be even larger in instances where the flux is poorly known.

The distinct advantage of cross section extraction in the real neutrino flux spectrum is that it minimizes the usage of the mapping between neutrino energy and visible kinematic variables in the extraction of cross sections. This makes these results less model-dependent. However, extractions in the real neutrino flux spectrum do not eliminate the dependence on the mapping from neutrino flux to observables but, rather, push it down the line to future analyzers of the data who are required to supply their own flux model and associated uncertainties when making comparisons. In other words, theorists must provide their own model of the mapping, which is usually just the prediction of the nominal flux and its uncertainties at the measurement location as reported by the experiment. However, in order to make a robust comparison between predictions and the data, these flux uncertainties should include correlations between the assumed neutrino energy spectrum and the extracted cross section result, which typically already include flux normalization uncertainties. 
These correlations are generally not reported by the experimental collaboration and may lead to incorrect conclusions about how well alternative models describe the data~\cite{Koch:2020oyf}.

If an analysis extracts cross sections in the nominal neutrino flux spectrum, model-dependence arises when the mapping between neutrino energy and the visible kinematic variable is used to extrapolate from the real to the nominal neutrino spectrum. In this approach, flux uncertainties are estimated by varying the assumed real flux, but not the well-defined nominal flux, when determining the impact of flux effects. Referring to Eq.~\ref{eq:extract}, this amounts to keeping $\Phi$ constant while re-evaluating $n_a$ based on the real flux in each flux systematic universe. This approach allows the covariance matrix to include the uncertainties of extrapolating the data from the unknown real neutrino flux to the nominal flux and corresponding $\Phi$ that the results are averaged over. In this case, comparisons to external theory or event generator predictions are straightforward because there is no need for future analyzers of the data to utilize additional flux uncertainties. Flux systematics are entirely accounted for by the experimentalist when estimating uncertainties for the cross section extraction. In this case, data-driven model validation can serve as a mechanism to quantify the bias introduced in the extrapolation from real to nominal flux and help avoid under-estimating or over-estimating flux uncertainties. This is similar to the utility of data-driven model validation when extracting derived quantities dependent on the neutrino flux spectrum, such as the cross section as a function of the neutrino energy or energy transferred to the nucleus, where it likewise serves to validate the mapping between reconstructed quantities and the incoming neutrino energy spectrum. The dependence on the model is greater for these derived quantities, hence a greater need for careful validation, but one should still be cautious in entirely ignoring this model dependence in the context of visible kinematic variables.

To reiterate, the primary difference between measurements in the real and nominal flux is the amount they depend on the mapping from neutrino energy to visible kinematics variables and their treatment of flux shape uncertainties, which are fully included in the covariance matrix extracted in a nominal flux measurements but not in a real flux measurement. The challenges of comparing predictions to measurements made in the real flux are described in Ref.~\cite{Koch:2020oyf}. Here, we illustrate those challenges with a toy example. We consider an ideal $\nu_\mu$CC cross section measurement as a function of the muon energy with perfect efficiency, no background, and perfect muon energy reconstruction. In this case, the smearing matrix $U_{\beta a}$, background $b_a$, and the efficiency $\varepsilon_\beta$ 
disappear from Eq.~\ref{eq:extract} leaving just the reconstructed signal event counts $n_a$, which are now exactly equal to the true signal event counts $n_\beta$, the integrated flux prediction $\Phi$, the number of target nuclei $T$, and the bin widths $\left(\Delta E_\mu\right)_\beta$:
\begin{equation}
\left(\frac{d\sigma}{dE_\mu}\right)_\beta=
\frac{n_\beta}
{\Phi\cdot T \cdot \left(\Delta E_\mu\right)_\beta}.
\end{equation}
The only uncertainties on the extracted cross section from this equation are statistical uncertainties on $n_\beta$ and an uncertainty on the integrated neutrino flux prediction $\Phi$. For the latter, we use the MicroBooNE $\nu_\mu$ flux with systematic uncertainties taken from Ref.~\cite{MiniBooNEFlux}. We then generate fake data distributions that fluctuate $n_\beta$ according to statistical and flux model uncertainties in the same 11 true muon energy bins used in Ref.~\cite{MicroBooNE:2021sfa}. This is performed by sampling the multivariate Gaussian distribution using a singular value decomposition, as well as Poisson statistical fluctuations.

In this toy study, we assume the true cross section is exactly the prediction from $\texttt{GENIE}$ with CCQE and CC2p2h parameters set according to~\cite{MicroBooNE:2021ccs}, which will be referred to as the ``MicroBooNE tune". The real flux used to produce the observed $n_\beta$ before additional statistical fluctuations is different in each fake data set. The flux prediction used in the cross section extraction is kept constant across all universes and corresponds to the central value (CV) MicroBooNE $\nu_\mu$ flux prediction. In this case, besides statistical fluctuations, any deviation in the results from the MicroBooNE tune CV is due to systematic fluctuations in the flux model.

Three different methods are utilized to compare the real flux-averaged cross section result to prediction; the ``Incorrect method", the ``Flawed method" and the ``Correct method". The ``Incorrect method" includes no additional flux uncertainties when comparing the extracted cross section with the prediction. This method is easy for analyzers to perform as it only requires them to supply a CV for the flux. The ``Flawed method" includes flux uncertainties on the prediction, but does not account for correlations between the extracted cross section results and the assumed neutrino energy spectrum. This is achieved by generating many alternative predictions, each of which uses a unique flux drawn according to the uncertainties on the flux model, and using them to build a covariance matrix. The ``Flawed method" is more difficult to perform, since it requires future analyzers of the data to have access to and utilize a published flux covariance matrix to derive an uncertainty for their prediction. The ``Correct method" is similar to the ``Flawed method" and also takes into account the flux uncertainty in the generator prediction by building a covariance matrix from predictions obtained with the flux model varied according to its uncertainties. The difference is that in the ``Correct method" one includes the correlations between the integrated flux normalization uncertainty in the extraction and the flux uncertainty in the prediction.

\begin{figure}[ht!]
\centering
 \begin{subfigure}{\linewidth}
    \includegraphics[width=0.88\linewidth]{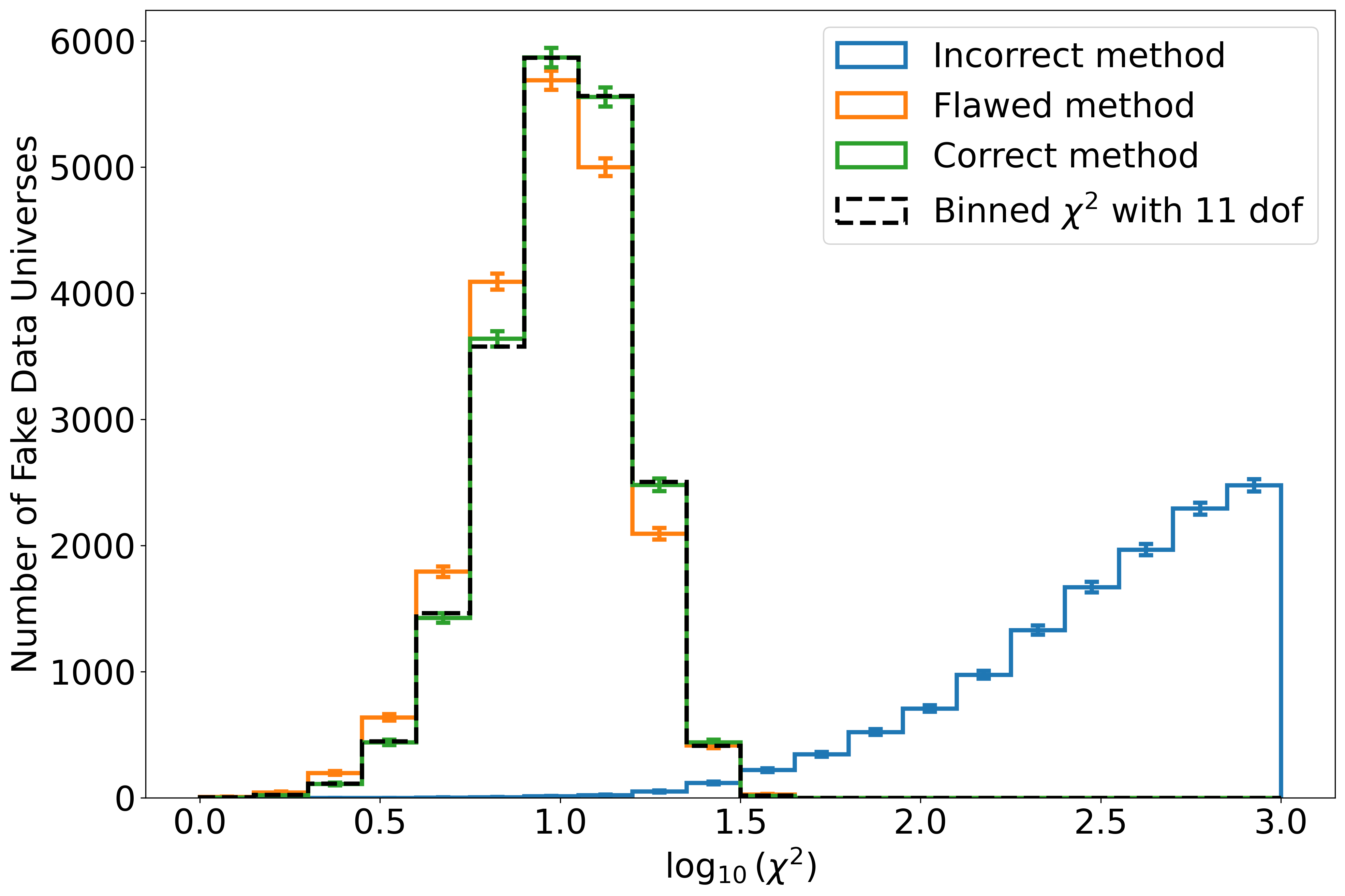}
    \caption{\label{fig:flux_toy_distibution}}  
\end{subfigure}
\begin{subfigure}{\linewidth}
    \includegraphics[width=0.88\linewidth]{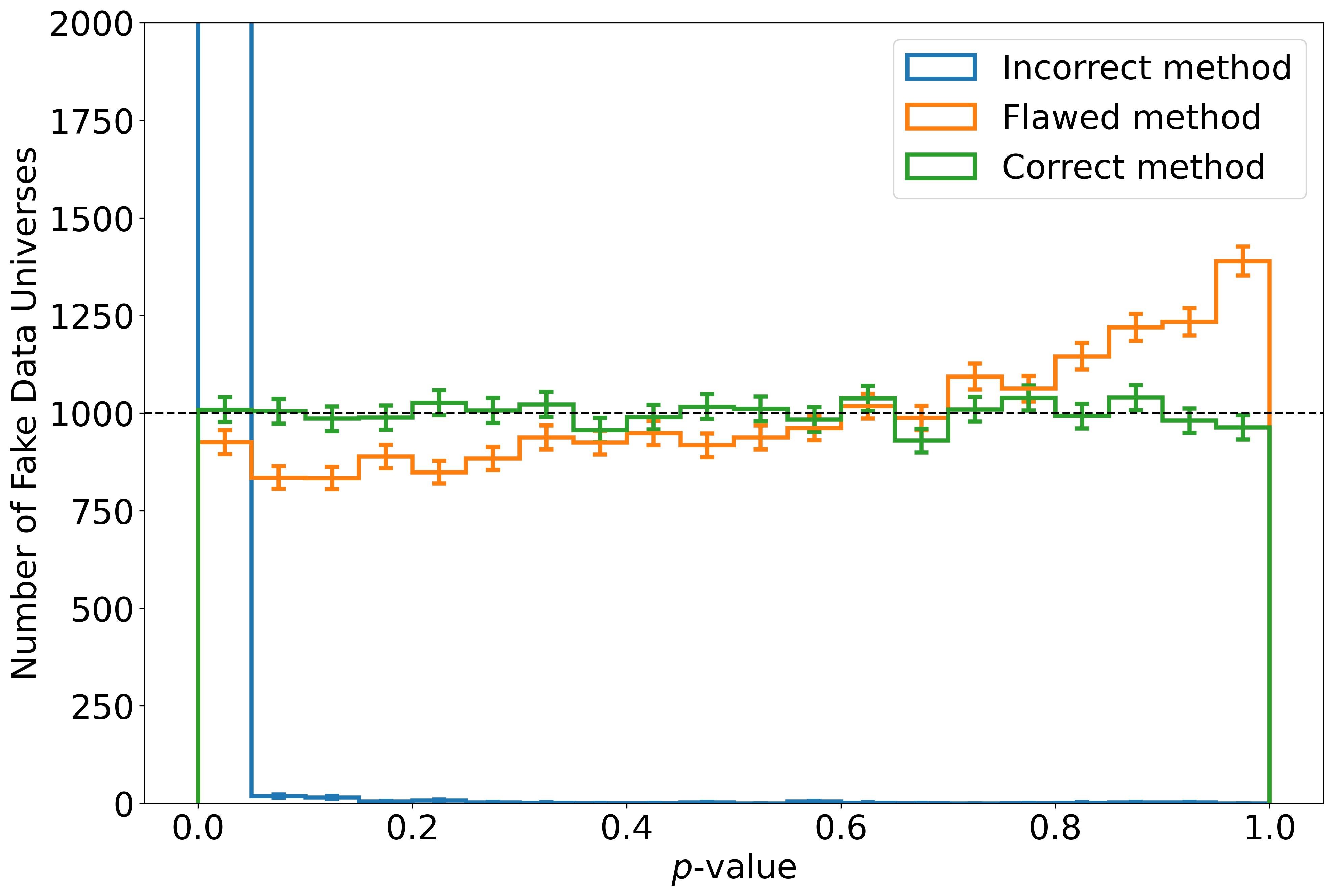}
    \caption{\label{fig:flux_toy_pvalues}}
\end{subfigure}
\begin{subfigure}{\linewidth}
    \includegraphics[width=0.88\linewidth]{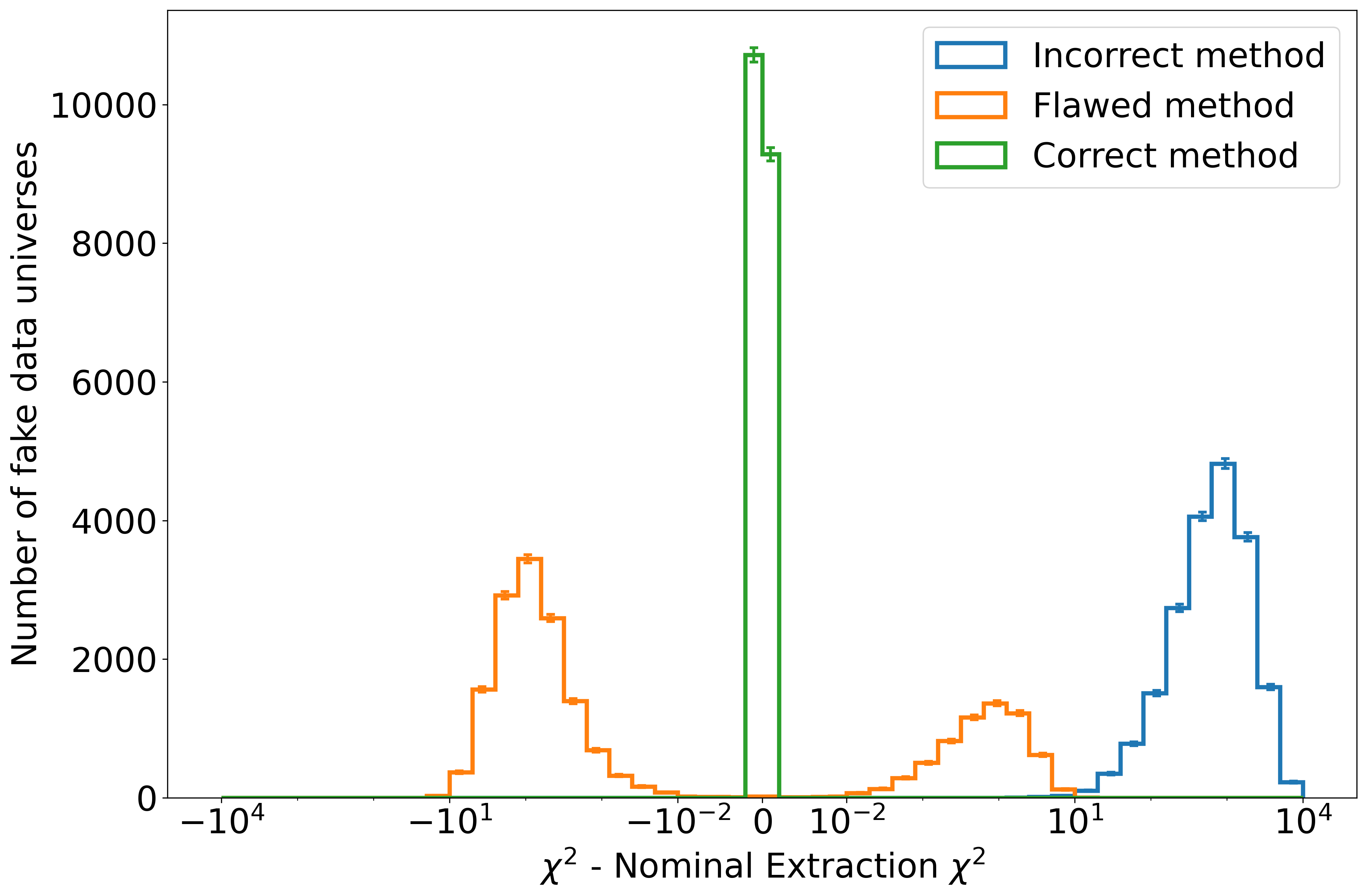}
    \caption{\label{fig:flux_toy_diff}}
\end{subfigure}
\caption{\label{fig:flux_toy}Toy study illustrating three methods of comparing a real flux measurement with prediction. The ``Incorrect method" neglects flux uncertainties on the prediction. The ``Flawed method" includes flux uncertainties on the prediction, but neglects flux correlations between the prediction and extraction. The ``Correct method" includes flux uncertainties on the prediction and correlations with the extracted result. The distributions of $\chi^2$ values obtained for each method is shown in (a) alongside a binned $\chi^2$ distribution. The corresponding distribution of $p$-values is shown in (b). The distribution of the difference between $\chi^2$ values and the nominal flux extraction $\chi^2$ obtained for each method is shown in (c).}
\end{figure}

In Fig.~\ref{fig:flux_toy}, we illustrate the three methods by comparing the extracted real flux-averaged cross sections to a generator prediction that assumes the CV of the MicroBooNE flux model. The associated uncertainties on the real flux measurement and central value of the extracted cross section are the same in each method. The differences arise from the treatment of flux uncertainties for the prediction. As an additional point of reference, we verify that the ``Correct" method gives identical $\chi^2$ values to those extracted with the nominal flux technique, within numerical errors. Figure~\ref{fig:flux_toy_distibution} shows the distribution of $\chi^2$ values obtained when the extracted cross section from each of 10000 fake data sets is compared to the MicroBooNE tune prediction, which corresponds to the truth for these fake data sets. Figure~\ref{fig:flux_toy_pvalues} shows the same, but with the $\chi^2$ values converted into $p$-values. Figure~\ref{fig:flux_toy_diff} expands this comparison further by showing the difference between $\chi^2$ values obtained with the various methods instead.

The ``Incorrect method", though easiest to perform, generates a distribution of $\chi^2$ values shifted towards larger values than expected from a $\chi^2$ distribution with 11 degrees of freedom. This arises from the fact that, in a given fake data set, there may be a sizable difference between the real flux and the CV flux, which produces a large $\chi^2$ value. The possibility of this difference is not accounted for due to the fact that flux uncertainties were neglected on the supplied flux model and thus this mis-modeling of the flux is wrongly attributed to defects in the MicroBooNE tune prediction.  

For the ``Flawed method", the distribution of $\chi^2$ values is significantly closer to what is expected from a $\chi^2$ distribution. However, it is still shifted to smaller values than the $\chi^2$ distribution with 11 degrees of freedom. This is the result of double counting the flux uncertainty, which is done on both the extraction and prediction side. 
 
The ``Correct method" method generates a distribution of $\chi^2$ values that agrees with a $\chi^2$ distribution with 11 degrees of freedom. However, in most current real-flux-averaged cross section data releases, this method is impossible for analyzers of the data to perform, since it requires additional information about how the extracted cross section is correlated with flux spectrum variations. This information is rarely provided by experimentalists, rendering the ``Correct method" of comparing cross sections extracted in the real neutrino flux to external prediction likely impossible.

The deviation between the three methods is further illistrated in Fig.~\ref{fig:flux_toy_diff}, which shows the distribution of the universe by universe differences between $\chi^2$ values obtained for the measurement reported in the nominal flux and those obtained for the measurement reported in the real flux. From this figure, it is apparent that  the ``Incorrect Method" consistently produces a larger than accurate $\chi^2$. Though the ``Flawed Method" produces a distribution of $\chi^2$ values shifted towards smaller values than a $\chi^2$ distribution with 11 degrees of freedom, it can still produce a larger $\chi^2$ value than obtained for the cross section extracted in the nominal flux. Since the nominal flux method is able to reproduce a $\chi^2$ distribution with 11 degrees of freedom, this suggests that the ``Flawed Method" does not strictly underestimate the $\chi^2$. The aforementioned phenomena are also described in~\cite{Koch:2020oyf}. These observations further complicate the use of real flux-averaged measurements as they prevent one from using the ``Flawed Method" and ``Incorrect Method" as lower and upper bounds on the GoF between data and prediction and necessitates the use of the ``Correct Method" for a rigorous comparison.

As demonstrated by this toy study, the requirements of reporting and using neutrino flux uncertainties and their correlations with cross section results make it very difficult, if not impossible, for theorists to properly compare their predictions with cross sections extracted using the real neutrino spectrum. Indeed, this may be related to some of the issues encountered in recent efforts to tune event generators to experimental data~\cite{MicroBooNE:2021ccs,PPP_nuwro,nu_A_tune_genie}. Because of this, we advocate for extracting cross sections in the nominal neutrino flux spectrum, which allows the results to be reported in a single well-defined flux. With this method, all flux uncertainties are included in the covariance matrix obtained in the unfolding. However, this method also introduces additional model dependence on the experimental side associated with the mapping between the neutrino energy spectrum and the measured kinematic variables. We thus advocate for data-driven model validation when extracting cross sections in any physics variable at the nominal neutrino flux spectrum. These methods are well suited to examine the entire uncertainty budget and thus help to ensure that the uncertainties associated with extrapolating from the true to the nominal neutrino flux spectrum are sufficient.

\section{Data-Driven Model Validation}\label{sec:validation}
Data calibration and data-driven model validation are closely related. Instead of using data to replace part of the model in a calibration procedure, the model validation procedure focuses on testing whether the model used for the extraction can describe the data in a self-consistent manner. When the validation indicates that the data falls within the allowed parameter space of the model, this builds confidence that the bias introduced in the cross section extraction will, in general, be within the quoted uncertainties of the measurement. This is demonstrated with several case studies in Sec.~\ref{sec:fds}. The data-driven methods we propose are in contrast with other approaches to model validation, which usually examine the variation between multiple different model predictions for backgrounds, efficiencies, or biases in the reconstruction of kinematic quantities. This is commonly done through FDSs used to inform additional uncertainties to be added to the primary model used for extracting the data cross sections. Differences between the role of FDSs in model validation based on comparisons with alternative models and the data-driven model validation we propose is discussed in more detail in Sec.~\ref{sec:fds_compare}. 

\subsection{Tools for Model Validation}

\subsubsection{Goodness of Fit Tests}

The data-driven model validation procedure is based on comparing the model prediction to the data with goodness of fit (GoF) tests that quantify the ability of the overall model to describe the data. Any GoF tests performed over a reconstructed space distribution should be evaluated in such a way that correlations between bins are accounted for. This can be achieved with a $\chi^2$ test statistic constructed via the covariance matrix formalism given by
\begin{equation}\label{eq:chi2}
\chi^2 = (M-P)^T \cdot V^{-1} \cdot (M-P),
\end{equation}
where $M$ is the measurement vector, 
$P$ is the prediction vector, and $V$ is the total covariance matrix which includes the uncertainties on the reconstructed distribution and bin-to-bin correlations. These $\chi^2$ values are interpreted by using the number of degrees of freedom, $ndf$, which corresponds to the number of bins, to obtain $p$-values. 

To obtain sufficient stringency, we require that all tests which probe the model in the phase space relevant for the cross section extraction yield a $p$-value greater than 0.05, which indicates that the model is able to describe the data at the $2\sigma$ significance level. If all tests pass this level of stringency, then the model is considered to be validated and may be used for the cross section extraction. In this case, given a set of models that pass validation, the discrepancy between the results extracted using different models are expected to be smaller than the total uncertainties. 

\subsubsection{$\chi^2$ Decompositions}

In an overall GoF test, it is possible that conservative uncertainties have hidden a significant discrepancy between data and model in some bins that may bias the cross section extraction in select regions of phase space. However, it is challenging to further evaluate the GoF in select regions of phase space because there are generally strong bin-to-bin correlations in the reconstructed distributions. To address this, the overall test statistic can be decomposed by diagonalizing the covariance matrix, thereby making all bins uncorrelated. In this transformed basis, the absence of correlations allows for a rigorous quantification of the GoF between the data and model prediction within a single bin given the uncertainties of the diagonalized covariance matrix. This allows one to test the local GoF of distributions in the decomposed space. Though it is challenging to draw physics conclusion from this decomposed space, the technique is a useful statistical tool in evaluating the compatibility between data and prediction when significant correlations in reconstructed distributions may render the overall $\chi^2$ test statistic insensitive to local discrepancies.

In particular, the symmetric covariance matrix can be decomposed into $V = \tilde{Q} \cdot \Lambda \cdot \tilde{Q}^T$ where $\Lambda$ contains the eigenvalues of $V$ along its diagonal and $\tilde{Q}$ has the corresponding eigenvectors as its columns. Defining $Q = \tilde{Q}^{-1}$ and $\Delta = (M-P)$ allows Eq.~\ref{eq:chi2} to be written as
\begin{equation}
\label{eq:dchi2_begin}
\chi^2 = (Q \cdot \Delta)^T \cdot (Q \cdot V \cdot Q^T)^{-1} \cdot (Q \cdot \Delta).
\end{equation}
By further defining $\epsilon_i = \Delta'_i/\sqrt{\Lambda_{ii}}$, where $\Delta' = Q \cdot \Delta$, the above expression can be written as 
\begin{equation}\label{eq:dchi2}
\chi^2 = \Delta'^T \cdot \Lambda^{-1} \cdot \Delta' = \sum_i \epsilon_i^2.
\end{equation}
Because $\Lambda$ is diagonal, the $\epsilon_i$ are all independent, and the $\chi^2$ is now written in terms of independent components, also known as the $\chi^2$ decomposition format. These $\epsilon_i$ are normally distributed and may be interpreted as the significance of the tension between data and prediction in the corresponding $i$th bin of the eigenvalue basis. Compared to the deviations on individual bins in the correlated reconstructed space distribution, the deviations between data and model in $\epsilon_i$ take into account the correlations and can be individually evaluated quantitatively in a consistent manner. 

A local $\chi^2$ and corresponding $p$-value, $\chi^2_{\mathrm{local}}$ and $p_{\mathrm{local}}$ respectively, can be computed from any number of large $\epsilon_i$ which indicate the presence of a local discrepancy with
\begin{equation}
\label{eq:chi2_local}
    \chi^2_{\mathrm{local}} = \sum_i^r \epsilon_i^2,
\end{equation}
where $r$ is the number of points summed over and the number of degrees of freedom in the $\chi^2_{\mathrm{local}}$ distribution. When computing local $p$-values in this way, a large number of tests can be examined, increasing the odds of randomly producing a larger value. As such, one must correct for the look-elsewhere effect~\cite{lookelsewhere1,lookelsewhere2} by converting the local $p$-value into a global $p$-value which describes the probability of observing such a discrepancy in any combination of the $\epsilon_i$. Several previous MicroBooNE analyses, such as~\cite{numuCC0pNp_PRD}, computed $p_{\mathrm{local}}$ from all $\epsilon_i$ above 2$\sigma$ and then preformed the $p_{\mathrm{local}}$ to $p_{\mathrm{global}}$ conversion according to
\begin{equation}
\label{eq:dchi2_end}
    p_{\mathrm{global}} = 1 - (1-p_{\mathrm{local}})^{n \choose r} = 1 - (1-p_{\mathrm{local}})^{\frac{n!}{(n-r)!r!}},
\end{equation}
where $n$ is the total number of bins and $r$ is the number of $\epsilon_i$ above the 2$\sigma$ threshold, which we refer to as ``extreme values". This accounts for the fact that for $n$ independent $\epsilon_i$ there are ${n \choose r}$ ways to chose $r$ with extreme values. However, calculating a $p_{\mathrm{local}}$ and converting it in a $p_{\mathrm{global}}$ in this manner does not produce an unbiased estimator, which is an undesirable attribute for a test statistic. Moreover, with this method, one must choose their definition of an extreme value and the choice of the threshold can impact the resulting $p_{\mathrm{global}}$. As an example, consider a case in which a $2\sigma$ threshold is chosen and a distribution with 10 bins shows one $\epsilon_i$ at $3\sigma$, another at $1.9\sigma$ and the rest all less than $1\sigma$. This produces $p_{\mathrm{global}}=1-(1-0.0027)^{10}=0.027$. However, if this distribution were to have its second most extreme $\epsilon_i$ at $2\sigma$ instead, this yeilds $p_{\mathrm{global}}=1-(1-0.0015)^{45}=0.065$. This is a undesired result; the worse agreement in the second most extreme $\epsilon_i$ increases the $p_{\mathrm{global}}$ rather than decreasing it. This property arises from the fact that Eq.~\ref{eq:dchi2_end} assumes that all observations are uncorrelated. Though each $\epsilon_i$ is uncorrelated, the observation of $\epsilon_1$ above threshold is correlated with the observation of both $\epsilon_1$ and $\epsilon_j$ above threshold, thereby violating the assumption behind Eq.~\ref{eq:dchi2_end}.

As such, we suggest two alternatives. Rather than examining all $\epsilon_i$ above an arbitrary threshold, one can instead select only the largest $\epsilon_i$. In this case $r=1$ and Eq.~\ref{eq:dchi2_end}, which is now valid, simplifies to 
\begin{equation}
\label{eq:dchi2_onebin}
    p_{\mathrm{global}} = 1 - (1-p_{\mathrm{local}})^{n}.
\end{equation}
If one still wished to examine multiple $\epsilon_i$ in accordance with Eq.~\ref{eq:chi2_local}, they could instead employ a frequentist method. This would entail simulating many pseudo-experiments with $n$ bins, then, in each pseudo-experiment, finding the minimum possible local $p$-value, $p_{\mathrm{local}}^{\mathrm{min}}$, out of all combinations of bins. The same quantity would then also be calculated for the observed data distribution. By computing the fraction of pseudo-experiments with a $p_{\mathrm{local}}^{\mathrm{min}}$ below the $p_{\mathrm{local}}^{\mathrm{min}}$ of the data, one obtains a rigorous $p_{\mathrm{global}}$ for the data distribution. 

Both of these alternative methods of calculating a $p_{\mathrm{local}}$ and converting it in a $p_{\mathrm{global}}$ produce an unbiased estimator that will not decrease the $p_{\mathrm{global}}$ if any of the $p$-values for individual $\epsilon_i$ increases. They also remove any dependence on an arbitrary threshold. Nevertheless, in most circumstances, the differences between these three methods will be small and we choose to utilize Eq.~\ref{eq:dchi2_onebin} in the fake data studies presented in Sec.~\ref{sec:fds}.

An example of utilizing the $\chi^2$ decomposition is illustrated in Fig.~\ref{fig:data_example} on MicroBooNE data. The distribution of interest in this figure is the $\nu_\mu$CC selection from~\cite{MicroBooNE:2021nxr} binned as a function of the reconstructed hadronic energy, which Fig.~\ref{fig:data_example_X} shows in reconstructed space. Figure~\ref{fig:data_example_dchi2} then shows the the $\chi^2$ decomposition and includes both the resulting distribution of $\epsilon_i$ values in the decomposition space and the matrix used to transform from the reconstructed space to the decomposition space. This matrix, though generally challenging to interpret, can provide some insight into the mapping between the two spaces and the types of discrepancies that would result in significant tension in individual $\epsilon_i$. Taking the first decomposition bin as an example, we see that this bin receives a negative contribution from the first three reconstructed bins and a positive contribution from all higher energy bins. As such, an effect that modifies the event rate differently for the  first three low energy bins and the higher energy bins would likely cause a large discrepancy in this decomposition bin which would likewise result in a large $\epsilon_i$ indicative of significant mis-modeling.

\subsubsection{Conditional Constraints}

The conditional constraint procedure~\cite{cond_cov} can be used to increase the stringency of the validation by providing an additional means of probing for relevant mis-modeling. In this procedure, a constraint from one set of data distributions is used to narrow the allowed model parameter space of a different set of distributions. This amounts to an updated central value and a reduced uncertainty band on the constrained prediction. More explicitly, consider two distributions, $X$ and $Y$, with model predictions $\mu^X$ and $\mu^Y$, and a covariance matrix containing these two channels ($X$, $Y$):
\begin{equation*}
    \Sigma = \begin{pmatrix}
\Sigma^{XX} & \Sigma^{XY} \\
\Sigma^{YX} & \Sigma^{YY} 
\end{pmatrix}.
\end{equation*}
Here, the distributions are assumed to be jointly Gaussian with $\Sigma^{XX}$ describing the uncertainties on channel $X$ as well as the correlations between its bins, and $\Sigma^{YY}$ aanalogously describing the uncertainties on channel $Y$ as well as the correlations between its bins. The correlations between the bins of $X$ and $Y$ are described by $\Sigma^{YX}$ and $\Sigma^{XY}$. If a data measurement of channel $Y$ results in the distribution $n^Y$, one can derive the prediction for $X$ given this observation in $Y$ to be
\begin{equation}
\label{eq:constraint}
\mu^{X,\text{const.}} = \mu^{X} + \Sigma^{XY} \cdot \left(\Sigma^{YY} \right)^{-1} \cdot \left( n^Y - \mu^Y \right), 
\end{equation}
\begin{equation}
\label{eq:constraint_cov}
\Sigma^{XX, \text{const.}} = \Sigma^{XX} - \Sigma^{XY} \cdot \left(\Sigma^{YY} \right)^{-1} \cdot \Sigma^{YX}.
\end{equation}
In this case, $Y$ is referred to as the \textit{constraining} channel and $X$ is referred to as the \textit{constrained} channel with its posterior prediction and uncertainties given by $\mu^{X,\text{const.}}$ and $\Sigma^{XX, \text{const.}}$, respectively. This method can be understood as deriving a conditional probability distribution for $X$ given $Y$. With this context, Eq.~\ref{eq:constraint} describes the conditional mean for $X$ and Eq.~\ref{eq:constraint_cov} describes the conditional covariance. 

The amount by which the conditional mean differs from the prior central value prediction will be dictated by the three factors in the second term on the right of Eq.~\ref{eq:constraint}. Smaller uncertainties on the constraining distribution results in a more powerful constraint and a larger update to the prediction, this enters through $(\Sigma^{YY} )^{-1}$. The correlations between the two channels enters through $\Sigma^{XY}$, with a higher level of correlation increasing the size of the update to the prediction. The amount the prediction in the constraining channels differs from the data observation enters through $( n^Y - \mu^Y)$, with a larger deviation from the prediction producing a larger update to the prediction in the constrained channel. The reduction in uncertainties on the constrained distribution is described by the second term on the right in Eq.~\ref{eq:constraint_cov}. Likewise, this term has factors that depend on the size of the uncertainties on the constraining distribution and the size of the correlation between the two channels, namely $(\Sigma^{YY} )^{-1}$ and $\Sigma^{XY}$, respectively. Smaller uncertainties on the constraining channel and larger correlations will increase the resulting reduction in uncertainties on the constrained channel. However, unlike the second term in Eq.~\ref{eq:constraint}, the second term in Eq.~\ref{eq:constraint_cov} does not depend on the data observation in the constraining channel; the amount of ``additional information" provided by the constraint remains constant regardless of the data observation.

The procedure described in Eqs.~\ref{eq:constraint}~and~\ref{eq:constraint_cov} allows a GoF test to be performed on $X$ after the constraints from $Y$. This allows the simultaneous examination of the correlated modeling of $X$ and $Y$, which provides additional information about the compatibility between the model and data. Typically, the constrained and constraining channels have highly correlated model predictions due to detector and physics effects that impact both sets of distributions. The model's description of the correlations is dictated by the modeling of these effects. The reduction in uncertainties for the prediction on the constrained distribution provides additional sensitivity by providing a means of exploring the modeling of these correlations. This direct incorporation of the correlations into the updated central value prediction, uncertainties, and subsequent GoF tests make it straightforward to perform a rigorous examination of the consistency of the modeling across channels and observables, which is a tangible advantage of this method. The modeling of $X$, $Y$, and the relationship between $X$ and $Y$ must all be sufficient for the model to pass a constrained validation test. 

\begin{figure*}[hp!]
\begin{subfigure}{0.9\linewidth}
\includegraphics[width=0.49\linewidth]{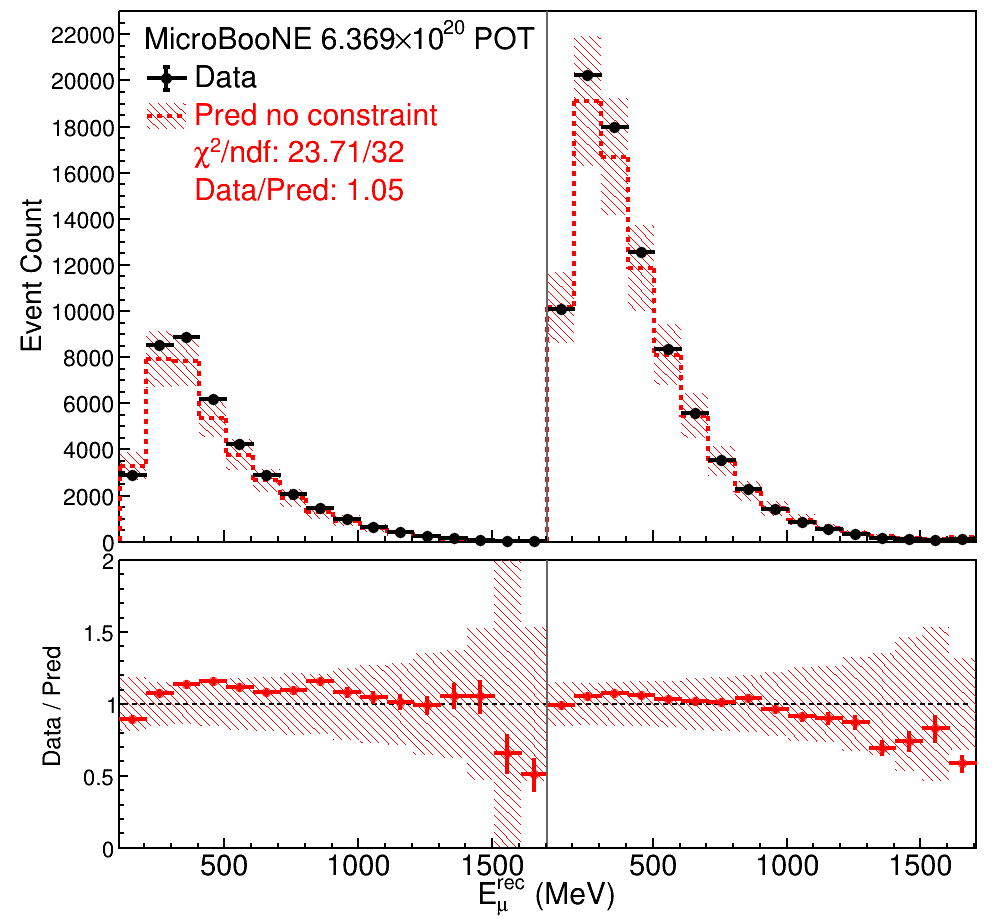}
\put(-150,120){FC}
\put(-50,120){PC}
\includegraphics[width=0.49\linewidth]
{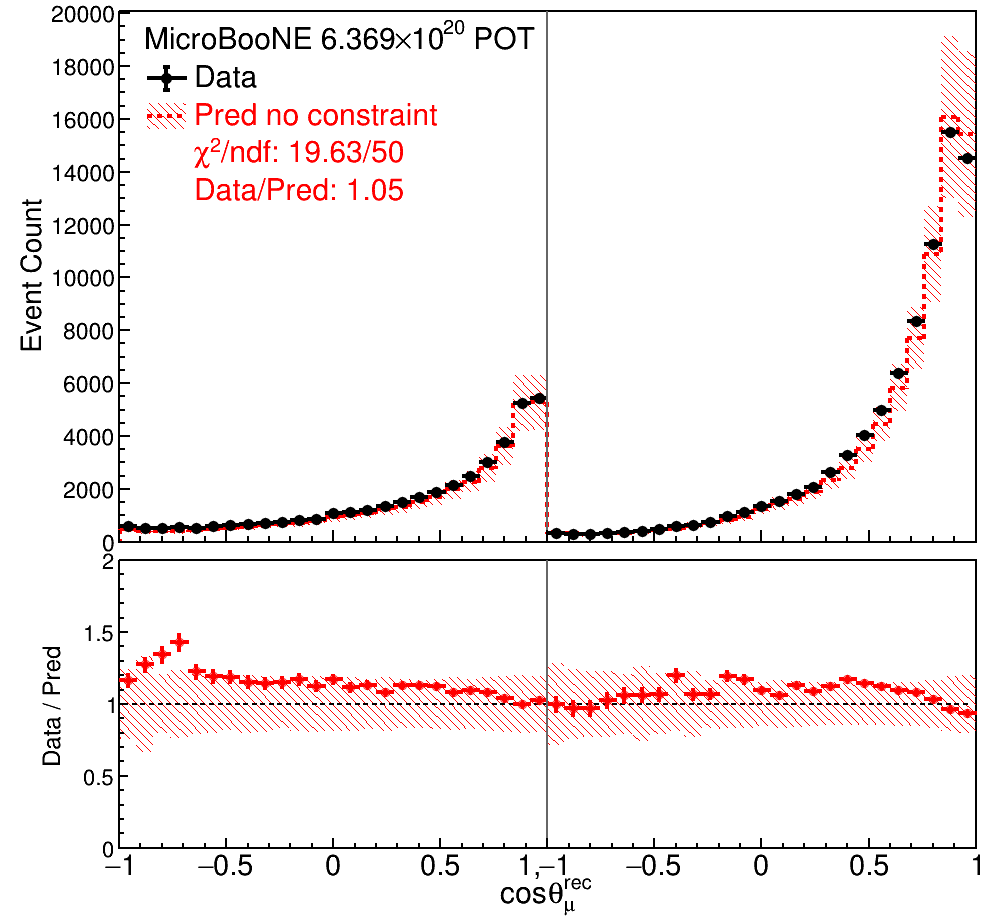}
\put(-150,120){FC}
\put(-50,120){PC}
\caption{The reconstructed muon energy and muon scattering angular distributions. These distributions are used to constrain the distribution shown in (b) and thus correspond to channel $Y$ in Eq.~\ref{eq:constraint}. Both distributions utilize separate bins for FC and PC events. The top panels show the data and MC reconstructed distributions and the bottom panels show the data to MC ratio. The uncertainties of the prediction are shown in the bands and the data statistical uncertainties are shown on the data points.}
\label{fig:data_example_Y}
\end{subfigure}
\begin{subfigure}{0.48\linewidth}
\includegraphics[width=0.95\linewidth]{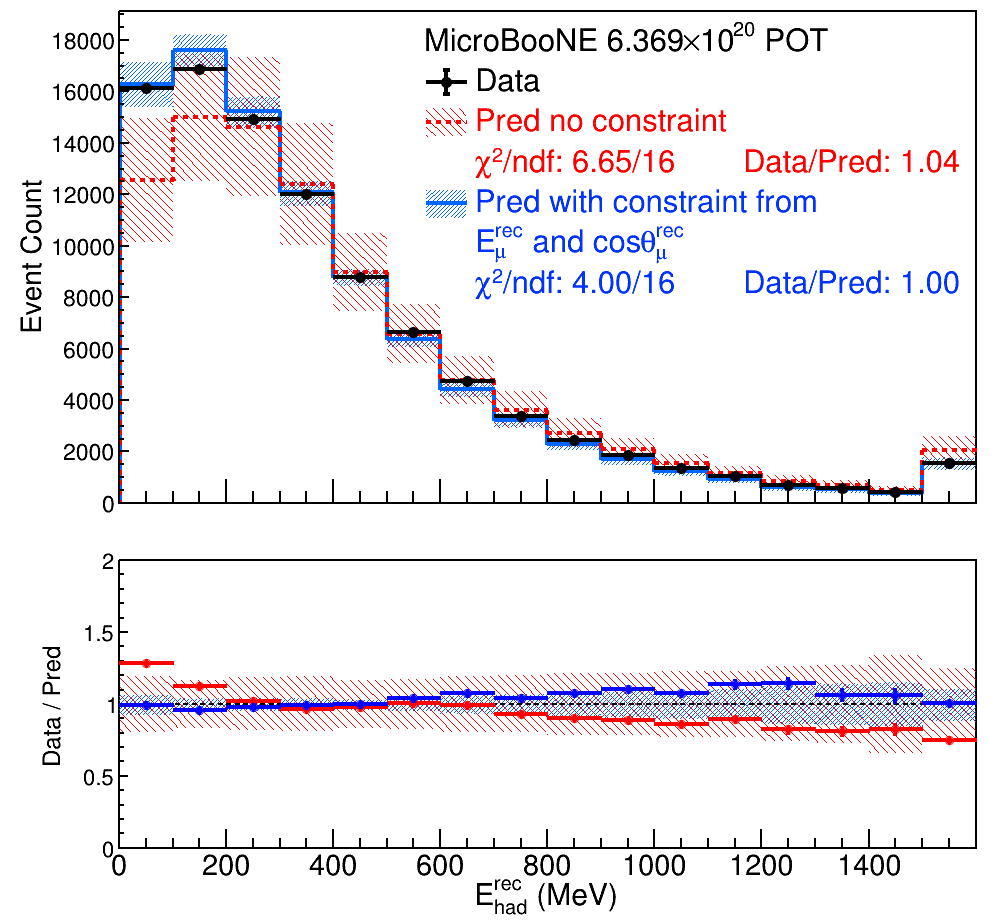}
\put(-50,125){PC}
\caption{The reconstructed hadronic energy distribution for PC events. This distribution is constrained by the distributions shown in (a) and thus corresponds to channel $X$ in Eq.~\ref{eq:constraint}. The MC prediction before (after) constraint from the observed muon energy and angle distributions is shown in red (blue). The top panel shows the data and MC reconstructed distributions and the bottom panel shows the data to MC ratio and its corresponding uncertainties both before and after constraint. The uncertainties of the prediction are shown in the bands and the data statistical uncertainties are shown on the data points. }
\label{fig:data_example_X}
\end{subfigure}
\begin{subfigure}{0.48\linewidth}
\includegraphics[width=0.8\linewidth]{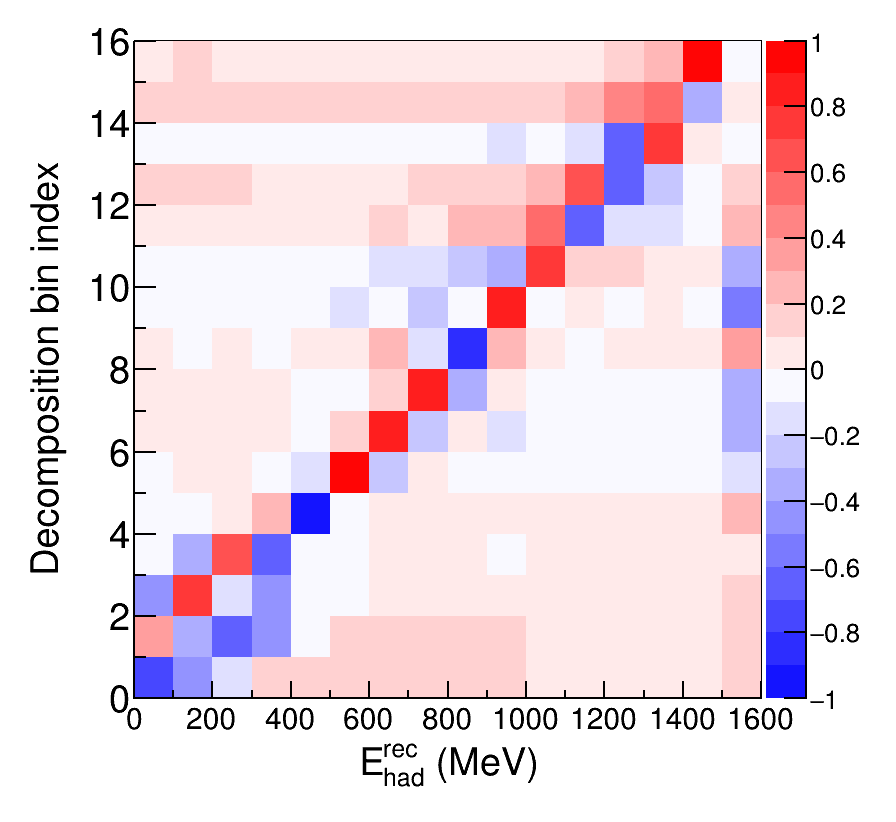}
\includegraphics[width=0.75\linewidth]{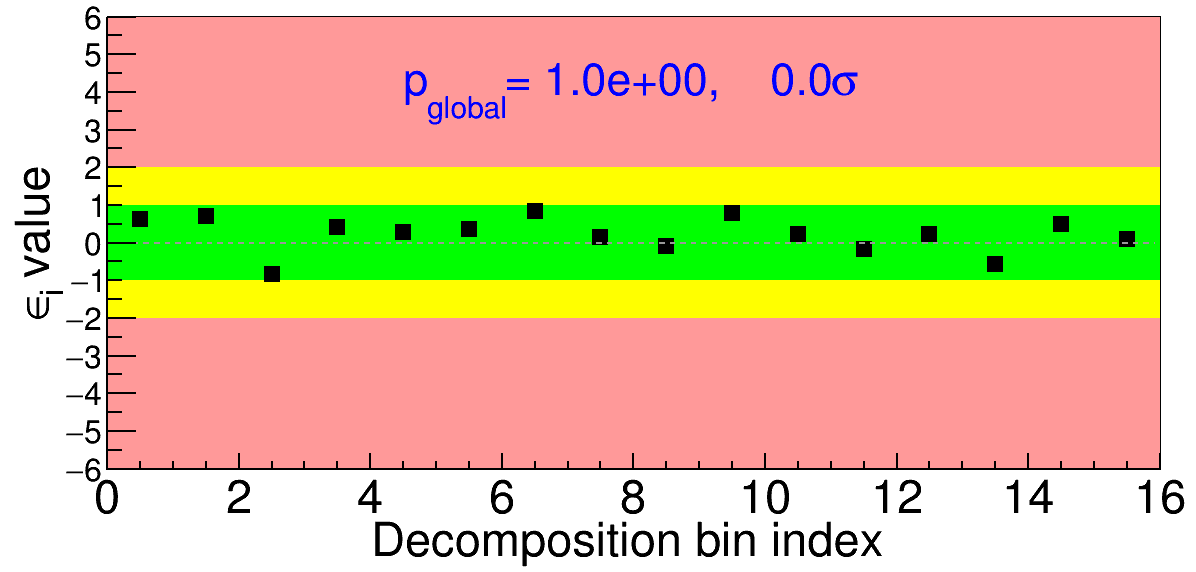}
\caption{Further examination of the GoF for (b) with the $\chi^2$ decomposition. The top panel shows the matrix used to transform the constrained reconstructed hadronic energy distribution into the eigenvalue basis of the covariance matrix. The bottom panel shows the significance of the tension between data and MC, otherwise known as $\epsilon_i$ values, in the eigenvalue basis. }
\label{fig:data_example_dchi2}
\end{subfigure}
\caption{Demonstration of the conditional constraint and $\chi^2$ decomposition techniques using MicroBooNE data. The constraining distributions are shown in (a), the constrained distribution is shown in (b) and the $\chi^2$ decomposition is shown in (c). All distributions utilize the $\nu_\mu$CC selection from~\cite{MicroBooNE:2021nxr} .}\label{fig:data_example}
\end{figure*}

An example of using the conditional constraint can be seen in Fig.~\ref{fig:data_example}. In this example, $Y$ is defined to be the distributions of the reconstructed muon kinematics, namely, the reconstructed muon energy, $E_{\mu}^\mathrm{rec}$, and the muon angle, $\cos\theta_\mu^\mathrm{rec}$, where $\theta_\mu^\mathrm{rec}$ is defined as the angle the muon scatters with respect to the incoming neutrino beam. These distributions are shown in Fig.~\ref{fig:data_example_Y}. Events fully contained (FC) and partially contained (PC) within the detector are placed into separate bins to better separate these two classes of events, which may be sensitive to different forms of mis-modeling. Distribution $X$ is chosen to be that of reconstructed hadronic energy, $E_\mathrm{had}^\mathrm{rec}$, for events partially contained (PC) within the detector. This distribution is shown in Fig.~\ref{fig:data_example_X}. Note that in this case, $X$ and $Y$ contain the same events, and therefore statistical correlations are present in the covariance matrix. The MicroBooNE Monte Carlo (MC) prediction before constraint is shown in red and the prediction after constraint is shown in blue. The shift between the red and blue predictions is dictated by the observation in data for the muon kinematics and by the model's predicted relationship between the muon kinematics and the reconstructed hadronic energy. This difference between the red and blue predictions indicates how large of a correction to the central value is induced by the constraint through the rightmost term in Eq.~\ref{eq:constraint}. The bands surrounding the prediction, which show the uncertainties before and after constraint, illustrate the large reduction in uncertainties and enhanced sensitivity to mis-modeling that is obtainable with the conditional constraint. In particular, this test is expected to provide sensitivity to the modeling of the missing hadronic energy as is described in more detail in Sec.~\ref{sec:modval_procedure}. Other examples of similar constraint tests can be found in~\cite{MicroBooNE:2021sfa,MicroBooNE:2023foc,numuCC0pNp_PRD}.

The conditional constraint procedure borrows principles from data calibration. It uses the data to reduce the uncertainty on the model prediction thereby allowing for more stringent examinations of the model. However, it should be emphasized that, in this context, the sole purpose of these constraints is model validation and they are not used in the cross section extraction. Since the constraining distributions are often using the same set of events as for the unfolding, this reduction of the systematic uncertainties would be superficial in the case of cross section extraction. As an extreme example, utilizing the exact same distribution for the constraining channel as the distribution to be unfolded results in a complete elimination of the uncertainties, which is obviously not realistic or useful for the cross section extraction. Nevertheless, as described in \cite{GardinerXSecExtract}, one could also utilize the conditional constraint formalism directly in the unfolding to employ background constraints from side-band channels. Though the mathematics of this technique is identical to the data-driven validation we describe, its utility is distinct and we advocate for still employing a model validation when a background constraint is part of the analysis strategy. 

\subsection{Developing a Model Validation Procedure}
\label{sec:modval_procedure}
Data-driven model validation focuses on the total model uncertainties, which includes flux, detector, and cross section effects as well as any other modeling or systematic uncertainties utilized for the cross section extraction. When the overall model passes data-driven validation, it suggests that the data is a suitable realization of the range of possibilities afforded by the model. This does not mean there is no under-estimation of individual components of uncertainty, but rather suggests that any under-estimation of individual components is small compared to the overall uncertainty budget. This approach is consistent with the evaluation of significance in discrepancies~\cite{093570275X,Kline1985ThePO}. As described in Sec.~\ref{sec:fds_compare}, this differs from more commonly used approaches to model validation that tend to focus on the cross section modeling. A data-driven approach naturally evaluates the overall model, thereby making the assessment of the total uncertainty band more straightforward than many traditional approaches. This can be seen as a distinct advantage of data-driven approaches.

However, this also brings about its own set of downsides. A combination of flux, detector and cross section effects could conspire in such a way to cancel out, thereby making reconstructed space tests less sensitive to mis-modeling that would still bias the extracted cross sections. The potential for such a situation is investigated in Sec.~\ref{sec:FDS_proton} and should be kept in mind when deciding which tests to include in the model validation procedure. In addition, it is important to note that the procedures we outline are more applicable when the background and signal models are both well defined and relatively well understood up to their uncertainties. If a signal significantly deviates from existing models or is completely unknown then these techniques cannot be properly applied as validation. In that case, other techniques should be employed to validate the background modeling and the signal efficiency within uncertainties before extracting a cross section. 

When developing a model validation procedure for a desired cross section extraction, one must make sure to design a set of tests that are sufficiently sensitive to mis-modeling in the phase space relevant for the cross section extraction. For example, validating only the overall event rate will not be sufficient for a differential cross section measurement. Such a test averages over the entire reconstructed phase space and will thus be insensitive to any mis-modeling related to the shape of the distribution that could bias the extraction of a differential cross section. Selecting an appropriately sensitive set of tests represents the key to a well-designed model validation procedure. To achieve this, one should consider what forms of mis-modeling may be capable of introducing bias into their cross section result and choose tests that address these possibilities. mis-modeling in distributions irrelevant to the extraction do not need to be examined. It may be useful to use FDSs to evaluate the ability of the selected set of tests to detect relevant mis-modeling before it begins to bias the extracted cross section results beyond stated uncertainties. A case study containing this type of FDS is presented in Sec.~\ref{sec:fds}. This study mirrors the analysis done in~\cite{MicroBooNE:2021sfa} and thus also provides an example of what a full suite of tests used in a data-driven validation may look like. However, such studies are not mathematical proofs that the intended validation procedure constructed for the given analysis is guaranteed to detect relevant mis-modeling. As such, it is pertinent to probe the model from a variety of different angles in order to maximize the probability of detecting problematic forms of mis-modeling that would bias the cross section results beyond their uncertainties.

The more challenging a variable or channel is to reconstruct or model, the more stringently it should be examined in the validation. For visible kinematic variables in more inclusive cross section measurements, a direct data to prediction comparison over the reconstructed distribution used in the unfolding is likely sufficient. However, more challenging distributions to model or reconstruct require additional validation. Examples include distributions containing events that are not fully contained within the detector volume, distributions used to map to quantities that cannot be fully reconstructed, or a more exclusive channel with kinematics cuts that are near detector thresholds. When selecting the constraining and constrained distributions, it is useful to keep in mind Eq.~\ref{eq:constraint} and Eq.~\ref{eq:constraint_cov}. The second term on the right of both these equations contain factors which dictate that smaller uncertainties on the constraining channel and larger correlations between the constraining and constrained channel will increase the power of the constraint.

Conditional constraints are particularly useful when validating aspects of the model that are impacted by substantial modeling or reconstruction challenges. One way to use this technique is to use a better understood and more easily reconstructed distribution to constrain the less understood and harder to reconstruct distribution that is more likely to show mis-modeling. For example, in the case of partially contained events, a constraint from the analogous distribution of fully contained events, which are correlated with the partially contained events through common physics and detector modeling, helps provides a more stringent test. In the case of measuring more exclusive channels, where the modeling of events near the detection threshold may become especially important, additional examination of the variables relevant to the detection threshold is warranted. In this situation, the constraint provides a means of examining multiple variables simultaneously, the relationship between which may contain important information on the sufficiency of the model.

One particularly prominent example of this is the reconstruction of low energy protons. Low energy nucleons are known to be modeled significantly differently among event generators~\cite{dytman_transparency,lowKp} thus leading to very different predicted selection efficiencies for protons near the detection threshold~\cite{GenCompare}. This can have a large impact on measurements related to the proton's energy, such as the kinetic energy of the leading proton or transverse kinematic imbalance variables~\cite{dolan_thesis,TKI_Ar,TKI}, which are projections of the lepton and proton momentum onto the plane perpendicular to the neutrino direction. In such a case, a conditional constraint from a related visible kinematic variable less susceptible to the threshold may be useful. Examples of this include the energy and angle of the outgoing lepton. These constraints will reduce the allowed parameter space through correlations due to shared detector, flux, and cross section modeling thereby allowing for a more thorough investigation of the modeling of the proton kinematics near the detection threshold. Including an additional constraint from a closely related channel, such as one without protons, can provide additional information on the modeling of backgrounds or signal events that did not pass selection cuts and may also be useful in probing for relevant mis-modeling. This can be seen as similar to studying side-bands, which are often used to evaluate the modeling of backgrounds, but intends to evaluate the modeling used in the cross section extraction more directly. The central role of the correlations in this procedure makes it straightforward to evaluate if the model can consistently treat multiple distributions which provides a means of identifying when an observation in a side-band distribution is inconsistent with the observation in the signal distribution. This is a distinct advantage over evaluating side-bands in isolation, which does not provide such sensitivity. Though one could also obtain much of the same information through other approaches, such as tuning a model to a side-band channel or alternative kinematic variable, the ease with which one can rigorously account for and evaluate correlations between distributions makes the conditional constraint an attractive option.

When attempting to extract variables that cannot be fully reconstructed, such as the the neutrino energy or energy transferred to the nucleus, a direct comparison between data and MC is likely insensitive to relevant mis-modeling. These quantities include both visible and missing portions, the latter of which cannot be examined with sufficient scrutiny in a direct comparison. In this case, the conditional constraint becomes a crucial tool in evaluating the mapping from reconstructed to true quantities. The choice of these constraints can be based upon physics arguments to help provide the required level of stringency. 

For the case of energy transfer $\nu$, one can use conservation of energy to design a test that is sensitive to the modeling of missing hadronic energy, $E_\mathrm{had}^\mathrm{missing}$, through the examination of the correlations between the leptonic and visible hadronic energy. The visible portion of the energy transfer, $E_\mathrm{had}^\mathrm{vis}$, can be measured through the reconstructed hadronic energy, $E_\mathrm{had}^\mathrm{rec}$, without needing the model to correct for contributions that are not directly measurable. The energy of the outgoing lepton, $E_{\ell}$, can likewise be measured through the reconstructed lepton energy, $E_{\ell}^\mathrm{rec}$, without such corrections. Together, these quantities account for the total energy of the incoming neutrino,
\begin{equation}\label{eqn:energy_conservation}
    E_{\nu} = E_{\ell} + \nu = E_{\ell} + E_\mathrm{had}^\mathrm{vis} + E_\mathrm{had}^\mathrm{missing}.
\end{equation}
In the rightmost equality any contributions from the initial state nucleus have been absorbed into $E_\mathrm{had}^\mathrm{missing}$, that is, $E_\mathrm{had}^\mathrm{missing}$ is defined as the difference between the initial state neutrino energy and the observable leptonic and hadronic energy present in the final state. Through the use of $E_{\ell}^\mathrm{rec}$ as a constraint, the simultaneous measurement of $E_\mathrm{had}^\mathrm{rec}$ and $E_{\ell}^\mathrm{rec}$ is able to validate the predicted relationship between these distributions. This is achieved by using $E_{\ell}^\mathrm{rec}$ as channel $Y$ and $E_\mathrm{had}^\mathrm{rec}$ as channel $X$ in Eq.~\ref{eq:constraint}:
\begin{multline}
\label{eq:detailed_constraint}
\mu^{E_\mathrm{had}^\mathrm{rec},\text{const.}} = \mu^{E_\mathrm{had}^\mathrm{rec}} \\ +  \Sigma^{E_\mathrm{had}^\mathrm{rec}E_{\ell}^\mathrm{rec}} \cdot \left(\Sigma^{E_{\ell}^\mathrm{rec}E_{\ell}^\mathrm{rec}} \right)^{-1} 
 \cdot \left( n^{E_{\ell}^\mathrm{rec}} - \mu^{E_{\ell}^\mathrm{rec}} \right),
\end{multline}
where $\Sigma^{E_{\ell}^\mathrm{rec}E_{\ell}^\mathrm{rec}}$ describes the uncertainties on $E_{\ell}^\mathrm{rec}$ and $\Sigma^{E_{\ell}^\mathrm{rec}E_\mathrm{had}^\mathrm{rec}}$ describes the correlations between $E_{\ell}^\mathrm{rec}$ and $E_\mathrm{had}^\mathrm{rec}$. The same is done for Eq.~\ref{eq:constraint_cov} to obtain the reduced uncertainty band, $\Sigma^{E_\mathrm{had}^\mathrm{rec}E_\mathrm{had}^\mathrm{rec}, \text{const.}}$, for the $E_\mathrm{had}^\mathrm{rec}$ prediction. 

From Eq.~\ref{eq:detailed_constraint}, we see that the modeling of $E_{\ell}^\mathrm{rec}$, $E_\mathrm{had}^\mathrm{rec}$, and the correlations between them all play a role in obtaining the posterior prediction. Following Eq.~\ref{eq:chi2}, these can then be simultaneously examined with a GoF test on the constrained $E_\mathrm{had}^\mathrm{rec}$ prediction:
\begin{multline}
\label{eq:chi2_detailed}
\chi^2 = (n^{E_\mathrm{had}^\mathrm{rec}}-\mu^{E_\mathrm{had}^\mathrm{rec},\text{const.}})^T \\
\cdot \big( \Sigma^{E_\mathrm{had}^\mathrm{rec}E_\mathrm{had}^\mathrm{rec}, \text{const.}} \big)^{-1}  
\cdot \big( n^{E_\mathrm{had}^\mathrm{rec}} - \mu^{E_\mathrm{had}^\mathrm{rec},\text{const.}} \big).
\end{multline}
Given a flux prediction with its associated uncertainties, the correlations that yield $\mu^{E_\mathrm{had}^\mathrm{rec},\text{const.}}$ will be dictated by the modeling of $E_\mathrm{had}^\mathrm{missing}$. Thus, if $E_{\ell}^\mathrm{rec}$ and $E_\mathrm{had}^\mathrm{rec}$ are measured, and a constraint from $E_{\ell}^\mathrm{rec}$ is applied to $E_\mathrm{had}^\mathrm{rec}$ to reduce uncertainties on the overall $E_{\nu}$ prediction through the high correlation in the flux prediction shared between the channels, a precise description of $E_\mathrm{had}^\mathrm{missing}$ becomes necessary for Eq.~\ref{eqn:energy_conservation} to be satisfied.
It is this narrowed parameter space for the correlated $E_{\ell}$, $E_\mathrm{had}^\mathrm{vis}$, and $E_\mathrm{had}^\mathrm{missing}$ predictions that provide sensitivity to mis-modeling in the missing energy through conservation of energy. Nevertheless, it should be noted that Eq.~\ref{eqn:energy_conservation} still allows for a conspiracy of flux mis-modeling (aka mis-modeling of $E_{\nu}$) and mis-modeling of $E_\mathrm{had}^\mathrm{missing}$ to cancel out and pass the test described by Eq.~\ref{eq:chi2_detailed}. However, the use of the constraint brings the examination of $E_\mathrm{had}^\mathrm{missing}$ through the correlations between $E_{\ell}^\mathrm{rec}$ and $E_\mathrm{had}^\mathrm{rec}$ onto more equal footing with the examination of a directly observable variable, like $E_{\ell}^\mathrm{rec}$, in an unknown neutrino flux spectrum. As in the case of $E_\mathrm{had}^\mathrm{missing}$, an $E_{\mu}$ prediction is always going to depend on the assumed $E_{\nu}$ and it is challenging to fully disentangle the two. 

The test described in Eq.~\ref{eq:detailed_constraint} is demonstrated on MicroBooNE data in Fig.~\ref{fig:data_example}. This technique of using the lepton energy to constrain reconstructed hadronic energy has been employed by MicroBooNE to enable the extraction of energy-dependent inclusive $\nu_\mu$CC cross sections on argon~\cite{MicroBooNE:2021sfa,MicroBooNE:2023foc,numuCC0pNp_PRD}. Similar constraints could be applied to other variables potentially sensitive to the modeling of missing hadronic energy.  Similarly, when extracting nominal flux-averaged cross sections, which requires extrapolating from the real to the nominal neutrino flux spectrum, examining the mapping between the true and reconstructed neutrino energy is a route to evaluate if the overall uncertainty budget is sufficient for propagating the reported result from the real to the nominal flux.

Constraints could also be used to explore other physics that is challenging to probe with a direct data to MC comparison. Examples include using the lepton energy distribution to constrain the lepton angular distribution. Since quasi-elastic (QE), resonance (RES), meson exchange current (MEC), and deep-inelastic scattering (DIS) processes have distinct predictions for lepton kinematics, a mis-modeling of the relative contribution of these events could be detected with such a constraint. This could also be explored by using the observed muon kinematics to constrain another variable that already shows some separation between interaction modes, such as the opening angle between the lepton and the leading proton. This is demonstrated in the fake data studies presented in Sec.~\ref{sec:add_fds}. 

When the model passes a well designed validation procedure, it suggests that the difference between the data and simulation are within the quoted total model uncertainty. When the model fails the validation procedure, it reveals that the difference between the data and simulation is beyond the model uncertainty and should trigger actions to improve the model for use in the analysis. This may include an expansion of the uncertainties or an updated central value prediction. These may be derived in a data-driven way~\cite{numuCC0pNp_PRD} similar to the overall model validation procedure or derived from alternative models or event generators. As long as the expanded model is able to pass all relevant model validation tests, it may be used in place of the original model to extract the desired cross section results.

\section{Fake Data Studies}\label{sec:fds}

\subsection{Usage of Fake Data Studies}\label{sec:fds_compare}
It is common practice in neutrino-nucleus cross section measurements to perform model validation on the primary interaction model used for cross section extraction through comparisons to alternative model or event generator predictions. Often, this comes in the form of FDSs designed to determine if the unfolding is able to recover the underlying true distribution when the ``data" are produced by an alternative model rather than by nature. This allows one to assess how the variations between model predictions for backgrounds, efficiencies, or biases in the reconstruction of kinematic quantities can impact the results. In the case of poor closure on the underlying true distribution, these studies can be used to inform additional uncertainties to be applied to the primary model prediction used for extracting the data cross sections. Using fake data studies in this way is thus akin to using the data-driven model validation that we advocate for, which likewise aims to verify that the model contains sufficient uncertainties for the intended measurement.  

In these FDSs, it is especially useful to consider alternative models that are expected to provide a particularly good description of the data or contain significantly different physics than the nominal model. Ideally, these FDSs should be conducted at statistics equal to, or higher than, that of the actual data. In many cases, the fake data is generated using the same detector and flux model as the nominal model. This makes these uncertainties superfluous and FDSs are therefore conducted removing these sources of uncertainty. This allows for a more direct test of the interaction model and the extent to which it may bias the unfolding.

The benefit of directly validating the interaction model used for unfolding with fake data is that this approach reduces the dependence on any given model. It helps ensure that the discrepancy in results extracted using the examined set of event generators is insignificant compared to the uncertainty on the results. While this method is a viable strategy for validating the interaction model used for unfolding, it has the following shortcomings:
\begin{enumerate}
\item There is no guarantee that the combined phase space covered by the set of tested models and event generators is able to completely describe nature. Therefore, this approach lacks a guarantee of sufficiency in determining the primary model's under-estimation of uncertainties.
\item It is also possible that the phase space covered by these event generators leads to a significant over-estimation of uncertainties. While over-estimating systematic uncertainties is better than under-estimating them, over-estimated uncertainties will reduce the power of the data. 
\item Since one can always invent new effective models or event generators, it is not clear when one should stop in evaluating the model uncertainty under this approach.
\end{enumerate}
It is even possible for both issues 1 and 2 to exist simultaneously in different regions of phase space and it is challenging to know where the spread of event generators lies relative to these two extremes. The fundamental issues of these shortcomings are related to the earlier discussion in Sec.~\ref{sec:generators} on the nature of event generators.

We propose utilizing a data-driven model validation procedure to reduce the reliance on alternative models to validate the model used for unfolding and address issues 1 and 2. As described in Sec.~\ref{sec:validation}, when a model passes a carefully designed data-driven validation procedure, it suggests that the difference between the data and simulation are within the quoted total model uncertainty in the phase space relevant for the cross section extraction. This approach also addresses issue 3 because it is based upon real data rather than alternative models, and is thus better suited to determine the adequacy of the model used for the extraction and whether an expansion of its uncertainties is appropriate. Nevertheless, the data-driven validation does not totally eliminate these issues. One can still run into issue 1 if one selects tests that do not probe the full phase space of the measurement, or issue 2 if one's suite of tests is overly sensitive to mis-modeling irrelevant to the desired measurement, and the choice of which tests to utilize is analogous to issue 3. However, having the validation centered around the data rather than alternative models minimizes the potential downsides of these issues.
 
When evaluating the data-driven model validation methods presented here, we use FDSs in a very different way. Instead of testing the robustness of the model, these FDSs aim to test the robustness of the model validation itself. In particular, FDSs in the case of data-driven model validation should aim to demonstrate the following points:
\begin{itemize}
\item The model validation procedure is indeed sensitive to mis-modeling in the phase space relevant for the cross section extraction. 
\item When the model passes validation, the potential bias in the extracted cross sections is small compared to the total uncertainties. 
\end{itemize}
These FDSs may additionally demonstrate that even in certain cases where the model fails validation, the biases in the extracted cross sections are still small compared to the total uncertainties. Though an analysis should not proceed to unfolding with a model that has not passed validation, observing such cases in these studies provides additional evidence that the validation is able to identify mis-modeling before bias is introduced. In short, FDSs employed in the context of data-driven model validation aim to examine the \textit{procedure} used to validate the model, whereas more typical FDSs aim to directly examine the \textit{model} used for the extraction. As was done in the Supplemental Material of Refs.~\cite{MicroBooNE:2021sfa,MicroBooNE:2023foc,numuCC0pNp_PRD,numuCC0pNp_PRL}, we illustrate how to test the sufficiency of a model validation procedure through FDSs presented in~\ref{sec:direct}.

When evaluating a data-driven model validation procedure with FDSs, it is perhaps more useful to utilize all systematic uncertainties rather than only the cross section uncertainties, even if some uncertainties are superficial in the context of fake data. Such studies provide a more realistic test of the stringency of the model validation as it is performed on the real data. Through the conditional constraint, the validation cancels shared systematics from all sources including neutrino flux, cross section, and detector effects. As seen in Eq.~\ref{eq:constraint_cov}, the way the constraint reduces these uncertainties on the model prediction is independent of the data observation and only depends on the uncertainties on and correlations between the model predictions for the different distributions. Because of this, the way the constraint reduces the uncertainties is exactly the same in these studies as in real data. This is not true when only the cross section uncertainties are included, in which case the constraint may behave very differently due to the significantly different treatment of the systematic uncertainties. Thus, a test with the complete set of systematics is required to demonstrate that the constraint is able to reduce uncertainties on the reconstructed distributions enough to detect relevant mis-modeling that will bias the extraction beyond stated uncertainties. Furthermore, many FDSs could be interpreted as a detector effect rather than a cross section effect, thereby testing the ability of the model validation to detect model discrepancies not attributable to the cross section modeling. 

If an analysis is employing data-driven validation, fake data studies that directly validate the model through comparisons to alternative event generators may still be useful to identify situations in which the extracted results are not biased beyond stated uncertainties but nevertheless show better agreement for the model used for unfolding over the truth. An example of this type of situation would be one in which the extracted cross section shows a mild discrepancy with the truth at 1.2$\sigma$ significance, but shows tension at only 0.5$\sigma$ significance with the model used for extraction. Such a situation can't be identified with data-driven model validation, as it only examines if the bias introduced is within stated uncertainties. This can be seen as an advantage of more typical FDSs. When employing data-driven validation, one should be cautious if this type of bias is identified in a FDS, but such bias does not necessarily invalidate the model used for unfolding if it still passes the data-driven validation. Unless the alternative generator used to produce the fake data is expected to be a particularly good description of the real data, this bias may not carry over into the extraction of the real data, and, even if it does, the extracted results are still expected to fall within the uncertainties of the true value in nature if the model passes validation by a well-constructed set of tests. 

\subsection{Illustration of Data-driven Model Validation Using Fake Data Studies}\label{sec:direct}
Using the model validation procedure and unfolding analogous to the methodology employed in~\cite{MicroBooNE:2021sfa}, we conduct two sets of FDSs that serve as case studies for data-driven model validation. The first set utilizes a fake data set produced from the nominal $\texttt{GENIE}$-based MicroBooNE model~\cite{GENIE,MicroBooNE:2021ccs} with a shift in the reconstructed proton energy intended to mimic a mis-modeling of the relative contribution of the missing hadronic energy to the total energy transfer. The second set of FDSs utilizes fake data sets produced from an alternative event generator; namely $\texttt{NuWro 19.02.2}$~\cite{nuwro}. Studies are performed with the full systematic uncertainties to probe the ability of the model validation to detect discrepancies under a treatment of systematic uncertainties akin to that of real data. A FDS utilizing the $\texttt{NuWro}$ fake data is also performed with only the cross section systematics in order to more directly probe the cross section model and its associated uncertainties. As will be shown throughout this section, the results of these two sets of FDSs are consistent with the expectation that the data-driven model validation is able to detect relevant mis-modeling before it begins to bias the extraction of cross sections. 

\begin{figure}[t]
\centering
\includegraphics[width=0.9\linewidth]{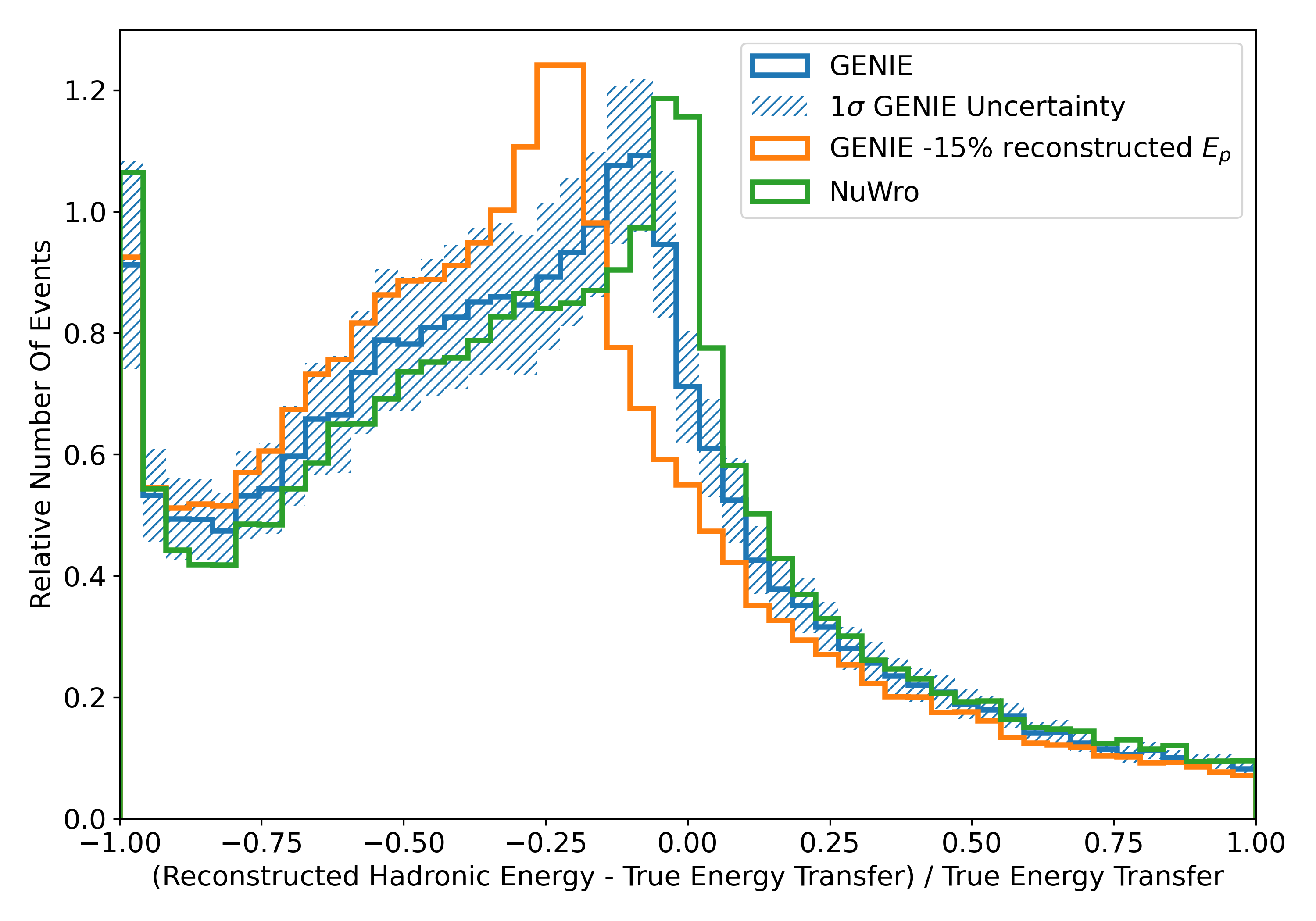}
\caption{\label{fig:fake_data_1d_illustration}The energy transfer resolution predicted by the $\texttt{GENIE}$-based MicroBooNE tune simulation and cross section uncertainties, the $\texttt{GENIE}$-based MicroBooNE tune simulation with 15\% reduction in reconstructed proton energy, and the $\texttt{NuWro}$ simulation. We show only the region between -1 and 1 for visual clarity.}
\end{figure}

These fake data sets represent significant deviations from the mapping between true and reconstructed energy transfer predicted by the nominal $\texttt{GENIE}$-based MicroBooNE MC. This is illustrated in Fig.~\ref{fig:fake_data_1d_illustration}, which shows the energy transfer resolution for the central value prediction of the MicroBooNE MC, the MicroBooNE MC prediction with a 15\% reduction in reconstructed proton energy, and the $\texttt{NuWro}$ prediction. The MicroBooNE MC prediction's cross section uncertainties are included on the nominal prediction without the proton energy scaling, but these uncertainties are not large enough to cover the $\texttt{NuWro}$ prediction. Furthermore, a 15\% reduction in the visible hadronic energy is a more significant deviation than changing from the nominal MicroBooNE MC prediction to $\texttt{NuWro}$ prediction. 

For both sets of FDSs, we employ the same $\nu_\mu$CC event selection and systematic uncertainties as in~\cite{MicroBooNE:2021sfa}, which are described in detail in~\cite{MicroBooNE:2021nxr,numuCC0pNp_PRD}. We then perform model validation designed to enable the extraction of the cross section as a function of $E_\nu$, the differential cross section with respect to the energy transfer $\nu$, and the differential cross section with respect to the muon energy $E_\mu$. The validation is described in more detail in the following paragraphs with a complete list of tests presented in Appendix~\ref{test_list}. Following the validation, cross section results are extracted using the Wiener-SVD unfolding technique~\cite{Tang:2017rob}. Results for $E_\mu$ are not extracted for the FDSs with the reconstructed proton energy scaling because these fake data samples do not change the underlying muon kinematics resulting in perfect closure in this variable for each FDS. 

The model validation utilizes a multitude of tests in order to increase the chances of detecting problematic forms of mis-modeling that would bias the cross section results beyond their uncertainties. One does not necessarily know a priori which test will be most sensitive, hence the numerous tests. In this case study, each test is performed on events fully contained (FC) and partially contained (PC) within the detector as well as jointly on all events (FC\&PC) with separate bins for FC and PC. The only exception is in tests where the FC distribution is used to constrain the PC one, in which case only the PC distribution is examined. In each test, the GoF is quantified using Eq.~\ref{eq:chi2} to determine how well the fake data distributions are described by the nominal MicroBooNE model used for the unfolding. The GoF is then further evaluated by decomposing the covariance matrix into linearly independent components via eigenvalue decomposition. This transformation to an uncorrelated basis allows a local $p$-value to be calculated from the decomposition bin that shows the largest discrepancy. The look-elsewhere effect is then corrected for by converting the local $p$-value into a global $p$-value via Eq.~\ref{eq:dchi2_onebin}. This procedure is described in more detail in Sec.~\ref{sec:validation}. These global $p$-values and the $p$-values obtained from the $\chi^2$ GoF test described in Eq.~\ref{eq:chi2} are the metrics used to evaluate the sufficiency of the model, and must be greater than 0.05 in all tests for the model to pass validation. 

The first set of tests examines the modeling of the muon kinematics in detail. These are not performed in the FDSs with the reconstructed proton energy scaling because these fake data samples do not modify the muon kinematics. The tests begin with evaluating the GoF on the muon energy distributions, $E_{\mu}^\mathrm{rec}$, and the muon angular distributions, $\cos\theta_\mu^\mathrm{rec}$. Next, the FC distributions are used to constrain the PC distributions in the same variable. These PC distribution are, in general, more susceptible to mis-modeling due to poorer reconstruction in events that escape the active volume of the detector, hence the additional tests. Then, to examine the muon kinematics more holistically, the $E_{\mu}^\mathrm{rec}$ distributions are used to constrain the $\cos\theta_\mu$ distributions. These tests provide a significantly reduced posterior uncertainty. This allows the muon kinematics to be examined in detail, potentially exposing mis-modeling of a variety of physics effects, such as the relative contribution of different interaction modes, which could bias the measurement of the muon kinematics or derived quantities such as $E_\nu$ or $\nu$. If the model passes this suite of tests, it builds confidence that it is sufficient to extract cross sections as a function of the muon kinematics.

With the muon kinematics validated, the focus shifts to the hadronic energy distributions in order to enable the extraction of cross sections as a function of $E_\nu$ and $\nu$. These tests begin the same as the ones on the muon kinematics. Events are binned as a function of the reconstructed hadronic energy, $E_\mathrm{had}^\mathrm{rec}$, or the reconstructed neutrino energy, $E_{\nu}^\mathrm{rec} = E_\mathrm{had}^\mathrm{rec} + E_{\mu}^\mathrm{rec}$, and GoF tests are performed both on the unconstrained distributions and on the PC distributions after constraints from the FC ones in the same variable. From here, the muon kinematics are used to constrain the $E_{\nu}^\mathrm{rec}$ and $E_\mathrm{had}^\mathrm{rec}$ distributions. These tests, which are described in Eq.~\ref{eq:chi2_detailed}, examine the correlated prediction between the hadronic and leptonic energy thereby providing sensitivity to the missing hadronic energy. If the model passes this second suite of tests, it builds confidence that the model can describe the leptonic and hadronic energy as well as the correlations between them and is sufficient to extract cross sections as a function of $E_\nu$ and $\nu$.

To evaluate how well the extraction has reproduced the underlying true distribution, a $\chi^2$ test statistic is computed between the underlying truth and the unfolded fake data. In the case of the FDSs conducted with the scaling of the reconstructed proton energy, the true distribution corresponds to the MicroBooNE MC. This is because neither the incoming neutrino energy, the energy transfer, nor the muon kinematics are modified by this scaling. These truth level distributions remain identical and only the relative contribution of missing and visible energy is modified. In the case of the FDSs conducted with the $\texttt{NuWro 19.02.2}$ fake data, the true distribution corresponds to a prediction from an independently produced high-statistics sample from the event generator. The $\chi^2$ test statistic is constructed using the covariance matrix obtained in the cross section extraction and is converted to a $p$-value assuming a $\chi^2$ distribution with degrees of freedom equal to the number of bins in the extracted result. To compare the stringency of the model validation to the amount of bias induced in the cross section extraction, this $p$-value is compared to the $p$-values used to evaluate the model validation. For ease of comparison, we convert all $p$-values to significance levels corresponding to multiples of the standard deviation $\sigma$ of a normal distribution. The significance level corresponding to a model validation test indicates the amount of mis-modeling detected by the validation and the significance level corresponding to a cross section extraction indicates the amount of bias induced in the unfolding. When a model validation test yields a larger significance level than the corresponding cross section extraction, this indicates that the model validation is more stringent than the cross section extraction. In this case, one is able to use the results of the model validation to identify situations in which the model is insufficient before such deficiencies become relevant to the unfolding. This allows for subsequent efforts to mitigate deficiencies in the overall model by using an updated CV prediction or expanded set of uncertainties before proceeding to the extraction.

\subsubsection{Proton Energy Scaling Fake Data Studies}
\label{sec:FDS_proton}

\begin{figure}[t]
\centering
  \includegraphics[width=\linewidth]{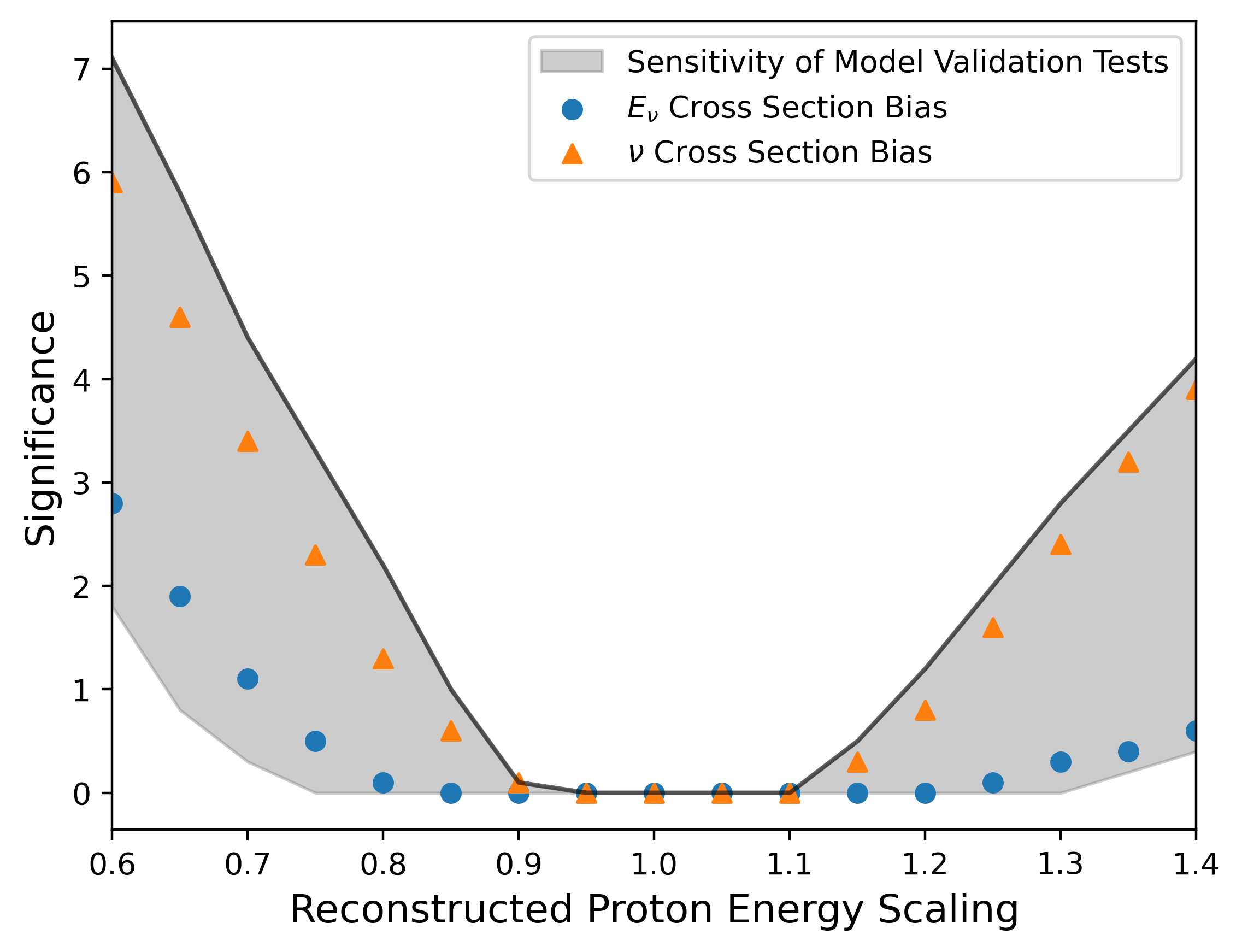}
\caption{Sensitivity of the model validation to discrepancies compared to the bias induced in the cross-section extraction in FDSs utilizing full-systematics. The fake data sets used for these FDSs each have their reconstructed proton energy scaled by different amounts to mimic mis-modeling of the missing hadronic energy. The x-axis corresponds to this scaling. The y-axis indicates the agreement between the fake data and nominal MicroBooNE MC for the cross-section extractions and model validation tests in terms of significance level. For the model validation, a large significance indicates high sensitivity to mis-modeling. For the extracted cross-section, high significance indicates more biased results and worse closure on the underlying true distribution.}
\label{FDS_full_sys}
\end{figure}

The results of these FDSs with different scalings of the visible proton energy are shown in Fig.~\ref{FDS_full_sys} and Table~\ref{table:64_full_sys}. In Fig.~\ref{FDS_full_sys}, the gray band corresponds to the range of the significance values obtained across the validation 
tests, and the blue and orange points indicate the significance of the bias between the cross sections extracted in $E_\nu$ and $\nu$ and MC truth. Theses FDSs are conducted with all systematic uncertainties at $6.4\times10^{20}$ POT, which corresponds to the exposure of the first three runs of MicroBooNE data taking that have been used for many recent MicroBooNE cross section measurements~\cite{MicroBooNE:2023foc,numuCC0pNp_PRD,numuCC0pNp_PRL}. In these studies, we do not account for correlations in the data and MC statistical uncertainties arising from the fact that the fake data and MC utilize the same set of events. However, this is treated identically between the validation and the cross section extraction making this a fair comparison between the sensitivity of the model validation and the bias induced in the cross section extraction.

\begin{table*}
\begin{tabular}{||c | c c | c@{\hskip 4mm} c@{\hskip 4mm} c | c@{\hskip 4mm} c@{\hskip 4mm} c | c@{\hskip 4mm} c@{\hskip 4mm} c | c@{\hskip 4mm} c@{\hskip 4mm} c||} 
\hline

     & \multicolumn{2}{c|}{} & \multicolumn{12}{c|}{Sensitivity of the Model Validation to Discrepancies} \\
 \hline   
   & \multicolumn{2}{c|}{Bias in Extraction} & \multicolumn{3}{c|}{$E_\mathrm{had}^\mathrm{rec}$ GoF} & \multicolumn{3}{c|}{$E_\mathrm{had}^\mathrm{rec}$ Decomposition} & \multicolumn{3}{c|}{$E_{\nu}^\mathrm{rec}$ GoF} & \multicolumn{3}{c||}{$E_{\nu}^\mathrm{rec}$ Decomposition} \\

  Visible $K_p$ scaling & $\sigma(E_\nu)$  & $d\sigma/d\nu$  & FC\&PC & FC  & PC & FC\&PC & FC  & PC & FC\&PC & FC  & PC & FC\&PC & FC  & PC\\ 
 \hline
0.6  & 2.8 & 5.9 & 5.9 & 2.0 & 6.3 & 6.0 & 3.6 & 6.6 & 5.3 & 1.8 & 6.4 & 6.6 & 3.6 & 7.1\\
\hline
0.65 & 1.9 & 4.6 & 4.2 & 1.3 & 4.9 & 4.9 & 2.8 & 5.5 & 3.3 & 0.8 & 4.6 & 5.3 & 2.9 & 5.8\\
\hline
0.7  & 1.1 & 3.4 & 2.7 & 0.6 & 3.7 & 3.8 & 2.0 & 4.4 & 1.5 & 0.3 & 2.9 & 3.9 & 2.2 & 4.4\\
\hline
0.75 & 0.5 & 2.3 & 1.3 & 0.2 & 2.4 & 2.7 & 1.1 & 3.3 & 0.2 & 0.0 & 1.3 & 2.5 & 1.4 & 3.1\\
\hline
0.8  & 0.1 & 1.3 & 0.2 & 0.0 & 1.2 & 1.6 & 0.4 & 2.2 & 0.0 & 0.0 & 0.2 & 1.1 & 0.6 & 1.8\\
\hline
0.85 & 0.0 & 0.6 & 0.0 & 0.0 & 0.3 & 0.4 & 0.0 & 1.0 & 0.0 & 0.0 & 0.0 & 0.0 & 0.1 & 0.4\\
\hline
0.9  & 0.0 & 0.1 & 0.0 & 0.0 & 0.0 & 0.0 & 0.0 & 0.1 & 0.0 & 0.0 & 0.0 & 0.0 & 0.0 & 0.0\\
\hline
0.95 & 0.0 & 0.0 & 0.0 & 0.0 & 0.0 & 0.0 & 0.0 & 0.0 & 0.0 & 0.0 & 0.0 & 0.0 & 0.0 & 0.0\\
\hline
1.05 & 0.0 & 0.0 & 0.0 & 0.0 & 0.0 & 0.0 & 0.0 & 0.0 & 0.0 & 0.0 & 0.0 & 0.0 & 0.0 & 0.0\\
\hline
1.1  & 0.0 & 0.0 & 0.0 & 0.0 & 0.0 & 0.0 & 0.0 & 0.0 & 0.0 & 0.0 & 0.0 & 0.0 & 0.0 & 0.0\\
\hline
1.15 & 0.0 & 0.3 & 0.0 & 0.0 & 0.1 & 0.1 & 0.0 & 0.5 & 0.0 & 0.0 & 0.0 & 0.1 & 0.1 & 0.3\\
\hline
1.2  & 0.0 & 0.8 & 0.0 & 0.0 & 0.5 & 0.5 & 0.2 & 1.2 & 0.0 & 0.0 & 0.0 & 0.7 & 0.6 & 1.1\\
\hline
1.25 & 0.1 & 1.6 & 0.3 & 0.1 & 1.3 & 1.1 & 0.8 & 1.9 & 0.0 & 0.0 & 0.2 & 1.6 & 1.3 & 2.0\\
\hline
1.3  & 0.3 & 2.4 & 1.1 & 0.4 & 2.2 & 1.7 & 1.4 & 2.5 & 0.0 & 0.0 & 0.7 & 2.5 & 2.0 & 2.8\\
\hline
1.35 & 0.4 & 3.2 & 2.0 & 0.9 & 3.0 & 2.2 & 2.1 & 3.0 & 0.2 & 0.2 & 1.3 & 3.3 & 2.6 & 3.5\\
\hline
1.4  & 0.6 & 3.9 & 2.8 & 1.5 & 3.8 & 2.7 & 2.7 & 3.4 & 0.6 & 0.4 & 2.1 & 3.9 & 3.2 & 4.2\\
\hline
\end{tabular}
\caption{\label{table:64_full_sys} Results of the model validation and 
cross-section extraction for each fake data set that scales the reconstructed proton energy. All sources of uncertainty are included at data statistics equivalent to 6.4$\times10^{20}$ protons on target (POT) of exposure. The significance values in the ``Bias in Extraction" columns indicate the level of closure the extracted fake data cross-sections have with the underlying truth. The significance values in the ``Sensitivity of the Model Validation to Discrepancies" columns indicate the level at which the fake data and the MC used for the cross section extraction disagree. The different columns correspond to different validation tests. All significance levels were obtained by converting the $\chi^2$ test statistic to a $p$-value assuming a $\chi^2$ distribution with degrees of freedom equal to the number of bins. These $p$-values are then interpreted as significance levels corresponding to multiples of the standard deviation $\sigma$ of a normal distribution.}
\end{table*}

For each fake data set, the significance of the bias between the extraction of the cross section as a function of $E_\nu$ is well below the significance of the discrepancy identified in the corresponding model validation tests. The significance of the bias for the differential cross section extracted in $\nu$ is always higher than it is for $E_\nu$ but is similarly always below the significance of the discrepancy identified in the most sensitive model validation test. Furthermore, even at large proton energy scalings past the point at which the validation indicates that there is relevant mis-modeling, the agreement between the underlying truth and extracted results remains quite good for the cross section as a function of $E_\nu$. These observations indicate that the stringency of the model validation is greater than the bias in the cross section extraction induced by mis-modeling. This allows mis-modeling of the missing hadronic energy to be detected before it becomes problematic to the extraction.

\begin{figure}[!htp]
\centering
  
 \begin{subfigure}{\linewidth}
  \includegraphics[width=0.75\linewidth]{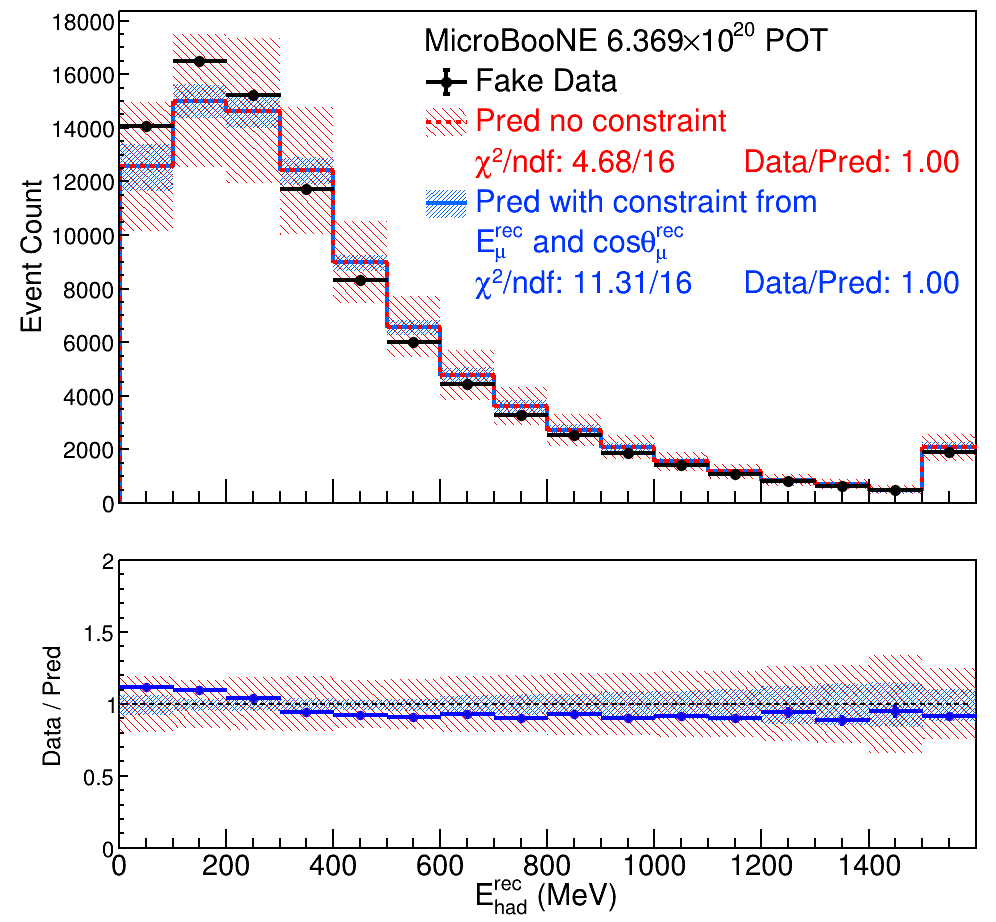}
  \put(-60,100){$K_p^\mathrm{rec}\times0.85$}
  \put(-60,90){$\sigma=0.3$}
  
    \includegraphics[width=0.65\linewidth]{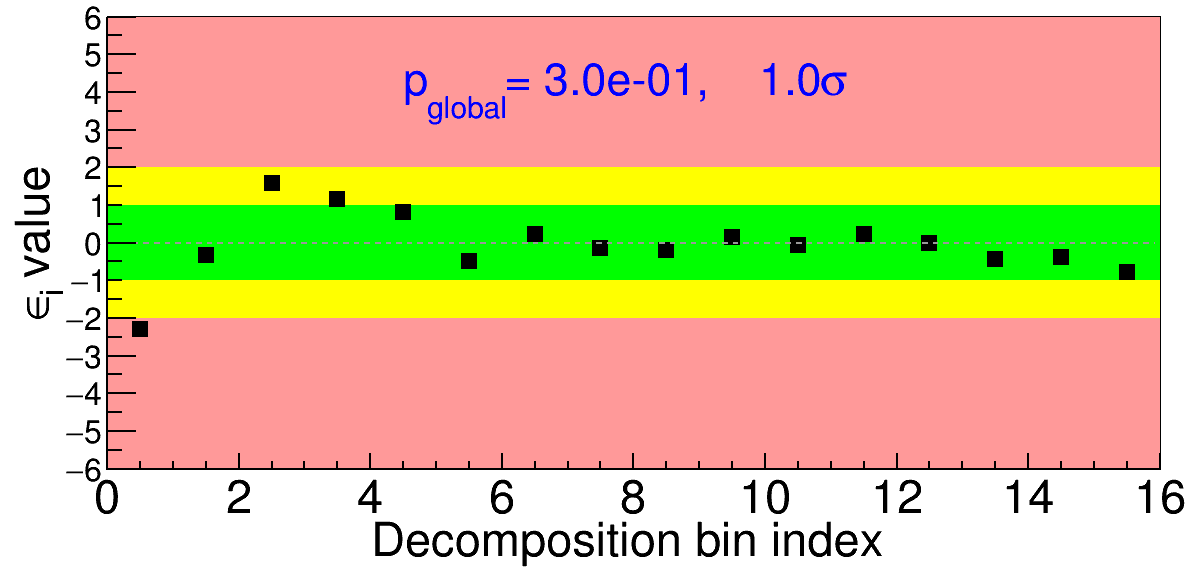}
  \caption{Model validation tests comparing the reconstructed hadronic energy for fake data and MC prediction for PC events. The top two pannels shows reconstructed space, with the red (blue) lines and bands showing the prediction without (with) the constraint from the muon kinematics. The uncertainties on the MC are shown in the bands and the statistical uncertainties on the data are shown on the data points. The bottom pannel shows the significance of the tension in each bin after the distribution has been constrained and transformed to the eigenvalue basis of the covariance matrix.}
  \label{PC_085}
  \end{subfigure}
  
 \begin{subfigure}[t]{0.49\linewidth}
  \includegraphics[width=\linewidth]{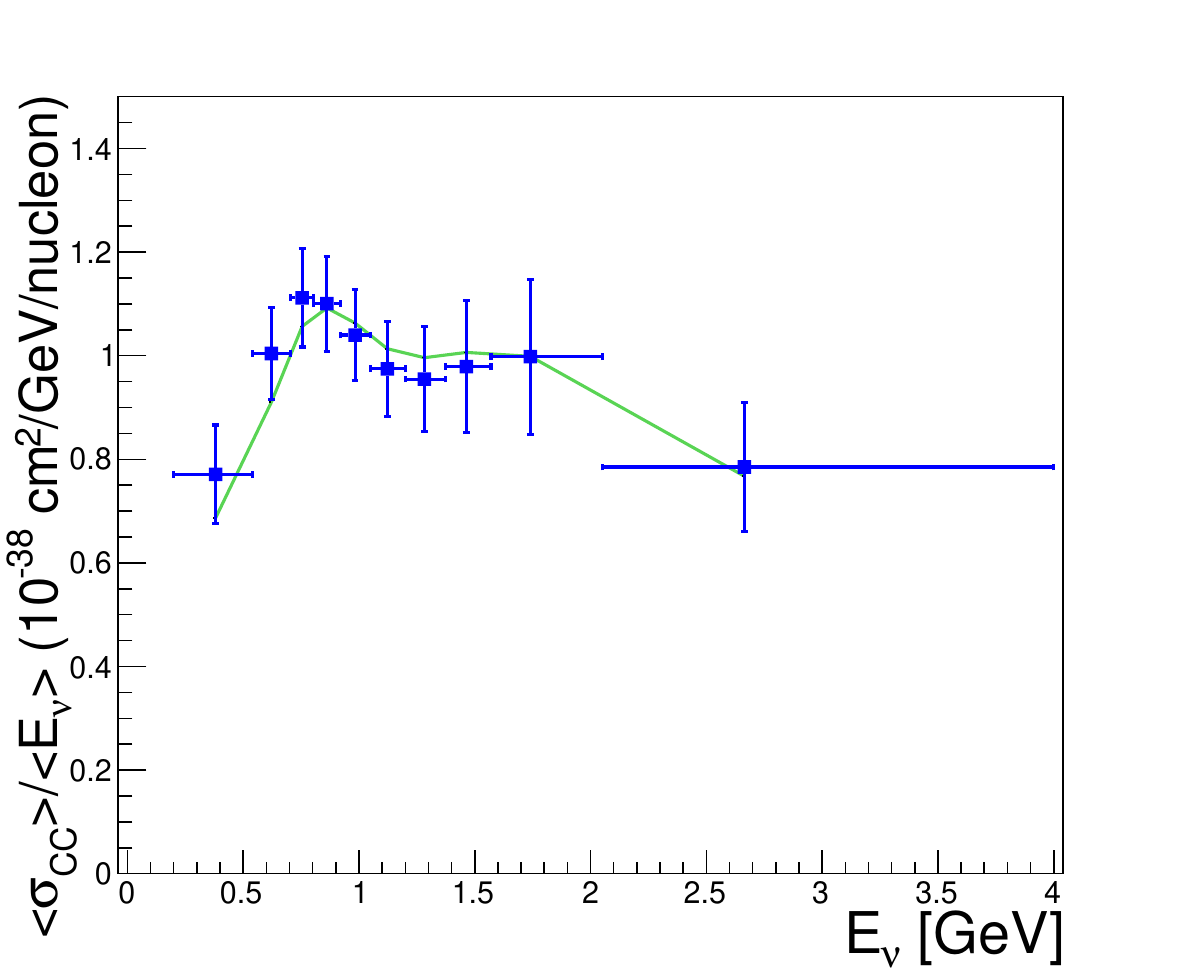}
    \put(-80,70){\includegraphics[width=0.5\linewidth]{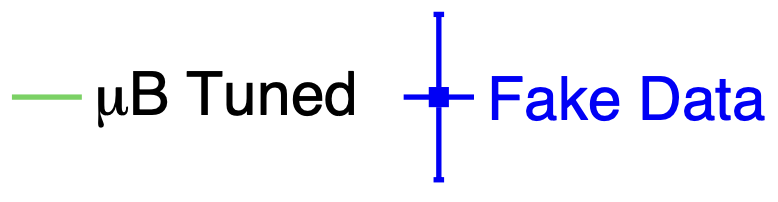}}
  \put(-85,37){\footnotesize{$K_p^\mathrm{rec}\times0.85$}}
  \put(-85,25){\footnotesize{$\chi^2/ndf=2.5/10$}}
  \put(-85,15){\footnotesize{$\sigma=0.0$}}
  
  \includegraphics[width=0.9\linewidth]{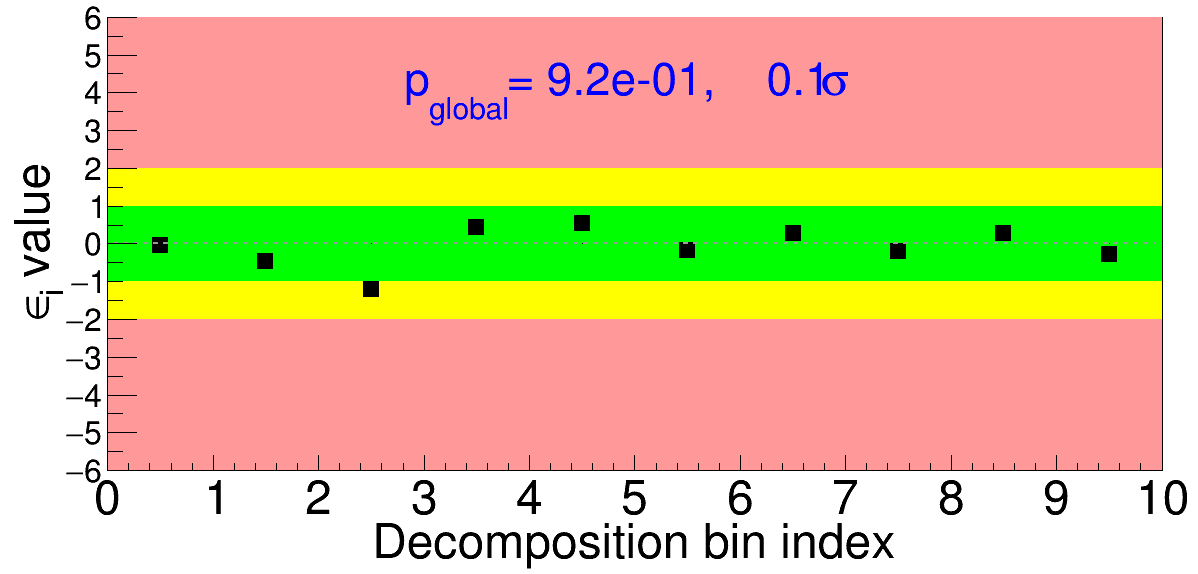}
  \caption{Extracted $\sigma(E_\nu)$.}
  \label{Enu_085}
  \end{subfigure}
  \begin{subfigure}[t]{0.49\linewidth}
  \includegraphics[width=\linewidth]{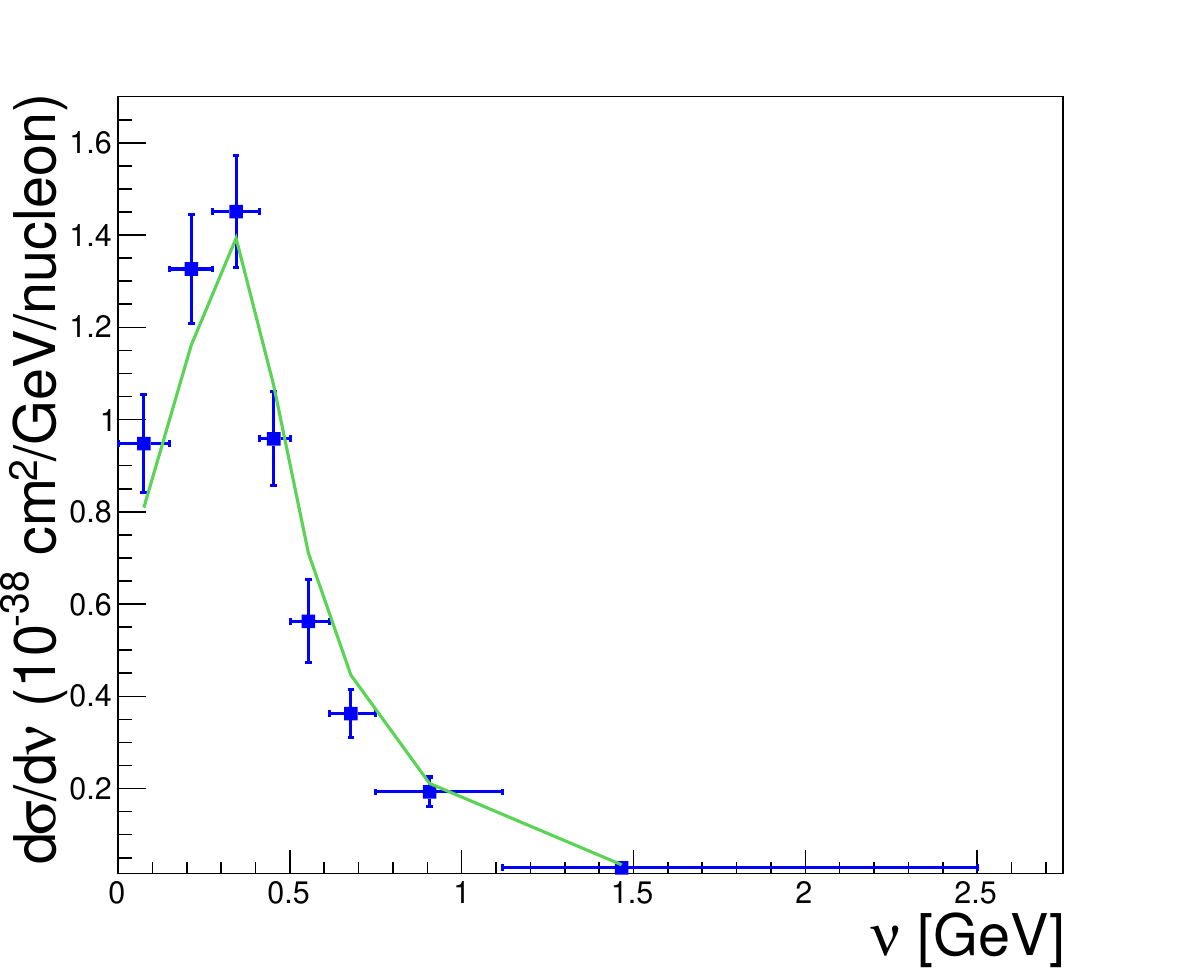}
      \put(-80,70){\includegraphics[width=0.5\linewidth]{fds_legend.png}}
  \put(-75,46){\footnotesize{$K_p^\mathrm{rec}\times0.85$}}
  \put(-75,34){\footnotesize{$\chi^2/ndf=6.8/8$}}
  \put(-75,25){\footnotesize{$\sigma=0.6$}}
  
  \includegraphics[width=0.9\linewidth]{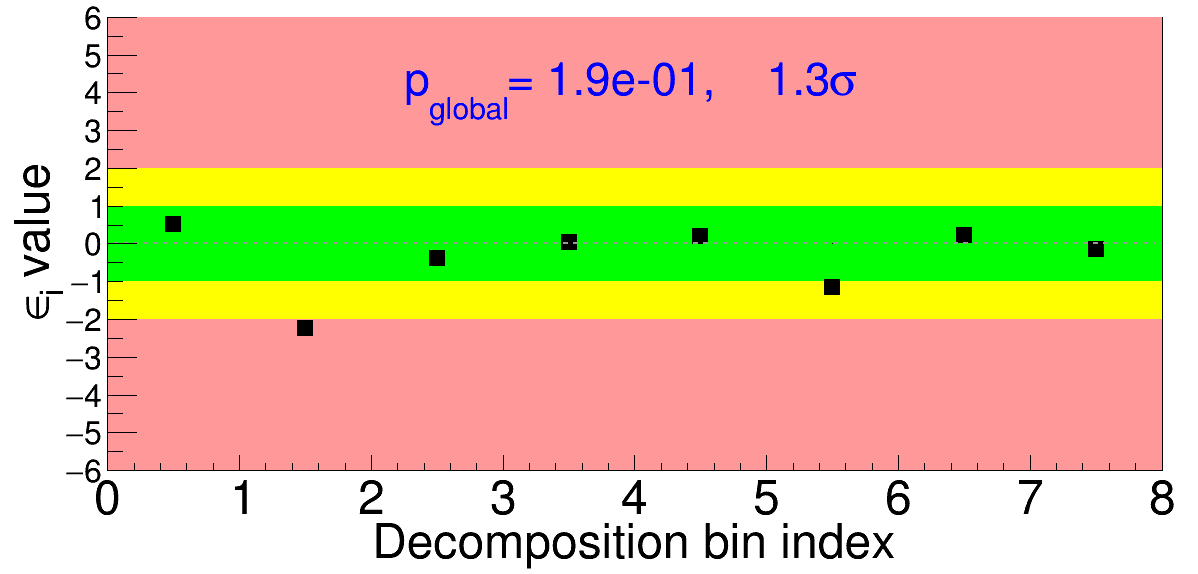}
  \caption{Extracted $d\sigma / d\nu$.}
  \label{nu_085}
  \end{subfigure}
 
\caption{The FDSs with the reconstructed proton energy scaled by 0.85. Select model validation tests are shown in (a). The extracted fake data cross section as a function of $E_\nu$ is shown in (b) and the extracted differential cross section as a function of $\nu$ is shown in (c). The $\chi^2$ displayed on these panels is calculated between the true distribution indicated by the green line and the extracted result. The top plots of (b) and (c) show the extracted result and the bottom panels show the significance of the tension in each bin after the distribution has been transformed to the eigenvalue basis of the covariance matrix. }
\label{FDS_085}
\end{figure}

\begin{figure}[!htp]
\centering
  
 \begin{subfigure}{\linewidth}
  \includegraphics[width=0.75\linewidth]{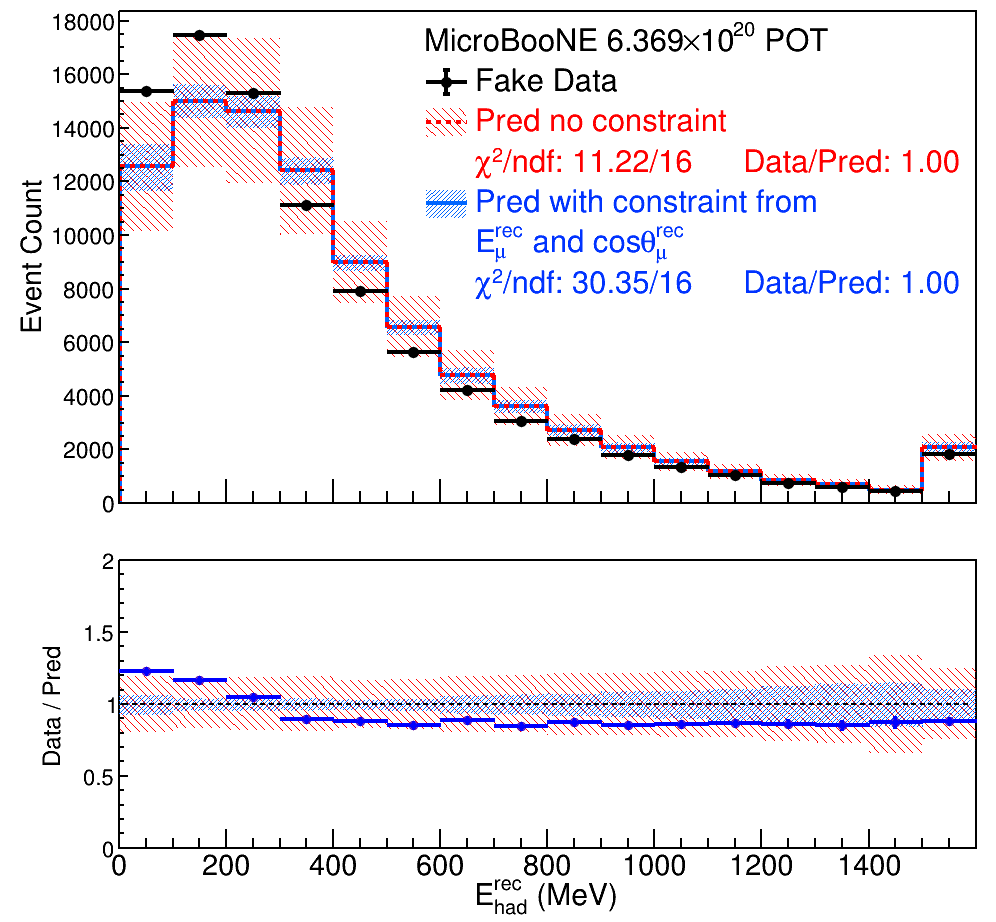}
  \put(-60,100){$K_p^\mathrm{rec}\times0.75$}
  \put(-60,90){$\sigma=2.4$}
  
    \includegraphics[width=0.65\linewidth]{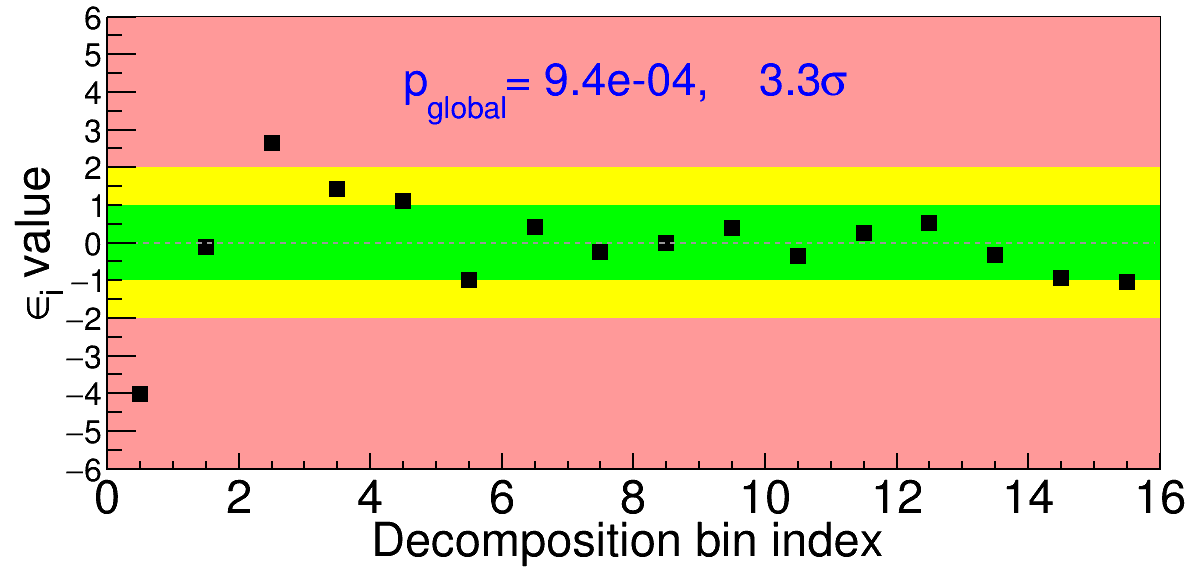}
  \caption{Model validation tests comparing the reconstructed hadronic energy for fake data and MC prediction for PC events.}
  \label{PC_075}
  \end{subfigure}
  
 \begin{subfigure}[t]{0.49\linewidth}
  \includegraphics[width=\linewidth]{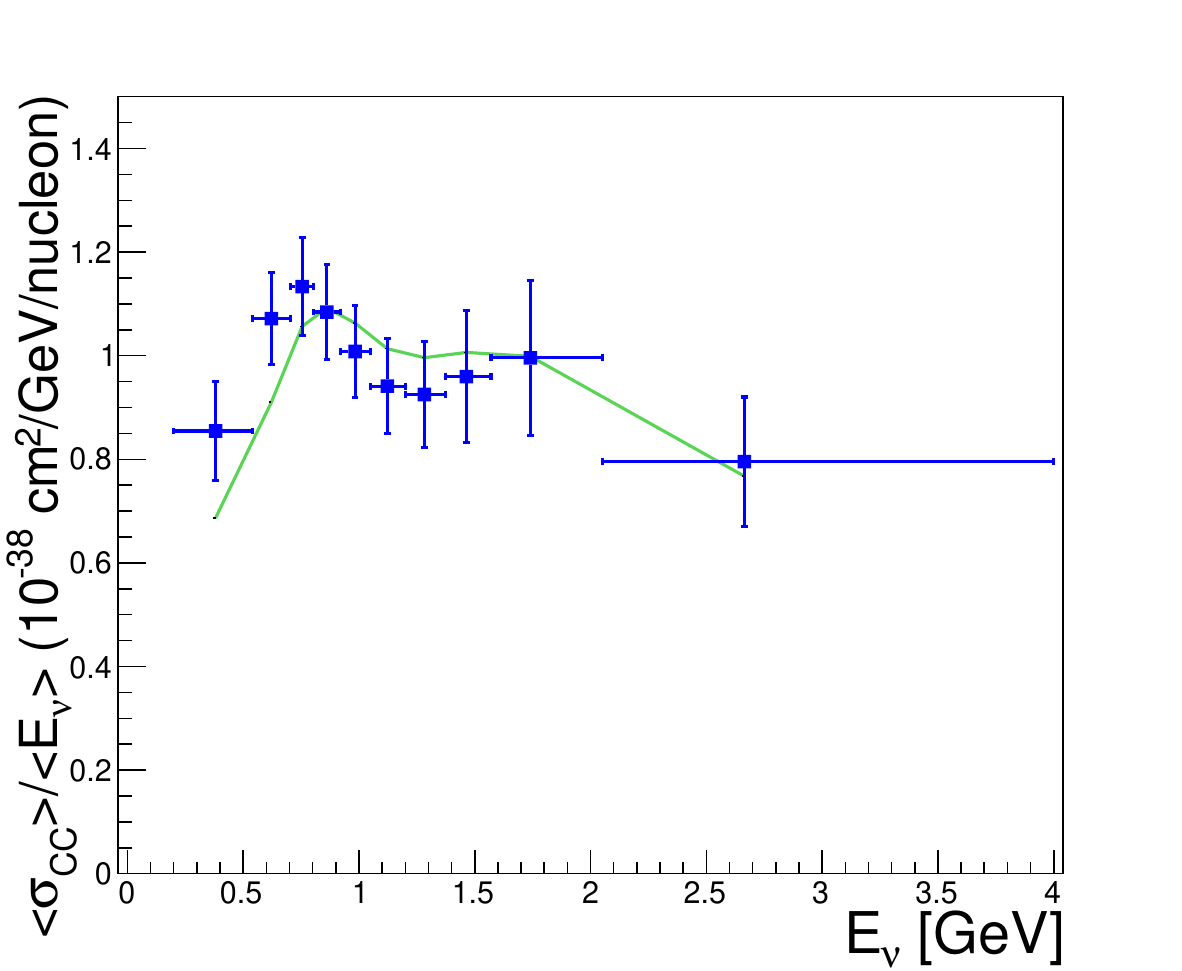}
    \put(-80,70){\includegraphics[width=0.5\linewidth]{fds_legend.png}}
  \put(-85,37){\footnotesize{$K_p^\mathrm{rec}\times0.75$}}
  \put(-85,25){\footnotesize{$\chi^2/ndf=8.1/10$}}
  \put(-85,15){\footnotesize{$\sigma=0.5$}}
  
  \includegraphics[width=0.9\linewidth]{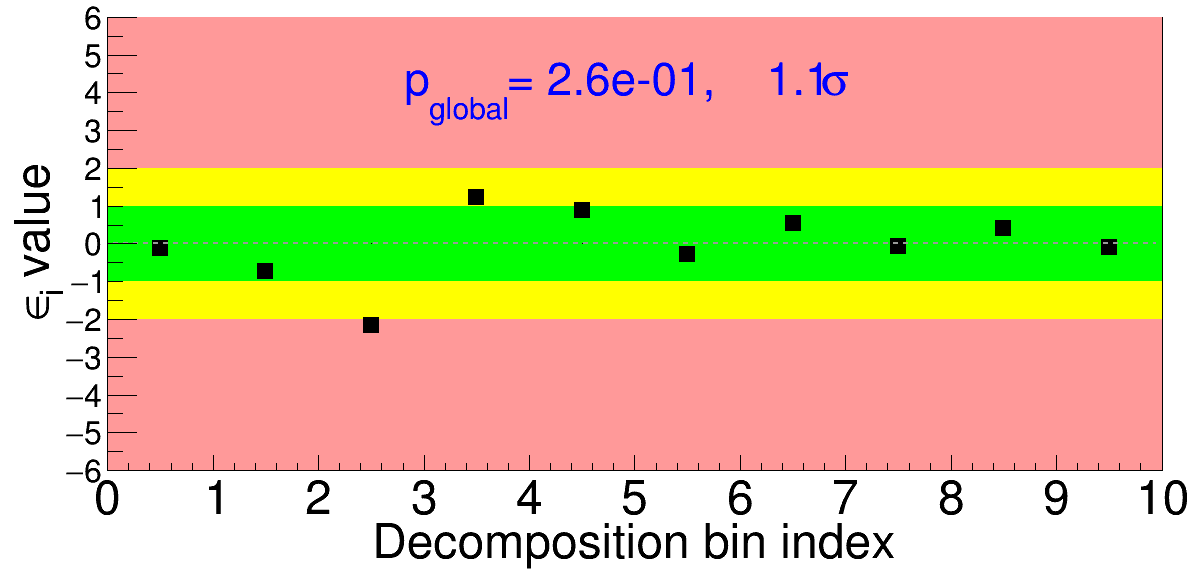}
  \caption{Extracted $\sigma(E_\nu)$.}
  \label{Enu_075}
  \end{subfigure}
  \begin{subfigure}[t]{0.49\linewidth}
  \includegraphics[width=\linewidth]{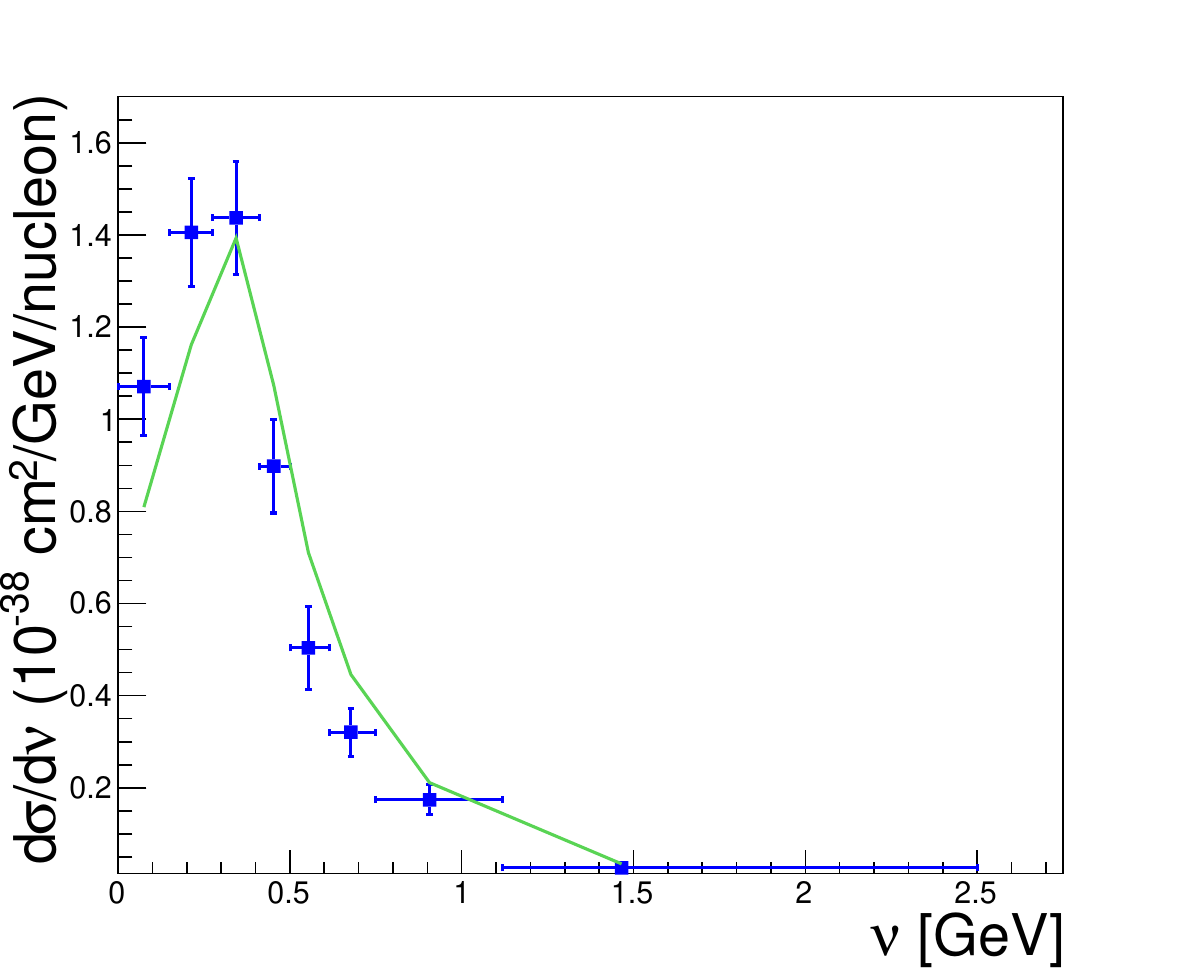}
      \put(-80,70){\includegraphics[width=0.5\linewidth]{fds_legend.png}}
  \put(-75,46){\footnotesize{$K_p^\mathrm{rec}\times0.75$}}
  \put(-75,34){\footnotesize{$\chi^2/ndf=17.7/8$}}
  \put(-75,25){\footnotesize{$\sigma=2.3$}}
  
  \includegraphics[width=0.9\linewidth]{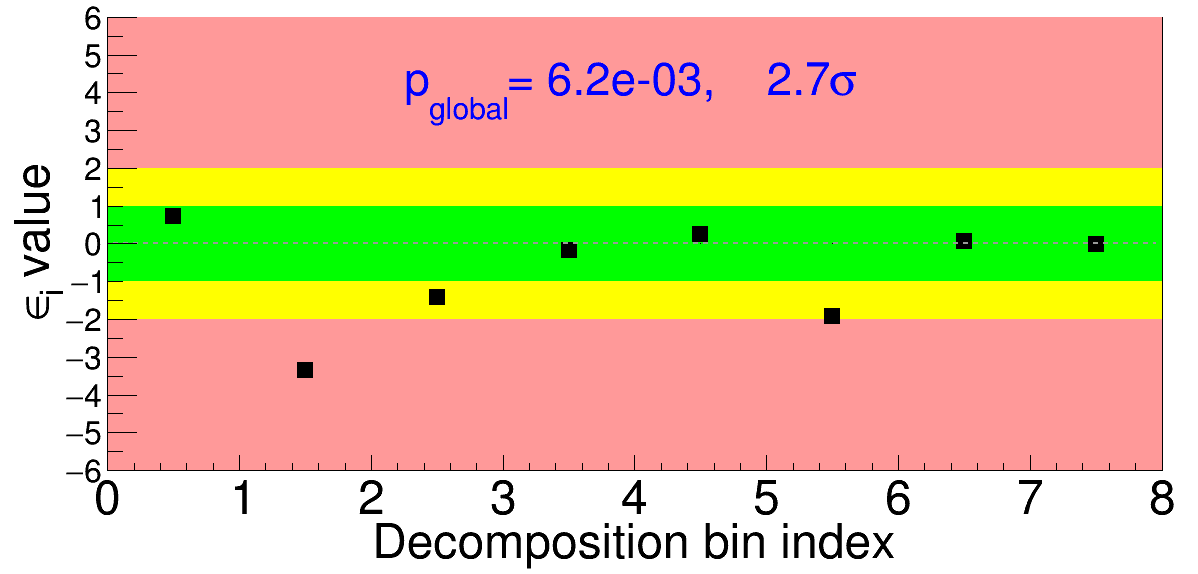}
  \caption{Extracted $d\sigma / d\nu$.}
  \label{nu_075}
  \end{subfigure}
 
\caption{Same as Fig.~\ref{FDS_085}, but for the fake data set with the reconstructed proton energy scaled by 0.75}
\label{FDS_075}
\end{figure}

As demonstrative examples, the extracted fake data cross sections and several of the more sensitive model validation tests for the fake data studies with visible proton energy scalings of 0.85 and 0.75 are shown in Figs.~\ref{FDS_085}~and~\ref{FDS_075}, respectively. Specifically, we show the model validation test which consists of evaluating  the GoF of the $E_\mathrm{had}^\mathrm{rec}$ distribution for PC events after the constraint from the muon kinematics. The corresponding $\chi^2$ decomposition is also shown. The combination of these two FDSs demonstrate the points outlined in Sec.~\ref{sec:fds_compare}. For the 0.85 scaling, the model passes the validation test depicted in Fig.~\ref{PC_085} as well as all other validation tests. This indicates that the uncertainties on the model are sufficient for this 15\% shift in the visible proton energy. The extracted cross sections likewise show little bias. This can be seen in Figs.~\ref{Enu_085}~and~\ref{nu_085}. The success of the cross section extraction is confirmed by examining the differences between the extracted results and the underlying truth in the eigenvalue basis of the covariance matrix, which eliminates correlations between bins. In this basis, the tension with the underlying truth is less than $1\sigma$ significance for the majority of the bins, with only a single bin of the extracted $d\sigma/d\nu$ decomposition falling slightly outside $2\sigma$ significance. This is illustrated in the bottom sub-panels of Figs.~\ref{Enu_085}~and~\ref{nu_085}, which show the significance of tension in this basis for each bin. 

The fake data set with the proton energy scaling of 0.85, shown in Fig.~\ref{FDS_085}, can be contrasted with what is seen for the proton energy scaling of 0.75, which is shown in Fig.~\ref{FDS_075}. Here, the model validation indicates the presence of discrepancies not covered by the systematics. In particular, the decomposition test on the PC distributions after constraint from muon kinematics shown in Fig.~\ref{PC_075} reveals disagreement at the $3.3\sigma$ significance level. If such a discrepancy were observed for real data, additional uncertainty or an otherwise expanded model would have been implemented to cover this difference and mitigate the potential for bias in the cross section extraction. Such a strategy was employed in Refs.~\cite{numuCC0pNp_PRD}~and~\cite{numuCC0pNp_PRL} when a modeling discrepancy related to the leading proton kinetic energy distribution relevant to the desired cross sections was identified. 

\begin{figure*}[ht!]
\centering
\begin{subfigure}{0.32\linewidth}
\includegraphics[width=\linewidth]{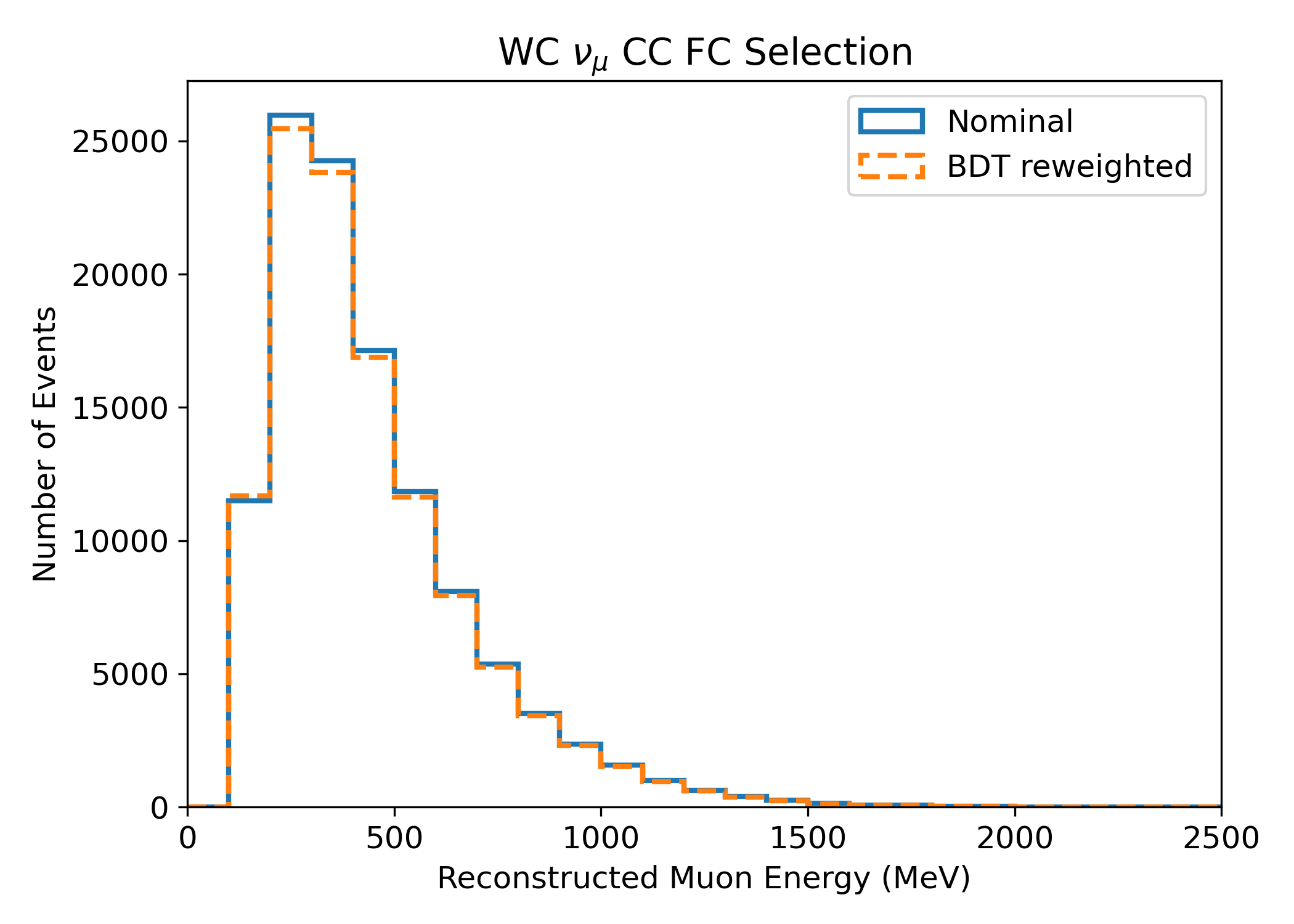}
\caption{\label{}}
\end{subfigure}
\begin{subfigure}{0.32\linewidth}
\includegraphics[width=\linewidth]{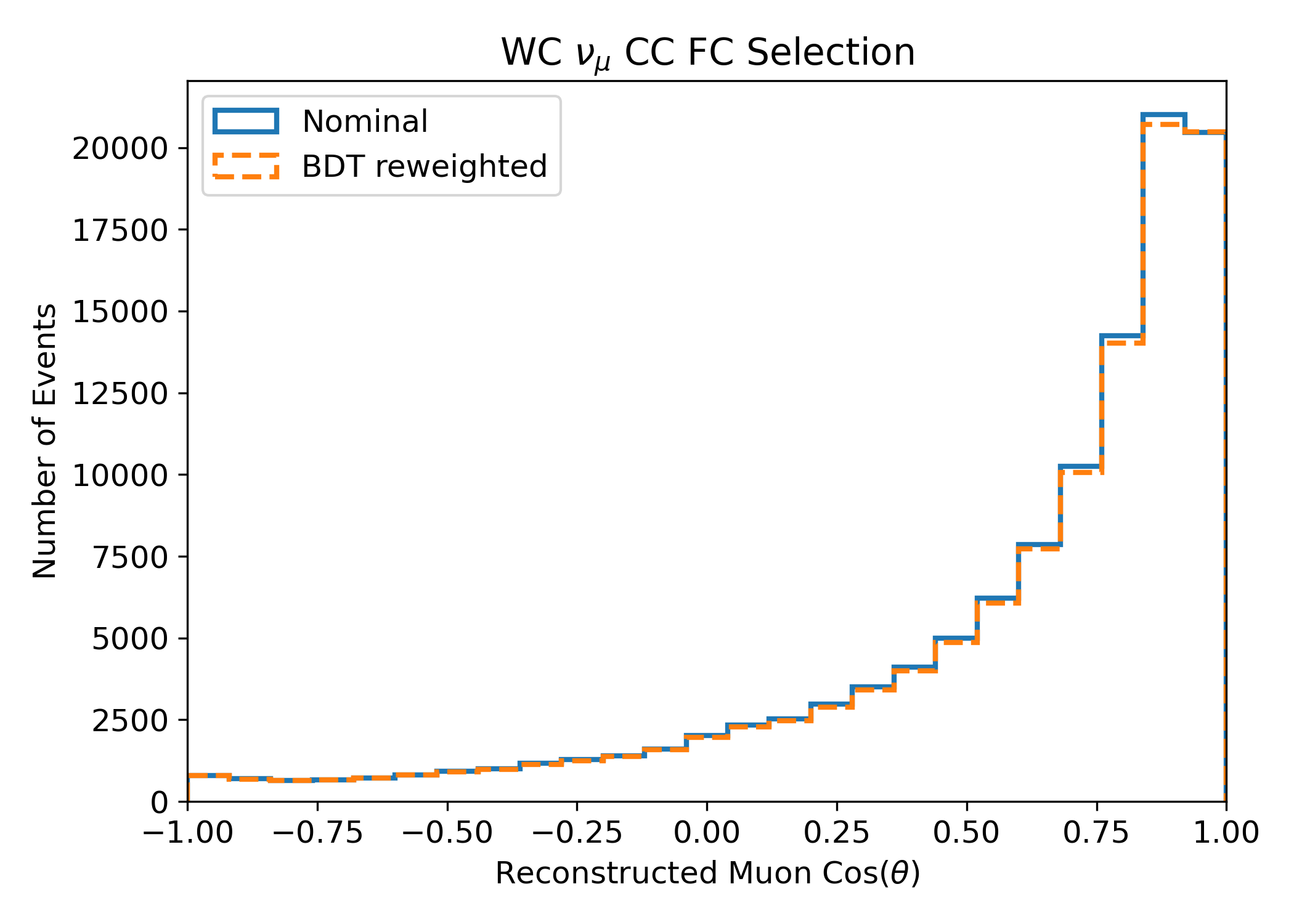}
\caption{\label{}}
\end{subfigure}
\begin{subfigure}{0.32\linewidth}
\includegraphics[width=\linewidth]{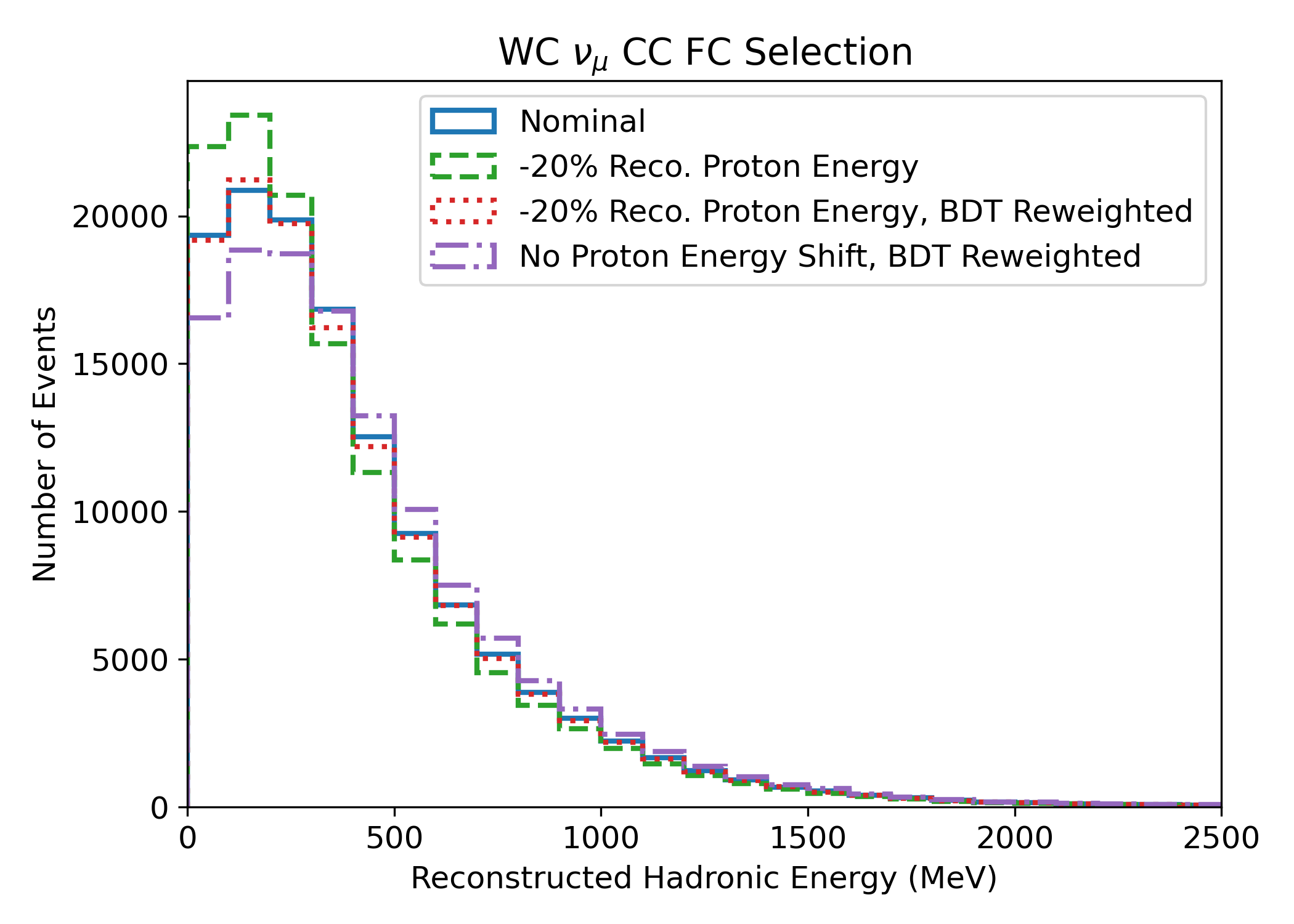}
\caption{\label{}}
\end{subfigure}
\caption{Reconstructed muon kinematic and hadronic energy distributions for events passing the $\nu_\mu$CC selection from~\cite{MicroBooNE:2021nxr} before and after the missing energy shift and multivariate event reweighting. The reweighting recovers good agreement in these variables despite the incorrect true-to-reconstructed energy mapping. Note that a proton energy shift has no impact on muon kinematic distributions.}\label{fig:prism_observations}
\end{figure*}

The level of tension identified in the model validation of the fake data set with the proton energy scaling of 0.75 indicates that this analysis should not proceed to the cross section extraction with the model and associated uncertainties in their present form. Nevertheless, for the purposes of this study, we proceed to cross section extraction in order to compare the sensitivity of the model validation to the bias induced in the cross section extraction. When the fake data cross section as a function of $E_\nu$ is extracted with the nominal model, a reasonable result is obtained. The extracted cross section and underlying truth only disagree at $0.5\sigma$ and $1.1\sigma$ significance in regularized truth space and in the eigenvalue basis, respectively. This is seen in the top and bottom panel of Fig.~\ref{Enu_075}, respectively. While the extracted $d\sigma/d\nu$ (Fig.~\ref{nu_075}) is in tension with the truth at $2.3\sigma$ significance, the discrepancy is less than is seen in the most sensitive model validation test, which shows tension at the $3.3\sigma$ level (Fig.~\ref{PC_075}). Furthermore, examination of the tension between the extracted fake data cross section and truth in the eigenvalue basis shows that all of the bins fall within $2\sigma$ significance, except for one, indicating a moderately successful unfolding despite the use of a model shown to be insufficient by the model validation.

In these FDSs, an interesting feature is illustrated by comparing the local $p$-values in the eigenvector basis of the covariance matrix obtained from the model validation to those of the cross section extraction. As can be seen by comparing Fig.~\ref{PC_085} to Fig.~\ref{nu_085} and Fig.~\ref{PC_075} to Fig.~\ref{nu_075}, the greatest tension seen in an individual bin is at a similar level in the extraction and validation. This trend is seen for all scalings and is perhaps unsurprising, as it is likely that the tension in these bins is originating from the same source of mis-modeling, which in the case of these FDSs, is the induced mis-modeling of the fraction of the energy transfer which is visible. This observation could potentially be utilized when, instead of performing model validation with data, FDSs and alternative event generator predictions are used to evaluate the sufficiency of the model. Rather than informing the need for additional uncertainties by evaluating the bias introduced in the extraction bin-by-bin in truth space, it may be useful to evaluate bias bin-by-bin in the eigenvalue basis instead. Evaluating the bias in truth space with highly correlated bins may be unable to account for discrepancies that are amplified or suppressed by bin-to-bin correlations. Such a scenario could be avoided via a transformation to the eigenvector basis where bins become uncorrelated and the totality of systematics can be taken into account while evaluating bias.

The proton energy scaling FDSs can be viewed as an extension to the ones shown in the Supplemental Material of~\cite{MicroBooNE:2021sfa,numuCC0pNp_PRD,MicroBooNE:2023foc} which were inspired by a similar DUNE FDS performed in Ref.~\cite{DUNE_ND_CDR}. The DUNE study utilized a fake data set with a 20\% reduction in the reconstructed proton energy for the purposes of studying the impact of mis-modeling on extracting oscillation parameters with a near and far detector. This fake dataset was then reweighted using a multivariate event reweighter to match the nominal MC in all the reconstructed distributions considered. However, as a result, there was still a significantly different mapping between the reconstructed and true neutrino energy in the reweighted fake data and the MC. Because of this, when a simultaneous near-far detector fit was performed on the reweighted fake data, the incorrect oscillation parameters were obtained despite the good agreement in the near detector.

Similar to the DUNE FDS, the FDSs presented in this section demonstrate that the total MicroBooNE MC uncertainties, of which the interaction model uncertainties are but a subset, cannot explain anything beyond a $\sim$20-25\% shift in the reconstructed proton energy. However, when comparing the conclusions of the DUNE FDS to the ones presented in this section, it is important to acknowledge the different set of assumptions in each. Though the multivariate event reweighter does not change reconstructed distributions and would indeed pass the model validation, the reweighting creates large shifts in the contributions from QE, RES, and DIS processes. As we will show, these modifications are likely beyond any reasonable event generator prediction. Since different interaction modes have distinct predictions for the muon kinematics, the multivariate event reweighter's freedom to make significant modifications to individual interaction modes would likely result in large tension when comparing the true $Q^2$ prediction from the event reweighter to any reasonable cross section model or event generator prediction and its associated uncertainties. 

\begin{figure}[!t]
\begin{subfigure}{0.52\linewidth}
\includegraphics[width=\linewidth]{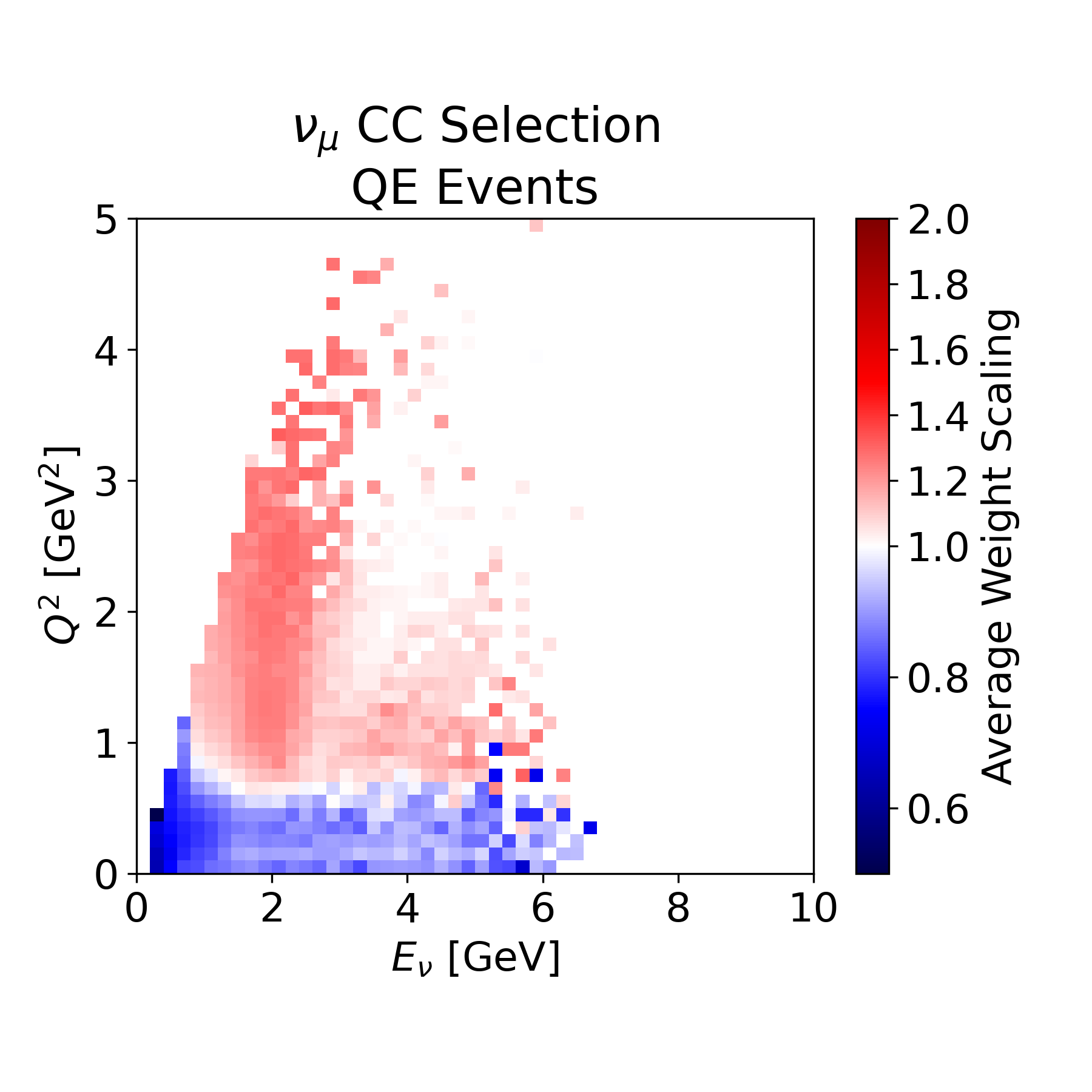}
\vspace*{-3.3mm}\caption{\label{}}
\end{subfigure}
\begin{subfigure}{0.465\linewidth}
\vspace{1.1mm}
\includegraphics[width=\linewidth]{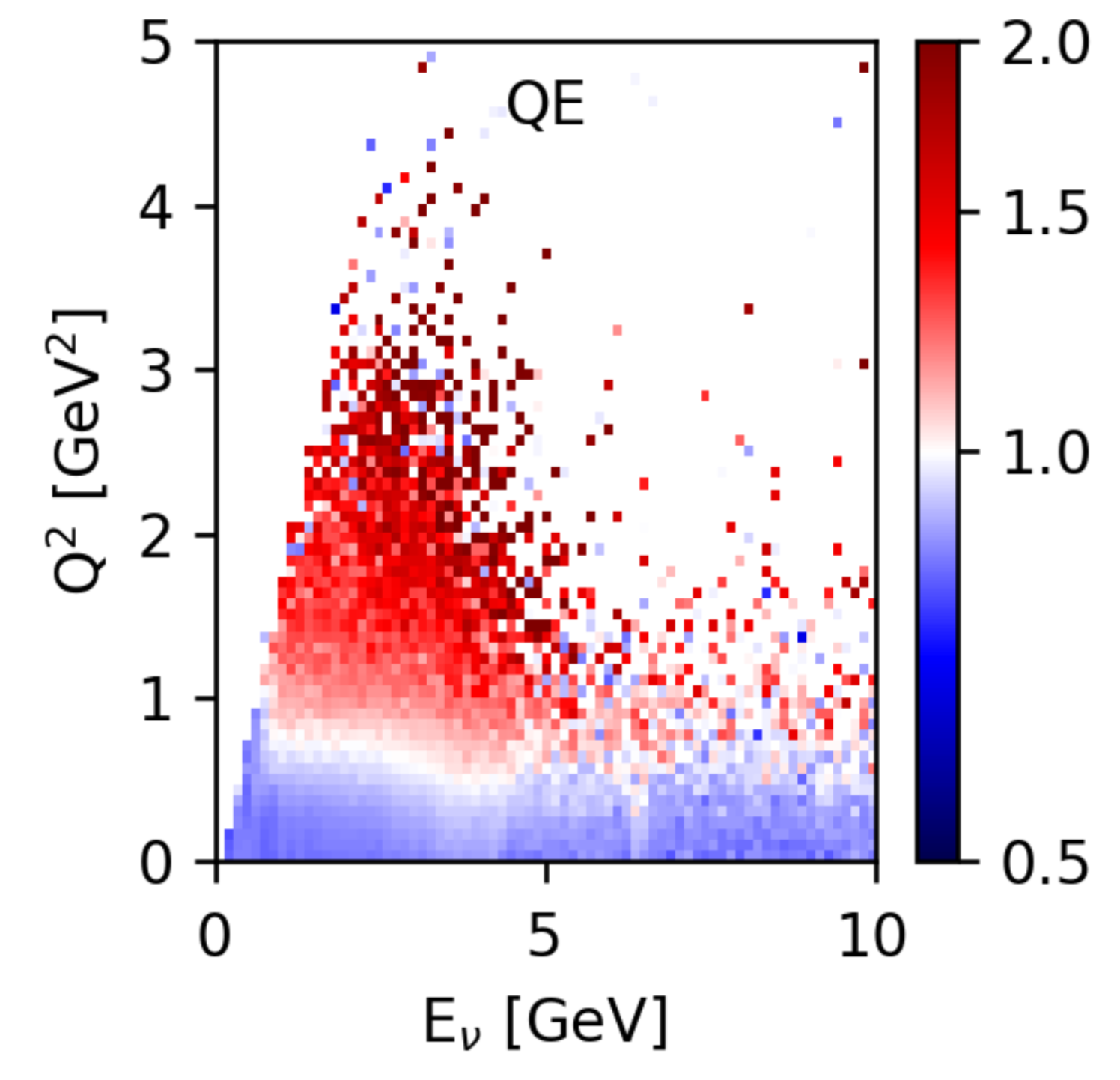}
\caption{\label{}}
\end{subfigure}
\caption{The (a) MicroBooNE BDT reweighted model and (b) DUNE BDT reweighted model from~\cite{DUNE_ND_CDR}.}\label{fig:prism_reweights_q2}
\end{figure}

To demonstrate this point, we emulate the DUNE FDS by training a boosted decision tree (BDT) multivariate event reweighter~\cite{bdt_reweight} to restore agreement between the fake data set with the $20$\% reduction in the proton energy and the MicroBooNE MC prediction. The reweighting is done in the true muon energy, true muon angle, and true transfer energy, which recovers good agreement between the MC and the fake data set with the $20$\% reduction in the proton energy in the analogous reconstructed distributions, as shown in Fig.~\ref{fig:prism_observations}. This allows the reweighted model to pass model validation despite a significantly different mapping between true and reconstructed distributions, which would create bias in the cross section extractions. For this particular example, good agreement is seen in all model validation tests consistent with tension at the $0\sigma$ level, but the extracted cross section as a function of the energy transfer shows bias at the 0.7$\sigma$ level when considering all systematic uncertainties. Though this is a relatively small amount of bias, it indicates that our model validation has not detected relevant mis-modeling. 

\begin{figure}[!t]
\centering
\includegraphics[width=0.85\linewidth]{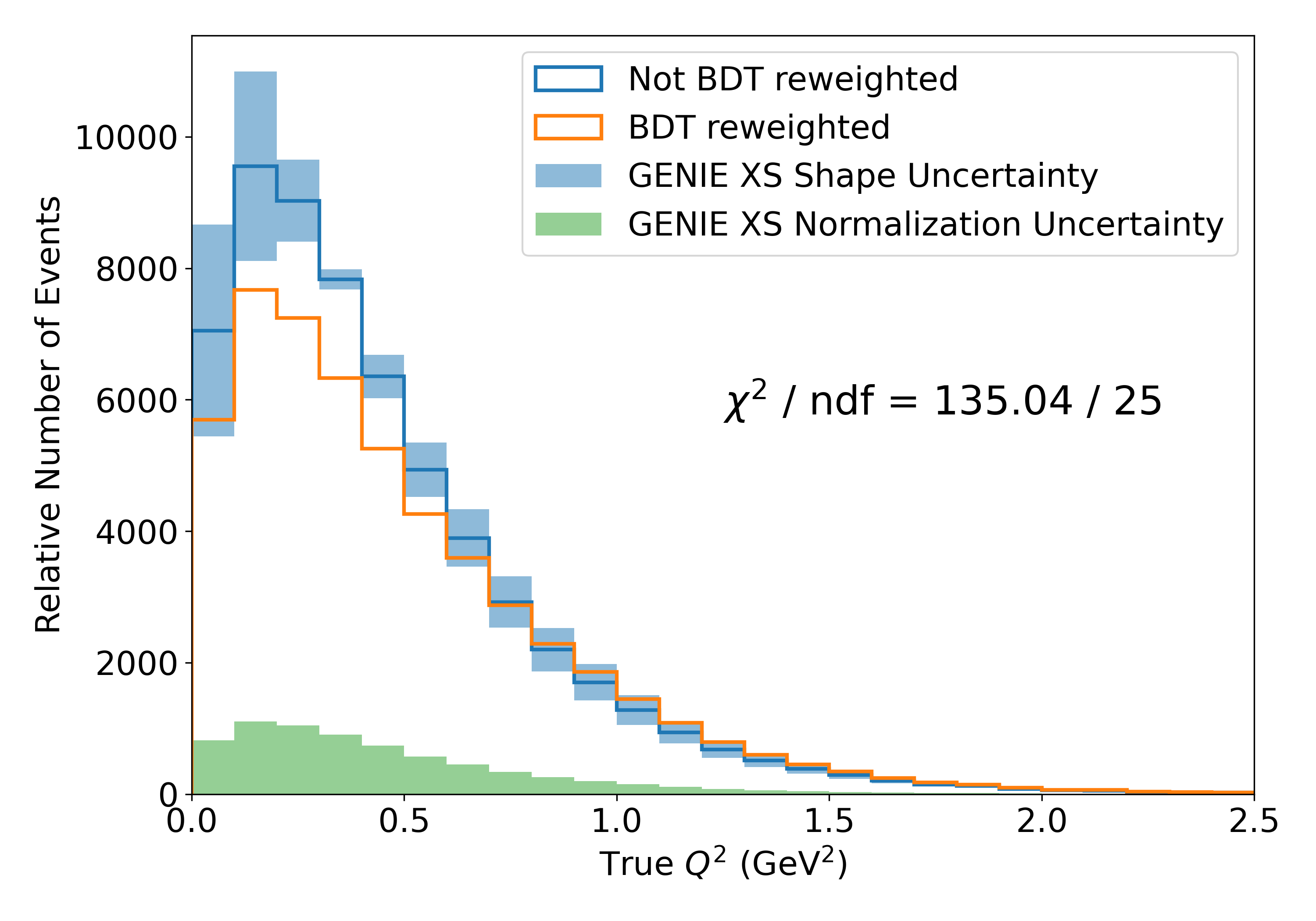}
\caption{Comparison between the BDT reweighted model and the GENIE MicroBooNE Tune prediction as a function of $Q^2$ for QE events. The normalization and shape components of the uncertainty bands are shown separately. The $\chi^2$ is calculated using all cross section uncertainties of the MicroBooNE tune.}\label{fig:prism_qe_q2_chi2}
\end{figure}

A comparison between the weights obtained in this study and those obtained in the DUNE FDS can be seen in Fig.~\ref{fig:prism_reweights_q2}. Similar behavior is observed with QE events at high $Q^2$ scaled up by as much as a factor of 2 and QE events at low $Q^2$ scaled down by as much as a factor of 0.6. As such, when the true $Q^2$ distribution for QE events is compared to the nominal model, a significant discrepancy is seen. This is demonstrated in Fig.~\ref{fig:prism_qe_q2_chi2}, where the $\chi^2/\mathrm{ndf}$ calculated between the reweighted distribution and the nominal model and its associated cross section uncertainties is $135.04/25$, indicating significant discrepancy. The basic lepton kinematics distributions for QE events are agnostic to final state interactions in the limit of complete factorization of the initial and final interactions, which is a well realized approximation employed by state-of-the-art neutrino event generators \cite{GenCompare,WP_NuScat}. As such, being supported by many neutrino and electron scattering experiments, these distributions are one of the better understood portions of cross section modeling, making such a large discrepancy with the nominal model hard to justify.

\begin{figure}[!t]
\centering
\includegraphics[width=0.85\linewidth]{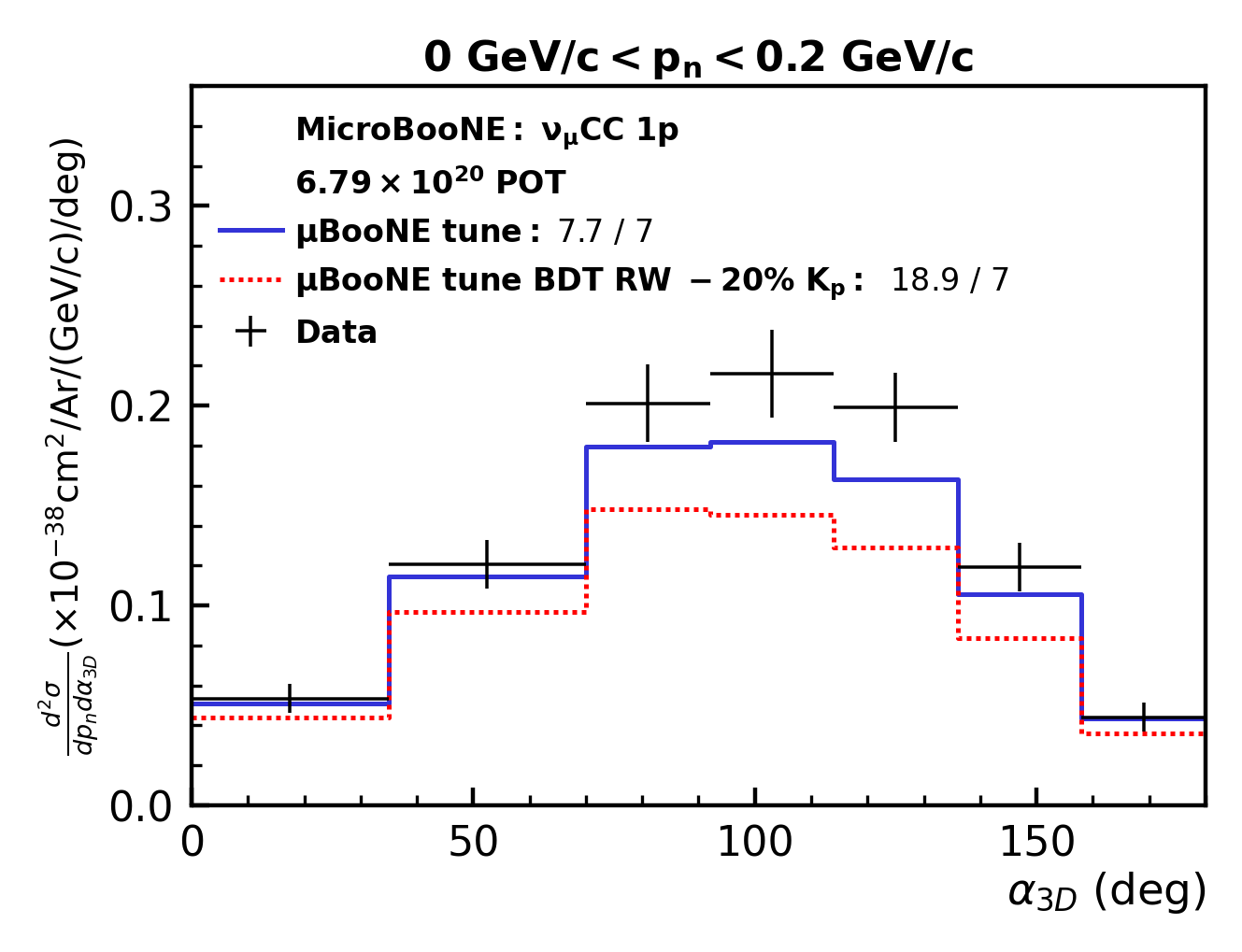}
\caption{Comparison of the BDT reweighted model prediction and various event generator predictions to MicroBooNE $\nu_\mu$CC1p data from~\cite{uboone_gki}. The $\chi^2/\mathrm{ndf}$ values in the legend are calculated using the reported covariance matrix and each prediction is smeared by the reported $A_C$ matrix.}\label{fig:prism_cc1p}
\end{figure}

To explore this discrepancy further, the BDT reweighted model was compared to QE-like MicroBooNE $\nu_\mu$CC1p data from~\cite{uboone_cc1p,uboone_gki}. This dataset is dominated by QE events, and, in select regions of phase space, it achieves an estimated $\sim$95\% pure QE sample. The impact of the BDT-reweighted model's significantly smaller QE prediction is illustrated in Fig.~\ref{fig:prism_cc1p}, where the MicroBooNE tune prediction with and without the BDT weights is compared to the extracted double-differential cross section as a function of $\alpha_{3D}$ in the $p_n<0.2$~GeV range. These variables describe the magnitude of the missing momentum ($p_n$) and the angle between the missing momentum vector and the momentum transfer vector ($\alpha_{3D}$). The low phase space $p_n$ region consists almost exclusively of QE events where the final state proton has not experienced significant final state interactions. The BDT reweighted model underestimates the data in this region and is in tension with the data at 2.6$\sigma$ significance. This is noticeably worse than the nominal model without the BDT weights, which shows tension at only 0.9$\sigma$ significance. Worsened agreement that approaches this level of increased tension is seen across the other distributions as well, indicating that the BDT model is not preferred by this data.

Encountering a situation in real data analogous to the multivariate event reweighter, which would pass the model validation but has an entirely different mapping between true and reconstructed neutrino energy, cannot be completely ruled out. However, such a scenario is unlikely as lepton kinematic distributions are well constrained by electron scattering data, particularly for QE events. The BDT reweighting we explore here shows noticeable tension with the QE prediction from the unreweighted model and performs worse when compared to the QE-like $\nu_\mu$CC1p dataset from~\cite{uboone_cc1p,uboone_gki}. The DUNE study demonstrates that, through the use of PRISM, one can still obtain a robust measurement of oscillation parameters even in such a scenario with large deviations from reasonable models, which would be required if one wished to perform an oscillation analysis with minimal assumptions on the cross section model. However, techniques such as PRISM are not applicable to many experiments, such as MicroBooNE, and other strategies like data-driven model validation are required. Likewise, we caution that the general conclusions about the model validation drawn from the FDSs present in this work may not extrapolate as well onto scenarios far outside our current understanding of neutrino-nucleus scattering. The same holds true if more ``tradition" FDSs are used to validate a model for a cross section measurement; one should be cautious extrapolating the results of successful closure on alternative models well beyond the bounds of reasonable models if these alternative models do not also extend well beyond these bounds. Indeed, this is directly related to the three issues raised earlier on ``tradition" FDSs. There is no guarantee that a model spread will cover nature (issue 1), but including too many extreme scenarios may result in overestimated uncertainties (issue 2), and it is not clear when one should stop examining closure on these alternative models, particularly when extreme deviations are included (issue 3). Given the extremity of the example, this DUNE FDS is not inconsistent with the notion that one is able to use data-driven validation to detect mis-modeling of the mapping between true and reconstructed neutrino energy to enable the extraction of energy dependent cross sections. 

\begingroup
\setlength{\tabcolsep}{8pt} 
\renewcommand{\arraystretch}{1.2} 

\begin{table*}[ht!]

\begin{tabular}{|| c c c | c c || c c c | c c  ||} 
\hline
\multicolumn{5}{||c||}{All Uncertainties} & \multicolumn{5}{c||}{XS Syst+Stat Uncertainties} \\
\hline
\multicolumn{3}{||c|}{Bias in Extraction } & \multicolumn{2}{c||}{Model Validation $\cos\theta_\mu^\mathrm{rec} | E_\mu^\mathrm{rec}$ } & \multicolumn{3}{c|}{Bias in Extraction } & \multicolumn{2}{c||}{Model Validation $\cos\theta_\mu^\mathrm{rec} | E_\mu^\mathrm{rec}$ } \\
 
$E_\nu$ & $E_\mu$ & $\nu$  & GoF & Decomposition & $E_\nu$ & $E_\mu$ & $\nu$ &  GoF & Decomposition \\ 
 \hline
 0.1 & 0.0 & 0.1 & 0.0 & 0.4 & 2.6 & 0.1 & 1.5  & 3.1 & 3.9 \\
 \hline
\end{tabular}
\caption{\label{tab:fakedata_summary}Results of the model validation and cross section extraction for the $\texttt{NuWro}$ FDSs. Two treatments of systematic uncertainties are considered: one with full systematic uncertainties (``All Uncertainties") and the other with only cross section induced systematic uncertainties (``XS Syst+Stat Uncertainties"). Statistical uncertainties are always included. The Bias in Extraction columns indicate  the significance of the tension between the extracted result and $\texttt{NuWro}$ prediction. The entries in the Model Validation columns indicate the significance at which the fake data and the MC used for the cross section extraction disagree in the test that shows the most tension. The significance levels were obtained using $p$-values calculated from the $\chi^2$ test statistic interpreted assuming a $\chi^2$ distribution with degrees of freedom equal to the number of bins.}
\end{table*}
\endgroup

\subsubsection{Alternative Event Generator Fake Data Studies}
~\label{sec:add_fds}
The results of the FDSs utilizing fake data produced by $\texttt{NuWro 19.02.2}$ are summarized in Table~\ref{tab:fakedata_summary}. This set of fake data was generated at 6.11$\times10^{20}$ POT, which is approximately equal to the exposure of the first three runs of data used for recent MicroBooNE cross section measurements~\cite{MicroBooNE:2023foc,numuCC0pNp_PRD,numuCC0pNp_PRL}. The fake data contains statistical fluctuations in addition to the difference in event generators. As in Sec.~\ref{sec:FDS_proton}, the amount of discrepancy between the fake data and MicroBooNE MC prediction identified in the model validation is compared with the deviations of the extracted cross sections from the truth. This is done by converting the $p$-values obtained in each test or extraction into significance levels corresponding multiples of the standard deviation of a normal distribution. Both the full systematic uncertainty and cross section systematic plus statistical uncertainty situations are considered, which are labeled ``All Uncertainties" and ``XS Syst+Stat Uncertainties", respectively. The results for each scenario are described throughout this section. 

\begin{figure}[t!]
\centering
\begin{subfigure}{\linewidth}
\includegraphics[width=0.85\linewidth]{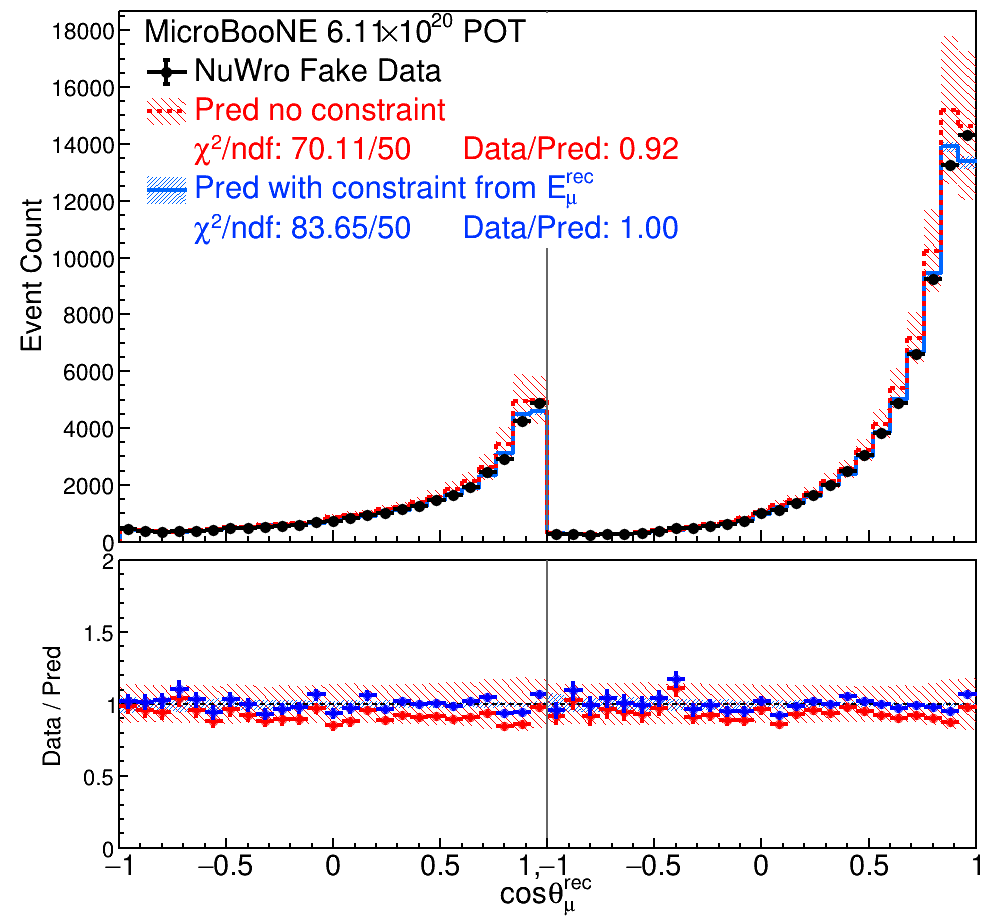}
\put(-179,127){\small{XS Syst+Stat}}
\put(-179,117){\small{Uncertainties}}
\put(-177,95){\small{FC}}
\put(-87,95){\small{PC}}
\end{subfigure}
\begin{subfigure}{\linewidth}
\includegraphics[width=0.85\linewidth]{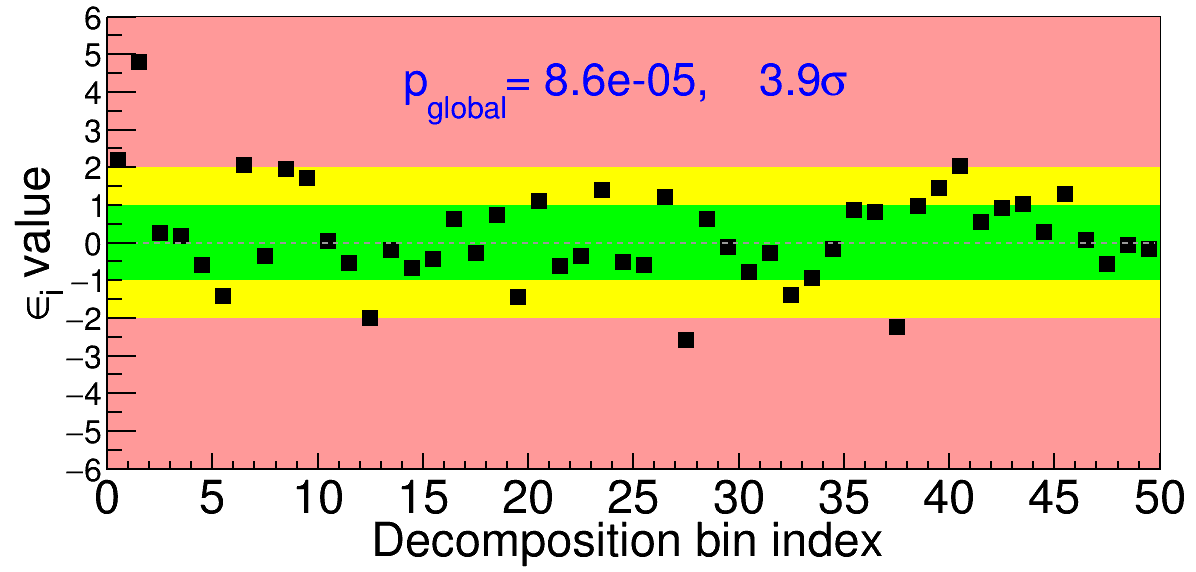}
\end{subfigure}
\caption{\label{fig:model_validation_nuwro_xs_only} Model validation tests comparing the $\texttt{NuWro}$ fake data sample to the nominal MC prediction for the reconstructed muon scattering angular distribution for FC\&PC events. In (a), the red (blue) lines and bands show the prediction without (with) the constraint from the observed muon energy distribution for FC\&PC events. The uncertainties of the prediction only include the cross section and statistical terms and are shown in the bands. The data statistical uncertainties are shown on the data points. In (b), the significance of the tension in each bin after the distribution has been constrained and transformed to the eigenvalue basis of the covariance matrix is shown.}
\vspace*{-4mm}
\end{figure}

\begin{figure*}[!ht]
\centering
\begin{subfigure}{0.3\linewidth}
\includegraphics[width=\linewidth]{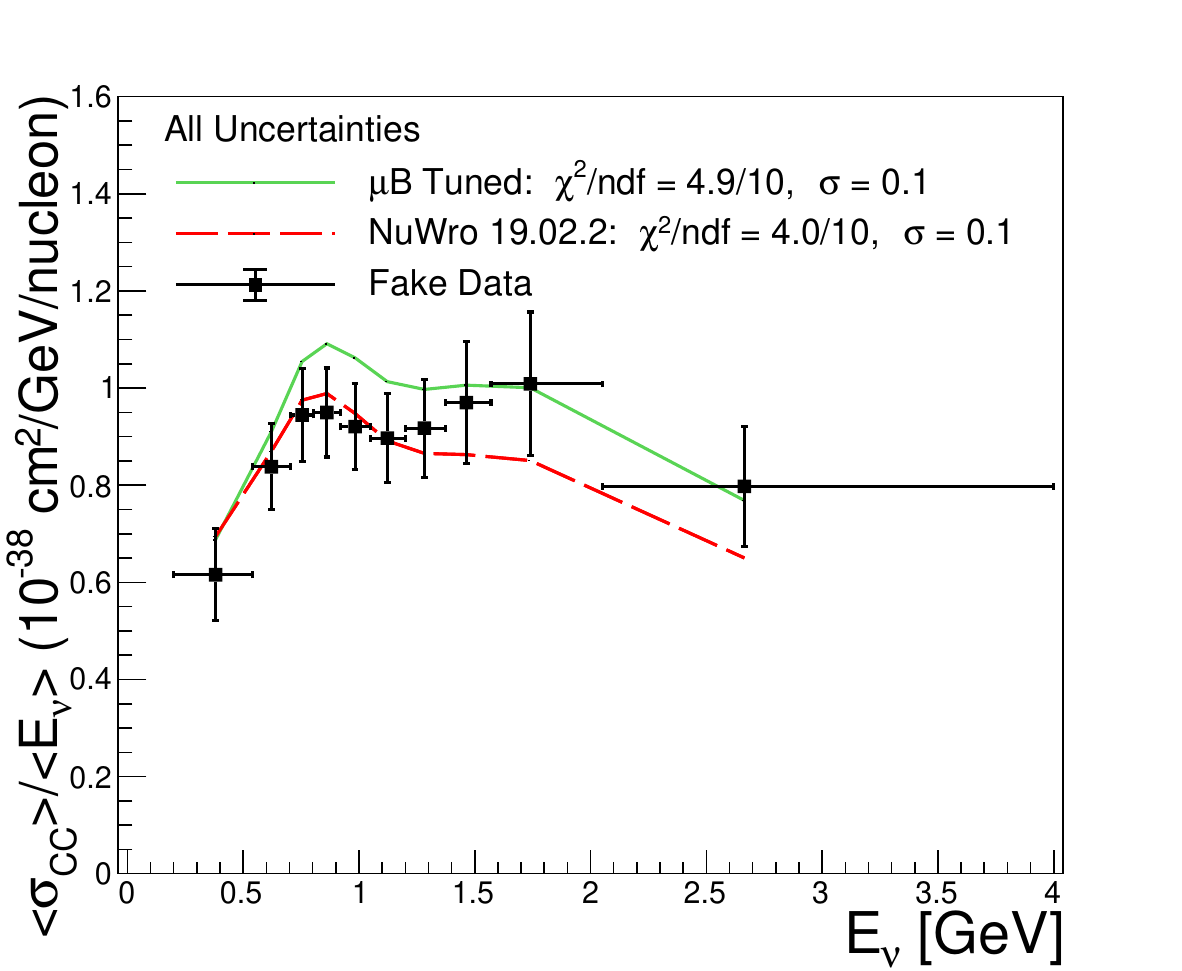}
\put(-35,25){(a)}
\phantomsubcaption
\label{fig:ccinc_enu_FDS_nuwro_all_sys}
\end{subfigure}
\begin{subfigure}{0.3\linewidth}
\includegraphics[width=\linewidth]{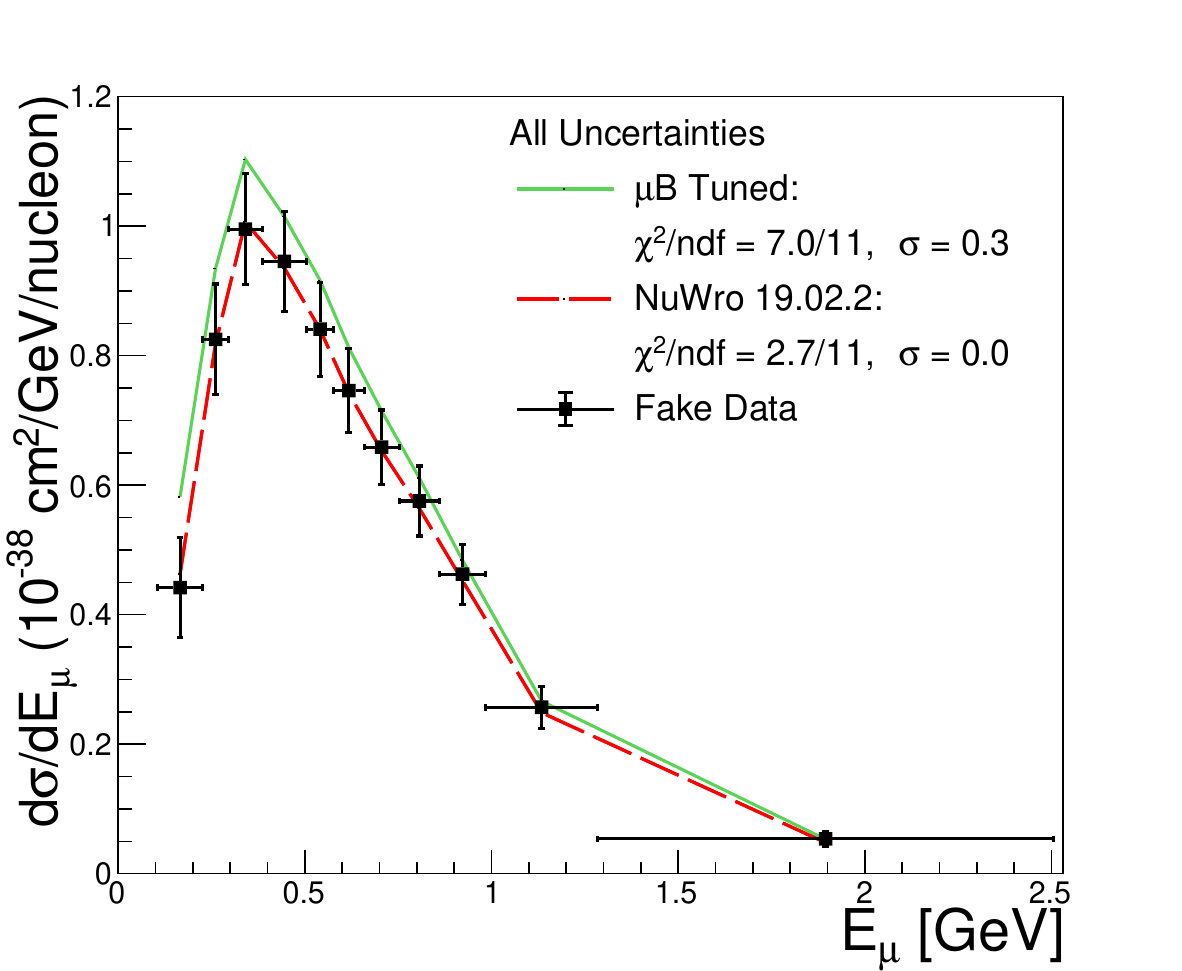}
\put(-35,25){(b)}
\phantomsubcaption
\label{fig:ccinc_emu_FDS_nuwro_all_sys}
\end{subfigure}
\begin{subfigure}{0.3\linewidth}
\includegraphics[width=\linewidth]{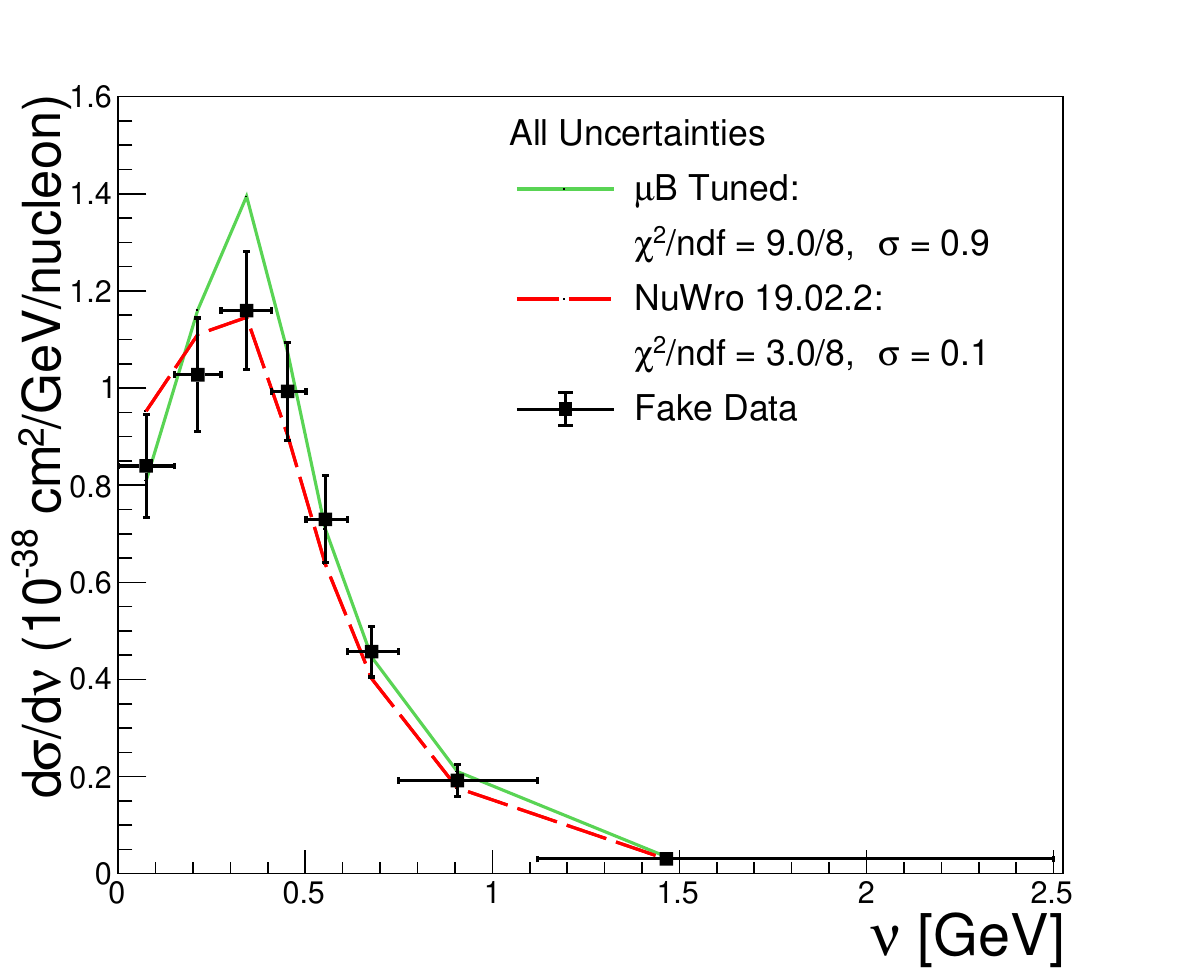}
\put(-35,25){(c)}
\phantomsubcaption
\label{fig:ccinc_nu_FDS_nuwro_all_sys}
\end{subfigure}
\begin{subfigure}{0.3\linewidth}
\includegraphics[width=\linewidth]{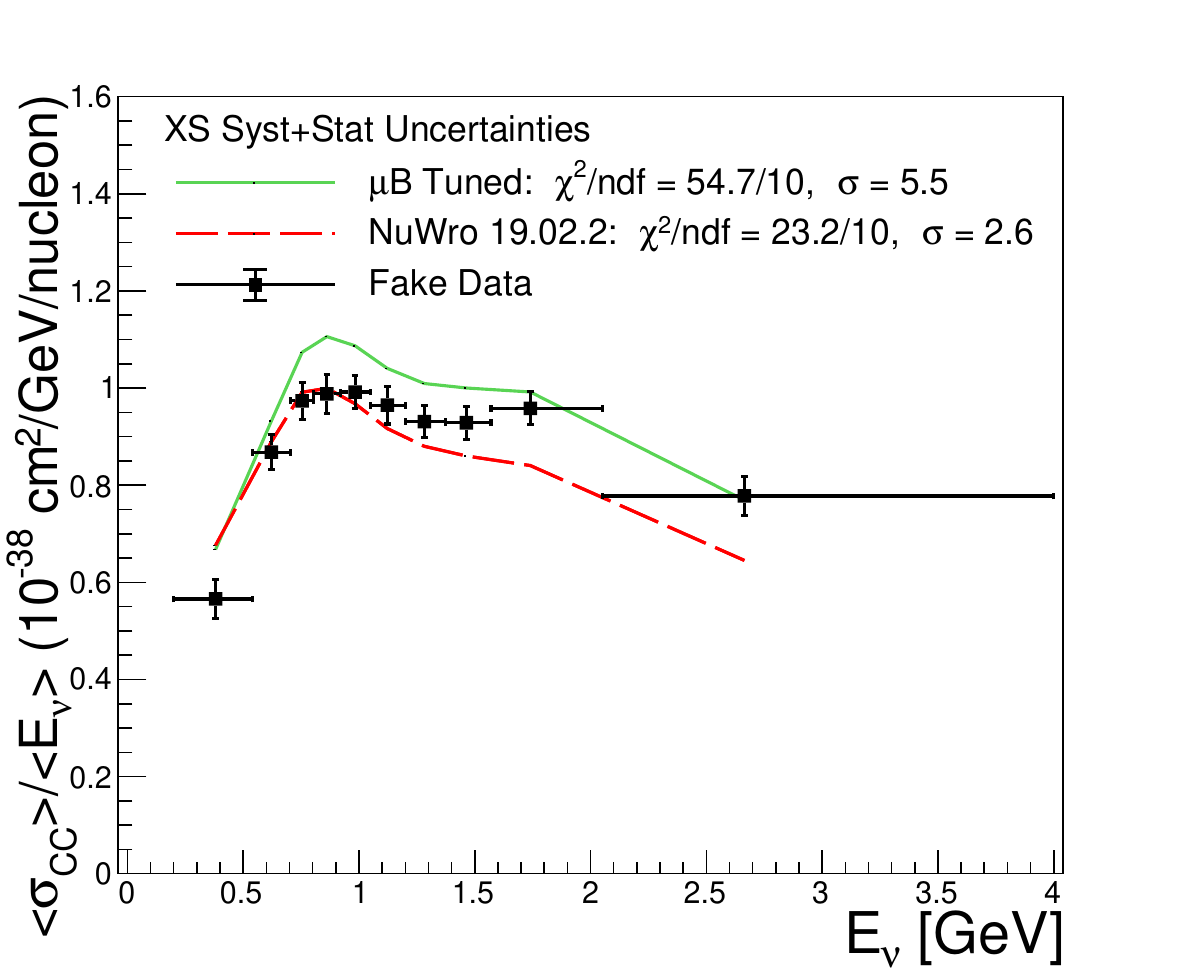}
\put(-35,25){(d)}
\phantomsubcaption
\label{fig:ccinc_enu_FDS_nuwro_xs_sys}
\end{subfigure}
\begin{subfigure}{0.3\linewidth}
\includegraphics[width=\linewidth]{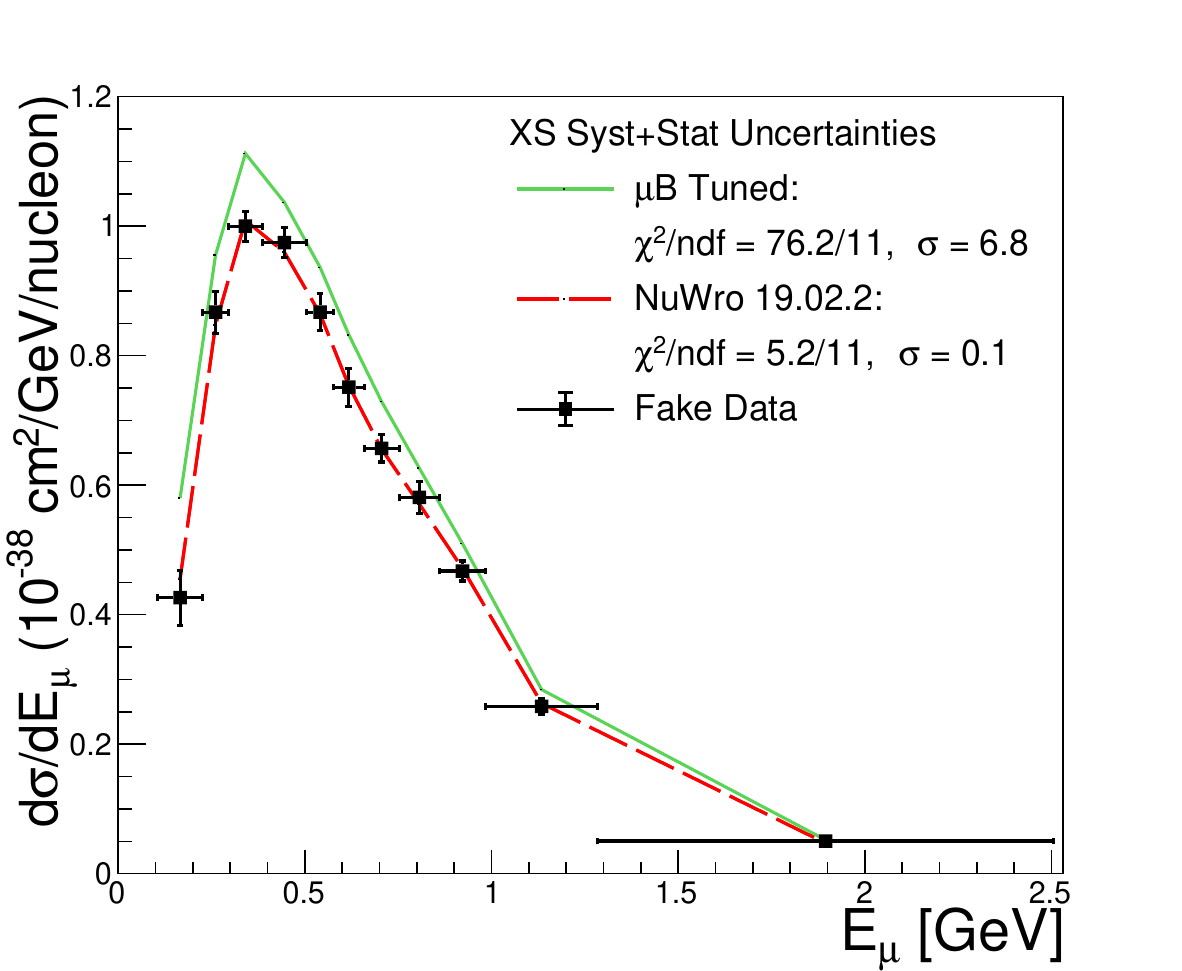}
\put(-35,25){(e)}
\phantomsubcaption
\label{fig:ccinc_emu_FDS_nuwro_xs_sys}
\end{subfigure}
\begin{subfigure}{0.3\linewidth}
\includegraphics[width=\linewidth]{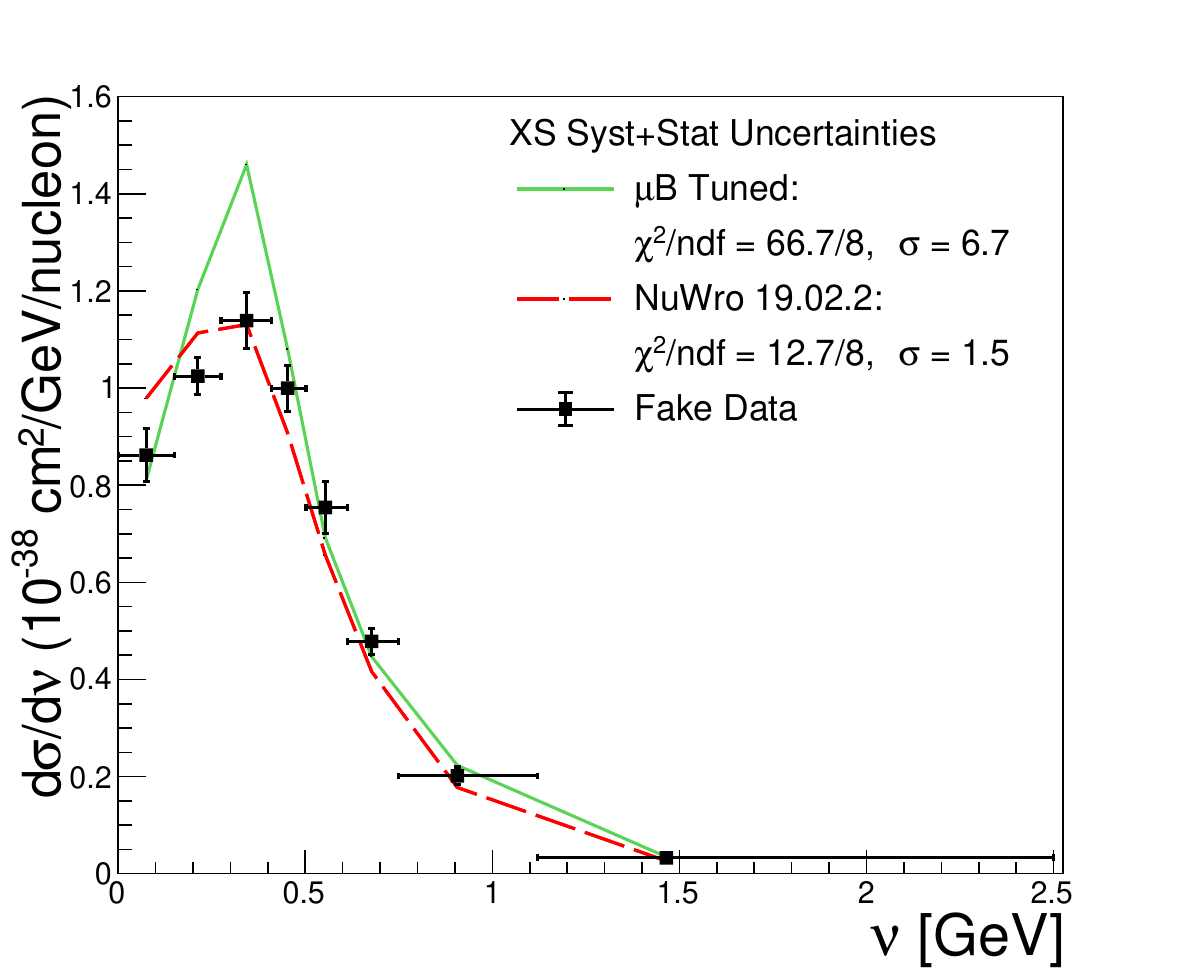}
\put(-35,25){(f)}
\phantomsubcaption
\label{fig:ccinc_nu_FDS_nuwro_xs_sys}
\end{subfigure}

\caption{\label{fig:ccinc_FDS_nuwro}The extracted $\nu_{\mu}$CC cross section from the
$\texttt{NuWro}$ fake data [(a),(d)] as a function of neutrino energy, [(b),(e)] as a function of muon energy, and [(c),(f)] as a function of energy transfer. The statically independent $\texttt{NuWro}$ prediction is compared to the measurements using full systematic and statistical uncertainties in (a)-(c), and only cross section related systematic and statistical uncertainties in (d)-(f). The nominal MicroBooNE MC prediction is also shown.}
\end{figure*}

Though the entire suite of tests described in Sec.~\ref{sec:direct} were performed, only the test that identified the most significant mis-modeling is shown in Table~\ref{tab:fakedata_summary}. In all cases, this test identifies tension that is equal to or greater than the tension between the extracted cross sections and $\texttt{NuWro}$ truth. This is illustrated in Fig.~\ref{fig:model_validation_nuwro_xs_only}, which shows the aforementioned most sensitive model validation test for the ``XS Syst+Stat Uncertainties" study, and~Fig.~\ref{fig:ccinc_FDS_nuwro}, which shows the extracted cross section for both sets of studies. These findings are consistent with the results shown in Sec.~\ref{sec:FDS_proton} and are consistent with the point that, in general, one may design a data-driven model validation procedure that is more sensitive to the mis-modeling than the extracted cross sections.

For the $\texttt{NuWro}$ study with the full systematic uncertainties, the nominal MicroBooNE model passes validation. Similarly, the comparison between the extracted cross sections and the $\texttt{NuWro}$ predictions yields $\chi^{2}/ndf$ values of 4.0/10, 2.7/11, and 3.0/8 for $E_{\nu}$, $E_{\mu}$ and $\nu$, respectively, indicating that minimal bias was induced by the unfolding. This can be seen in Figs.~\ref{fig:ccinc_enu_FDS_nuwro_all_sys},~\ref{fig:ccinc_emu_FDS_nuwro_all_sys}~and~\ref{fig:ccinc_nu_FDS_nuwro_all_sys}. In these figures, the extracted cross sections are compared to predictions from the $\texttt{NuWro}$ truth and the nominal MicroBooNE MC. Though both predictions show good agreement with the extracted results, $\texttt{NuWro}$ is slightly favored in all cases.

\begin{figure}[!th]
\centering
\begin{subfigure}{0.49\linewidth}
\includegraphics[width=\linewidth]{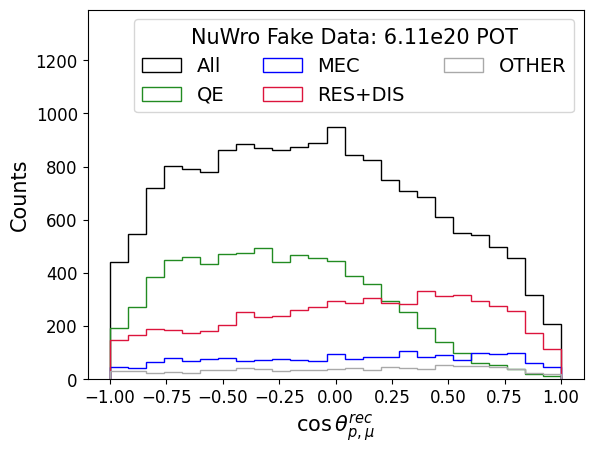}
\put(-16,60){(a)}
\end{subfigure}
\begin{subfigure}{0.49\linewidth}
\includegraphics[width=\linewidth]{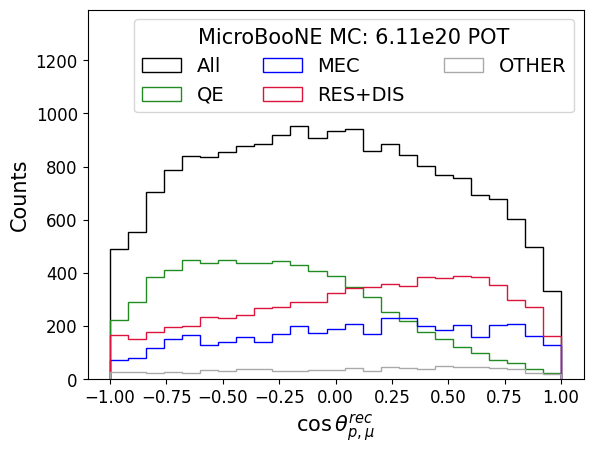}
\put(-16,60){(b)}
\end{subfigure}
\caption{The (a) $\texttt{NuWro}$ and (b) MicroBooNE MC $\cos\theta_{\mu,p}^\mathrm{rec}$ predictions for events passing the semi-inclusive $\nu_mu$CCNp selection broken down by interaction mode.}\label{fig:p_mu}
\end{figure}

\begin{figure}[h!]
\centering
\includegraphics[width=\linewidth]{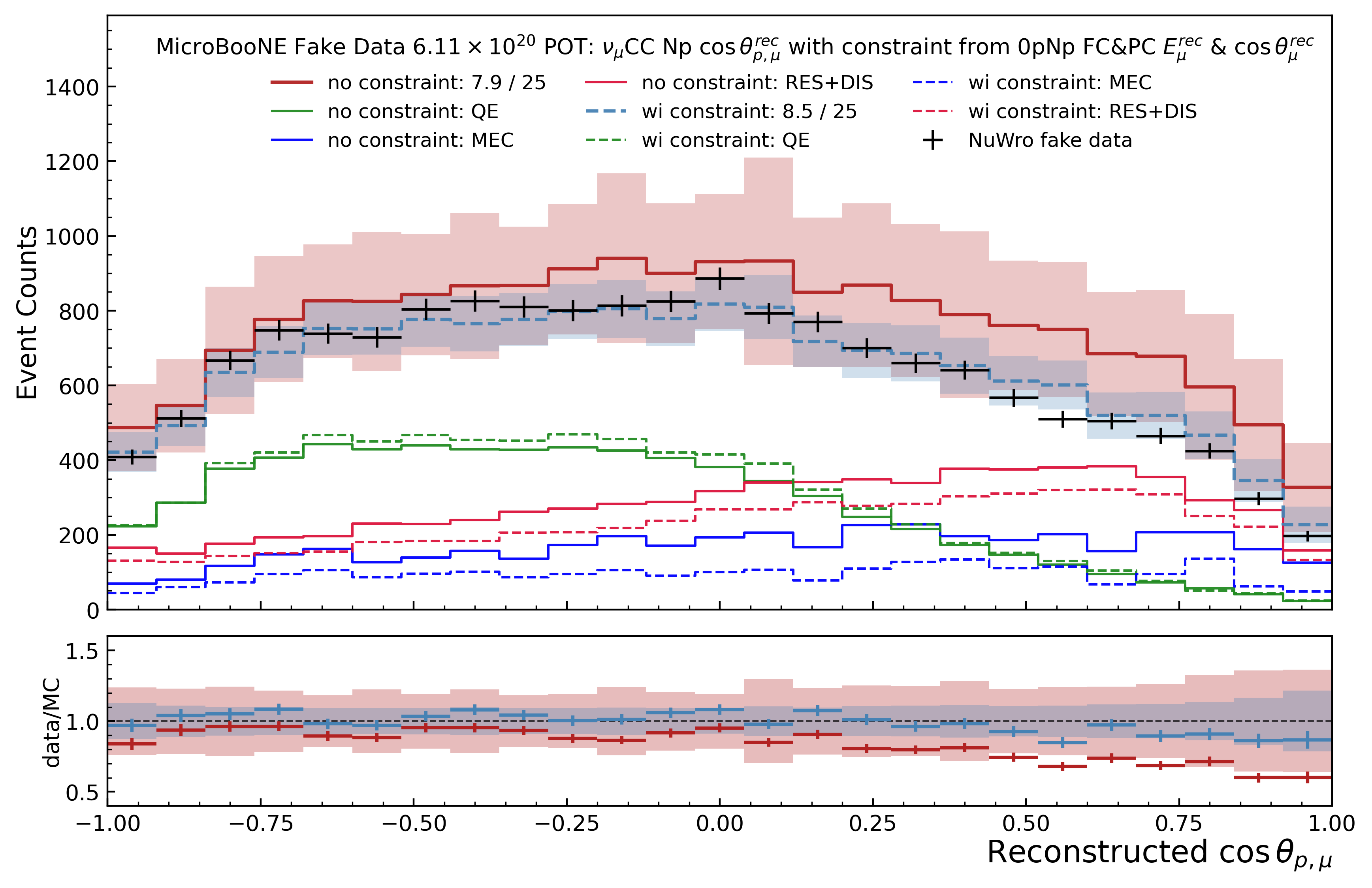}
\caption{\label{fig:const_p_mu} Model validation test comparing the $\texttt{NuWro}$ fake data sample to the nominal MC prediction for the reconstructed muon-proton opening angle distribution formed from events passing the semi-inclusive $\nu_\mu$CCNp selection. The red (blue) lines and bands show the prediction without (with) the constraint from muon kinematics for the distribution formed from events passing the $\nu_\mu$CCNp selection and the distribution formed from events passing the $\nu_\mu$CC0p selection. The uncertainties of the prediction include all systematic and statistical terms. These are shown in the bands on the prediction. The statistical uncertainties on the fake data are shown on the data points.}

\end{figure}

\begin{table*}
\begin{tabular}{||c | c@{\hskip 6mm} c@{\hskip 6mm} c ||}
\hline
& \texttt{NuWro} Fake Data & MicroBooNE MC & MicroBooNE MC with constraint \\
\hline
QE & 0.449 & 0.366$\pm$0.058 & 0.456$\pm$0.025 \\
MEC & 0.116 & 0.218$\pm$0.104 & 0.145$\pm$0.032 \\
RES+DIS & 0.375 & 0.367$\pm$0.093 & 0.352$\pm$0.035 \\
\hline
\end{tabular}
\caption{\label{table:int_mode_frac} Fractional contribution of $\nu_\mu$CC events from different interaction modes to the total $\nu_\mu$CC Np selection. The remaining $\sim$5\% of events are mostly background non-$\nu_\mu$CC events that pass the selection.}
\end{table*}

For the study with only the cross section systematic uncertainties, the nominal MicroBooNE model does not pass validation. The aforementioned model validation test that demostrates the most sensitivity to mis-modeling shows this explicitly in Fig.~\ref{fig:model_validation_nuwro_xs_only}. In this test, the FC\&PC $\cos\theta_\mu^\mathrm{rec}$ distribution is constrained by the FC\&PC $E_{\mu}^\mathrm{rec}$ distribution. 
Shape differences in the forward scattering bins are present even after constraint, and the decomposition of the $\chi^2$ reveals tension at the 3.9$\sigma$ significance level. Given the choice of 2$\sigma$ significance stringency for the model validation, this level of tension would prompt further investigation if it were encountered in real data and an alternative central value or expanded uncertainty budget would be implemented before unfolding. For the purposes of this study, the cross sections were extracted with the nominal model despite the tension identified in the validation. These results can be seen in Figs.~\ref{fig:ccinc_enu_FDS_nuwro_xs_sys},~\ref{fig:ccinc_emu_FDS_nuwro_xs_sys}~and~\ref{fig:ccinc_nu_FDS_nuwro_xs_sys}. The comparison between these extracted cross sections and $\texttt{NuWro}$ predictions yields $\chi^{2}/ndf$ values of 23.2/10, 5.2/11, and 12.7/8 for $E_{\nu}$, $E_{\mu}$, and $\nu$, respectively. The corresponding significance of this bias is 2.6, 0.1, and 1.5, indicating that a moderate amount of bias was induced in the extraction of $\sigma(E_\nu)$ and, to a lesser extent $d\sigma/d\nu$. However, the significance of this bias is still less than the tension identified in the model validation procedure, which reaches the 3.9 significance level. This observation is again consistent with the previous result that the model validation is, in general, more sensitive than the extraction of cross sections in any of these variables. Furthermore, we note that the $\texttt{NuWro}$ truth is noticeably favored by the data for all three results and the large $\chi^2$ values obtained for the MicroBooNE MC prediction indicate the stringency of this FDS.


\subsubsection{Further exploration of the conditional constraint}
In what follows, we explore the potential utility of conditional constraints beyond their role in model validation for cross section extraction. To do so, we further investigate the behavior of the constraint from the muon kinematics by examining its impact on the distribution of the reconstructed muon-proton opening angle, $\theta_{\mu,p}^\mathrm{rec}$, for a semi-inclusive selection of $\nu_\mu$CC events with protons ($\nu_\mu$CC Np). This selection is identical to the $\nu_\mu$CC Np selection described in Ref.~\cite{numuCC0pNp_PRD} and consists of all events passing the inclusive $\nu_\mu$CC selection that also have a primary reconstructed proton with greater than 35~MeV of kinetic energy. The $\theta_{\mu,p}^\mathrm{rec}$ distribution is particularity interesting because it is sensitive to the breakdown of different interaction modes, with QE interactions dominating at negative $\cos\theta_{\mu,p}^\mathrm{rec}$ and MEC and RES+DIS events more prominent at positive $\cos\theta_{\mu,p}^\mathrm{rec}$. Figure~\ref{fig:p_mu}, which displays the $\texttt{NuWro}$ and MicroBooNE MC $\cos\theta_{\mu,p}^\mathrm{rec}$ predictions broken down by interaction mode, illustrates this separation. 

Figure~\ref{fig:p_mu} illustrates that there is a noticeable difference in the strength of the MEC predictions from $\texttt{NuWro}$ and the MicroBooNE MC. In addition, the QE prediction from the MC is somewhat larger than $\texttt{NuWro}$, while the RES+DIS contributions are similar. This is quantified in Table.~\ref{table:int_mode_frac}, which contains the fraction contribute to the total selection for $\nu_\mu$CC events from each interaction mode. These differences are also apparent when the $\texttt{NuWro}$ fake data $\cos\theta_{\mu,p}$ distribution is compared to the MicroBooNE MC prediction with all its associated uncertainties. This is shown in Fig.~\ref{fig:const_p_mu}. Even with the full uncertainty band, the more MEC and RES+DIS rich region ($\cos\theta_{\mu,p}^\mathrm{rec}>0$) is overpredicted to the point that the $\texttt{NuWro}$ fake data is not completely covered by the uncertainties on prediction. The fake data is also overpredicted in the QE rich region ($\cos\theta_{\mu,p}^\mathrm{rec}<0$), but to a lesser extent and still falls within the uncertainties of the MC.

To further examine the data to MC differences in this observable, we expand the typical procedure of applying the constraint to the total prediction by also appling it to the prediction for each individual interaction mode. The constraint utilizes the muon kinematics distributions of the FC\&PC $\nu_\mu$CCNp selection and the FC\&PC $\nu_\mu$CC0p selection, which consists of the subset of events in the inclusive $\nu_\mu$CC selection without a final state proton or with a final state proton below the 35~MeV threshold. The latter selection is likewise equivalent to the $\nu_\mu$CC0p selection described in Ref.~\cite{numuCC0pNp_PRD}. This application of the constraint to individual interaction modes rather than the prediction as a whole is only done in this study for the purposes of interpreting the action of the constraint, this procedure is not used in any of the model validation described throughout the rest of this work. As expected based on the difference in size of the MEC contribution between the two generators, the constraint produces a noticeable decrease in the MicroBooNE MC MEC prediction. A decrease in the RES+DIS prediction is also observed. Nevertheless, the fake data still falls at the lower edge of the resulting total posterior prediction's uncertainties for $\cos\theta_{\mu,p}^\mathrm{rec}>0.5$ where the contribution from QE interaction modes is small and MEC and RES+DIS interaction modes dictate the strength of the prediction. For $\cos\theta_{\mu,p}^\mathrm{rec}<0.5$, a small increase in the QE prediction offsets the decrease in the other interaction modes. This causes the total prediction to decrease slightly such that it agrees with the fake data within uncertainties across the QE dominated region of phase space.

Quantitatively, the posterior QE prediction corresponds to 0.456$\pm$0.025 of the selection. This is in good agreement with the NuWro fake data, for which 0.449 of the selection is from QE $\nu_\mu$CC interactions. This indicates that the constraint has utilized the information provided by the muon kinematics, with the only hadronic information provided by splitting the distribution based on the presence of a $>35$ MeV kinetic energy proton, to increase the QE such that it falls in line with the $\texttt{NuWro}$ fake data. Similarly, the prediction for the fraction of MEC events decreases from 0.218$\pm$0.104 to 0.145$\pm$0.032 and it also agrees with the fake data for which the fraction of MEC events is 0.116. However, despite the initially similar RES+DIS agreement, the constraint reduces this component such that the posterior prediction of 0.352$\pm$0.035 is further from the $\texttt{NuWro}$ value of 0.375, but is nevertheless still within uncertainties. These metrics are summarized in Table.~\ref{table:int_mode_frac}.

This study demonstrates how the model validation and the conditional constraint can be used to probe the modeling of different interaction modes and their composition. The constraint greatly reduces the uncertainties on the rates ($> 50\%$) for each mode and tends to brings their CVs closer to the truth from the fake data. More generally, this study demonstrates how these methods allow one to further examine where the data lies within the allowed model space with significantly more sensitivity. The versatility of these methods means they can be utilized to provide additional insight about the models in such a way to be complementary to the actual cross section extraction. This opens up additional avenues of investigation for future analyses. For example, one could use the conditional constraint to fit various free parameters within a given model more directly. This can be easily achieved by denoting the free parameter as the constrained variable and using the distribution of a given observable as the constraining channel. From then, the rest of the procedure follows from Eqs.~\ref{eq:constraint}~and~\ref{eq:constraint_cov}, which updates the free parameter to its posterior prediction based on an assumed prior and the given model. However, we note that interpreting these results in this way is by construction, model dependent and care must be taken to do so across different models, given that there is likely a complex interplay between implementation of these free parameters and other parameters of that model. Nevertheless, this can be quite enlightening and also complementary to the extracted cross section which, though potentially not providing as direct of insight, describes a well-defined physics quantity and will be interpretable in the context of any other model prediction.

\section{Summary}\label{sec:summary}
Neutrino-nucleus cross section measurements are motivated by the desire to improve interaction modeling to meet the precision needs of modern neutrino experiments. More robust interaction models are needed for these experiments to reach their desired level of sensitivity in measurements of oscillation parameters and searches for physics beyond the Standard Model. The difficulties associated with modeling neutrino-nucleus interactions at this level of precision result in the reliance on event generators using effective models in both neutrino oscillation experiments and cross section measurements.  

The heavy reliance on event generators makes model validation important when extracting neutrino-nucleus interaction cross sections. To this end, we have presented a set of data-driven validation procedures and demonstrated that they can be powerful for detecting deficiencies in the models used for cross-section extraction. This validation utilizes techniques based on goodness-of-fit tests and the conditional constraint procedure to determine if the overall model can describe the data within uncertainties in the phase space relevant for the unfolding. Since this approach is based upon real data, it is well-grounded to appropriately validate the unfolding model and to evaluate the need for additional uncertainties. If a carefully selected set of data-driven tests are passed by the model, this builds confidence that any bias introduced in the cross section extraction will be within the quoted uncertainties of the measurement. This applies to both measurements of visible kinematic variables, such as the outgoing muon kinematics, and to derived quantities, such as the energy transferred to the nucleus. We demonstrate the efficacy of these data-driven methods with fake data studies aimed at comparing the sensitivity of the validation to the amount of bias introduced in the cross section extraction, in which we find that the validation is able to detect mis-modeling before it impacts the cross section extraction. These data-driven methods based upon the conditional constraint are highly versatility and can be utilized to provide additional insight that complements the actual cross section extraction.

These validations are particularly important in the case of extracting nominal flux-averaged cross sections, which we advocate for due to the additional challenges associated with flux uncertainties created for the future analyzers of the data when extracting cross sections in the real flux. Producing robust cross section measurements with proper treatment of uncertainties is essential to tuning efforts and event generator improvements, which will enable the desired precision in neutrino experiments in the near future. Utilizing data-driven model validation to extract nominal flux-averaged cross sections represents a reliable strategy to achieve this goal. 

\appendix
\section{Fake Data Model Validation Test}
\label{test_list}
In this appendix, we present the complete list of model validation tests used in the fake data studies presented in Sec.~\ref{sec:direct}.
\begin{itemize}
\item Evaluation of the $E_\mu^{\mathrm{rec}}$ FC, PC and FC\&PC distributions through overall $\chi^2$ GoF tests and $\chi^2$ decompositions (6 total tests).
\item Evaluation of the $E_\mu^{\mathrm{rec}}$ PC distribution after constraint from the analogous FC distribution. The overall $\chi^2$ GoF test and $\chi^2$ decomposition are examined (2 total tests).
\item Evaluation of the $\cos\theta_\mu^{\mathrm{rec}}$ FC, PC and FC\&PC distributions through overall $\chi^2$ GoF tests and $\chi^2$ decompositions (6 total tests).
\item Evaluation of the $\cos\theta_\mu^{\mathrm{rec}}$ PC distribution after constraint from the analogous FC distribution. The overall $\chi^2$ GoF test and $\chi^2$ decomposition are examined (2 total tests).
\item Evaluation of the $\cos\theta_\mu^{\mathrm{rec}}$ FC, PC and FC\&PC distributions after constraint from the FC\&PC $E_\mu^{\mathrm{rec}}$ distribution. The overall $\chi^2$ GoF test and $\chi^2$ decomposition is examined for each distribution (6 total tests).
\item Evaluation of the $E_\nu^{\mathrm{rec}}$ FC, PC and FC\&PC distributions through overall $\chi^2$ GoF tests and $\chi^2$ decompositions (6 total tests).
\item Evaluation of the $E_\nu^{\mathrm{rec}}$ PC distribution after constraint from the analogous FC distribution. The overall $\chi^2$ GoF test and $\chi^2$ decomposition are examined (2 total tests).
\item Evaluation of the $E_\nu^{\mathrm{rec}}$ FC, PC and FC\&PC distributions after constraint from the FC\&PC $E_\mu^{\mathrm{rec}}$ and $\cos\theta_\mu^{rec}$ distributions. The overall $\chi^2$ GoF test and $\chi^2$ decomposition is examined for each distribution (6 total tests).
\item Evaluation of the $E_{\mathrm{had}}^{\mathrm{rec}}$ FC, PC and FC\&PC distributions through overall $\chi^2$ GoF tests and $\chi^2$ decompositions (6 total tests).
\item Evaluation of the $E_{\mathrm{had}}^{\mathrm{rec}}$ PC distribution after constraint from the analogous FC distribution. The overall $\chi^2$ GoF test and $\chi^2$ decomposition are examined (2 total tests).
\item Evaluation of the $E_{\mathrm{had}}^{\mathrm{rec}}$ FC, PC and FC\&PC distributions after constraint from the FC\&PC $E_\mu^{\mathrm{rec}}$ and $\cos\theta_\mu^{\mathrm{rec}}$ distributions. The overall $\chi^2$ GoF test and $\chi^2$ decomposition is examined for each distribution (6 total tests).

\end{itemize}

\begin{acknowledgments}
\vspace*{-4mm}
This document was prepared by the MicroBooNE collaboration using there sources of the Fermi National Accelerator Laboratory (Fermilab), a U.S. Department of Energy, Office of Science, HEP User Facility. Fermilab is managed by Fermi Research Alliance, LLC (FRA), acting under Contract No. DE-AC02-07CH11359.  MicroBooNE is supported by the following: the U.S. Department of Energy, Office of Science, Offices of High Energy Physics and Nuclear Physics; the U.S. National Science Foundation; the Swiss National Science Foundation; the Science and Technology Facilities Council (STFC), part of the United Kingdom Research and Innovation; the Royal Society (United Kingdom); the UK Research and Innovation (UKRI) Future Leaders Fellowship; and the NSF AI Institute for Artificial Intelligence and Fundamental Interactions. Additional support for the laser calibration system and cosmic ray tagger was provided by the Albert Einstein Center for Fundamental Physics, Bern, Switzerland. We also acknowledge the contributions of technical and scientific staff to the design, construction, and operation of the MicroBooNE detector as well as the contributions of past collaborators to the development of MicroBooNE analyses, without whom this work would not have been possible. For the purpose of open access, the authors have applied a Creative Commons Attribution (CC BY) public copyright license to any Author Accepted Manuscript version arising from this submission.

\end{acknowledgments}
\typeout{}
\bibliography{wcresp}

\end{document}